\documentclass[traditabstract]{aa} 

\usepackage{graphicx}
\usepackage{txfonts}
\usepackage{natbib}
\bibpunct{(}{)}{;}{a}{}{,} 
\usepackage{multirow}
\usepackage{subfigure}

\begin{document}

\title{\textsc{Long-Term Evolution of Three Light Bridges Developed on the Same Sunspot}}

\author{A. B. Griñón-Marín\inst{1,2,3}, 
        A. Pastor Yabar\inst{4},
        R. Centeno\inst{5},
        \and
        H. Socas-Navarro\inst{1,2}}

\institute{Instituto de Astrof\'{\i}sica de Canarias,
           V\'{\i}a L\'actea, 38205 La Laguna, Tenerife, Spain \\
         \and
           Universidad de La Laguna, Departamento de Astrof\'{\i}sica, 
              38206 La Laguna, Tenerife, Spain \\
         \and
           W. W. Hansen Experimental Physics Laboratory, 
           Stanford University, Stanford, 
           CA 94305-4085, USA \\
         \and
           Institute for Solar Physics, Department of Astronomy, Stockholm University,
           Albanova University Centre, SE-106 91 Stockholm, Sweden \\
         \and
           High Altitude Observatory (NCAR),
           3080 Center Green Dr. Boulder CO 80301 \\}

\date{February 2021}

\abstract
{One important feature of sunspots is the presence of light bridges. These structures are elongated and bright (as compared to the umbra) features that seem to be related to the formation and evolution of sunspots. In this work, we studied the long-term evolution and the stratification of different atmospheric parameters of three light bridges formed in the same host sunspot by different mechanisms. To accomplish this, we used data taken with the GREGOR Infrared Spectrograph installed at the GREGOR telescope. These data were inverted to infer the physical parameters of the atmosphere where the observed spectral profiles were formed of the three light bridges. We find that, in general, the behaviour of the three light bridges is typical of this kind of structure with the magnetic field strength, inclination, and temperature values between the values at the umbra and the penumbra. We also find that they are of a significantly non-magnetic character (particularly at the axis of the light bridges) as it is deduced from the filling factor. In addition, within the common behaviour of the physical properties of light bridges, we observe that each one exhibits a particular behaviour. Another interesting result is that the light bridge cools down, the magnetic field decreases, and the magnetic field lines get more inclined higher in the atmosphere. Finally, we studied the magnetic and non-magnetic line-of-sight velocities of the light bridges. The former shows that the magnetic component is at rest and, interestingly, its variation with optical depth shows a bi-modal behaviour. For the line-of-sight velocity of the non-magnetic component, we see that the core of the light bridge is at rest or with shallow upflows and clear downflows sinking through the edges.}

\titlerunning{short title}
\authorrunning{name(s) of author(s)}
\maketitle

\section{Introduction}
\label{intro}
In recent years, our understanding of the formation and evolution of sunspots has improved thanks to, among other things, the advent of spectropolarimetric observations and/or high temporal, spatial, and/or spectral resolution. Light bridges (LBs; \citealt{brayley1869}) play an important role in our understanding of the evolutionary stages of sunspots. LBs are irregular, bright, elongated structures that are seen in umbra during the formation and/or decay of sunspots or pores (\citealt{sobotka2003}, \citealt{thomas2004}). They can indicate the break-up of sunspots in the decay or the formation phases of complex active regions, as if they were `seams' where a sunspot forms or decays \citep{garciaDeLaRosa1987}. Their lifetimes are shorter than those of the sunspots that host them, and they are very dynamic. LBs have different shapes and sizes, varying from less than $0\farcs5$ to several seconds of arc in width (\citealt{berger2003}; \citealt{sobotka2013}; \citealt{toriumi2015a}). 
 
These structures may be categorised according to geometrical shape, brightness, or the magnetic polarity of surrounding umbral cores. They may show filamentary or granular shapes, depending on their physical properties. They may develop as an intrusion of a penumbral filament into the umbra \citep{katsukawa2007}, or they may be of a granular nature (\citealt{lagg2014}; \citealt{falco2017}). The latter type are present during the formation of a sunspot and often reappear during the decaying phase of sunspots \citep{vazquez1973}, as though LBs preserved their identity throughout the life of the sunspot.
 
Light bridge classification schemes have varied over time (\citealt{korobova1966}, \citealt{muller1979}, \citealt{sobotka1997}, \citealt{thomas2004}, \citealt{lagg2014}), but the most complete one is probably Sobotka's classification. This is based on i) the morphology related to the sunspot configuration (if the LB splits umbral cores it is called strong, and if it is an intrusion into the umbra it is called faint), and ii) the internal structure (the LB can be granular or filamentary). 

Some decades later, \cite{schlichenmaier2016} found a different kind of LB not represented in Sobotka's classification system. This new type is too thick to be called filamentary and does not present granules. This type of LB is called a `plateau light bridge'. Also, \cite{Kleint2013} studied some filamentary structures observed within the umbra of an active region that are characterised by a horizontal flow opposite to the Evershed flow (\citealt{evershed1909}) along them at the photospheric level. Because these structures do not have the same properties as penumbral filaments or LBs, they called them umbral filaments (UFs).
 
Currently, there is no established consensus on how LBs are formed, but some authors have studied possible causes. \cite{vazquez1973} proposed a first interpretation of photospheric LBs and reported that they were produced by sunspot decay preceding the restoration of the granular surface. In another scenario, \cite{katsukawa2007} observed a mature sunspot and found that many umbral dots appear in the umbra of the sunspot to form an LB that cuts across the umbra. \cite{toriumi2015b} proposed that LBs of convective origin are formed due to the emergence of the horizontal magnetic field inside an umbra.
 
One physical property that makes these structures different from the umbra is the magnetic field, which is weaker and more horizontal in LBs than in their surrounding umbra (\citealt{beckers1969}; \citealt{lites1991}; \citealt{leka1997}; \citealt{jurcak2006}; \citealt{louis2015}). \cite{leka1997} proposed the presence of a magnetic canopy structure above LBs by carrying out the first systematic study analysing 11 LBs with full spectropolarimetric data. Later, \cite{jurcak2006} confirmed this model by giving the stratification of plasma parameters with optical depth. They suggested that the field-free plasma intrudes vertically into the umbra from below the photosphere and forms a magnetic field canopy configuration (see Figure 7 from \citealt{jurcak2006}). This was the first study to analyse the atmospheric parameter stratifications of two LBs. They found that the magnetic field strength is weaker in the deep layers of LBs, and that the inclination of the magnetic field lines decreases with optical depth, acquiring similar values of the surrounding umbra at greater optical depths. Finally, \cite{toriumi2015b} supported their interpretation using the numerical results of the radiative magnetohydrodynamics simulation of a large-scale flux emergence.
 
Regarding the LB velocity structure, many studies have reported blueshifts (\citealt{beckers1969}; \citealt{hirzberger2002}; \citealt{schleicher2003}; \citealt{katsukawa2007}; \citealt{rimmele2008}) with respect to the umbral velocities, others have measured redshifts (\citealt{ruedi1995}; \citealt{shimizu2011}), and yet others a combination of both blueshifts and redshifts (\citealt{leka1997}; \citealt{rimmele1997}; \citealt{bharti2007}; \citealt{giordano2008}; \citealt{rouppe2010}; \citealt{toriumi2015a}; \citealt{guglielmino2017}; all these papers reported upflows in the middle part of the LB and downflows at the edge). \cite{louis2009} reported, for the first time, supersonic downflows with values of up to 10~km/s in the photosphere. They analysed complex Stokes $V$ spectra with double red lobes with two-component inversions. Furthermore, \cite{lagg2014} measured supersonic downflows at the edges of an LB granule with subsonic upflows in the middle of it. \cite{rimmele1997} found a correlation between the brightness and upflow velocities in strong granular LBs, considering this evidence of the magnetoconvective origin of this kind of LB. 
 
One feature discovered in recent decades related to LBs is the presence of dark lanes running parallel to their axis. \cite{lites2004} found that the optical depth of this lane is in the range of 200--450~km above the horizontal plane defined by the umbral floor. Moreover, \cite{berger2003} measured the width of a dark lane and obtained a value of 380~km. \cite{falco2016} calculated the main axis of the LB to have an average width of $\approx$225~km and a length of about 6000~km. These dark lanes can be explained by the same principle that produces dark cores in bright penumbral filaments (\citealt{heinemann2007}; \citealt{ruiz2008}), and they are a common feature of strong LBs \citep{berger2003}. \cite{giordano2008} and \cite{rouppe2010} measured upflow velocities surrounded by downflows along the central dark lane. In the case of Plateau LBs, the dark lane has a small, transverse, Y-shaped dark lane similar to dark-cored penumbral filaments \citep{schlichenmaier2016}. This feature is not seen in granular LBs.

In May 2014, four main LBs were formed in the leading sunspot of the NOAA active region 12049. We took advantage of one of the new 1.5-m class solar telescopes to address, for the first time, a temporal evolution (the timescale of the life of a sunspot, as some previous studies had already performed such a temporal study of an LB but on scale of hours, i.e. \citealt{shimizu2011}) of some atmospheric parameters, both in space and the optical depth stratification of three of them, using spectropolarimetric data.

\section{Observations and methodology}
\label{observations}

\subsection{Observations}
\begin{figure}
  \begin{center}
    \includegraphics[width=0.5\textwidth]{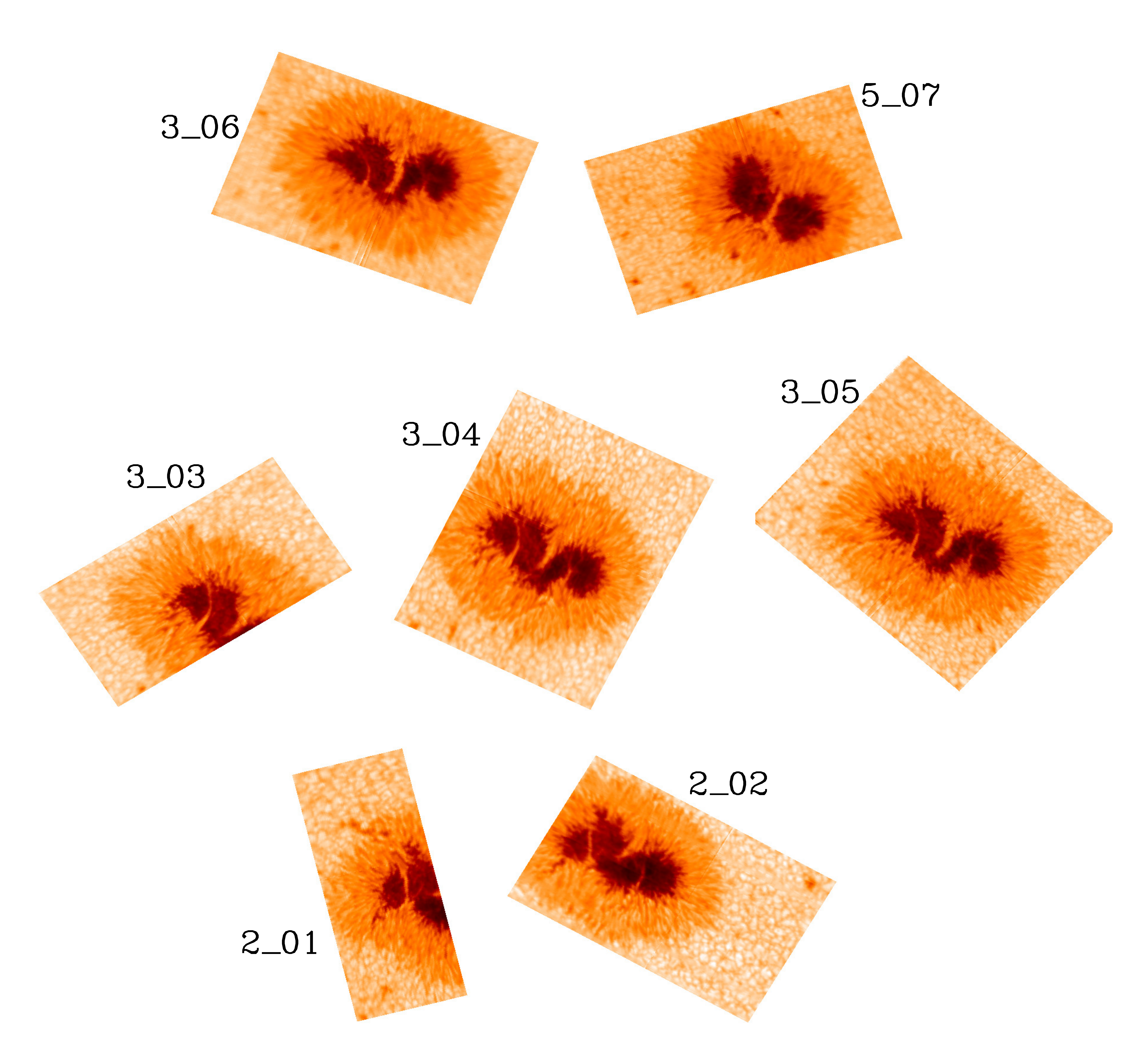}
    \caption{Data observed by GRIS spectrograph. The details of these observations are summarised in Tab. \ref{Tab:observationsInfo}. The labels indicate the observation day and the number of the scan of the third column of Tab. \ref{Tab:observationsInfo} (observationDay\_scan).}
    \label{Fig:grisSunspots}
  \end{center}
\end{figure}

We present an analysis of three LBs observed in the leading sunspot of NOAA AR12049. This sunspot was observed from 2014 May 2 to 2014 May 5. The data were taken using the GREGOR Infrared Spectrograph (GRIS; \citealt{collados2012}) installed on the GREGOR (\citealt{volkmer2010}; \citealt{schmidt2012}; \citealt{soltau2012}) solar telescope at Observatorio del Teide (Spain). The observed spectral range was centred on the \ion{Fe}{I} 15648 \AA\ line with a spectral sampling of approximately 40~m$\AA$. We used an exposure time of 30 ms per accumulation with a varying number of repetitions. The scanning step size and the image scale along the slit were 0\farcs135. These observational parameters are summarised in Tab. \ref{Tab:observationsInfo}. The telescope derotator was not installed during this observing run, so each map was taken with a different orientation. Additionally, the image (slightly) rotates from the beginning to the end of the scan. Since the time needed for the acquisition of the maps were short and no data were taken when the image rotated the most, the rotation from the beginning until the end of the scan was kept below a few degrees (see last column of Tab. \ref{Tab:observationsInfo}). Figure \ref{Fig:grisSunspots} shows the continuum intensity maps of all the scans analysed. In this figure, all the scans are orientated such that north is at the top of the page, using the apparent X and Y solar coordinates calculated by Dr. M. Franz,\footnote{http://archive.leibniz-kis.de/pub/gris/yyymmdd/context\_data/ where yyyy is year, mm is the month, and dd is the day of the observation.} who aligned them using Helioseismic and Magnetic Imager (HMI; \citealt{scherrer2012}, \citealt{schou2012}) data.
 
\begin{table*}[h]
  \caption{Details of the observations taken with the GRIS spectrograph.}
  \label{Tab:observationsInfo}
  \centering
\begin{tabular}{ccccccccc}
\hline
\hline
Date & Hour (UT) & Scan \# & Helio. Angle & Accum. & Area ($arcsec^{2}$) & Max. Rot. Ang. (deg) \\
\hline
\hline
2014 May 2& 11:35 & 2\_01 & 12.365 & 10 & 27.00 x 63.45 & 5.34 \\
2014 May 2& 15:05 & 2\_02 & 9.701 & 10 & 40.50 x 63.32 & 2.83 \\
2014 May 3& 09:59 & 3\_03 & 4.620 &  5 & 33.75 x 63.45 & 0.54 \\
2014 May 3& 10:49 & 3\_04 & 5.431 &  5 & 50.09 x 63.45 & 3.13 \\
2014 May 3& 14:05 & 3\_05 & 5.957 &  5 & 54.00 x 63.72 & 7.18 \\
2014 May 3& 17:41 & 3\_06 & 8.085 &  5 & 43.20 x 63.32 & 0.43 \\
2014 May 5& 09:07 & 5\_07 & 29.05 & 10 & 40.50 x 63.59 & 0.11 \\
\hline
    \end{tabular}
    \tablefoot{The rows represent the different observed maps. The first two columns correspond to the date and time of the observations, the third column is the number of the scan (hereinafter, this is the notation used to refer to each scan), the fourth is the heliocentric angle, the fifth and sixth columns correspond to the number of accumulations and the number of the steps used for each observation, respectively, and the last column corresponds to the rotation angle between the first and last slit positions during the scan.}
\end{table*}

\begin{figure*}
  \begin{center}
    \includegraphics[width=0.85\textwidth]{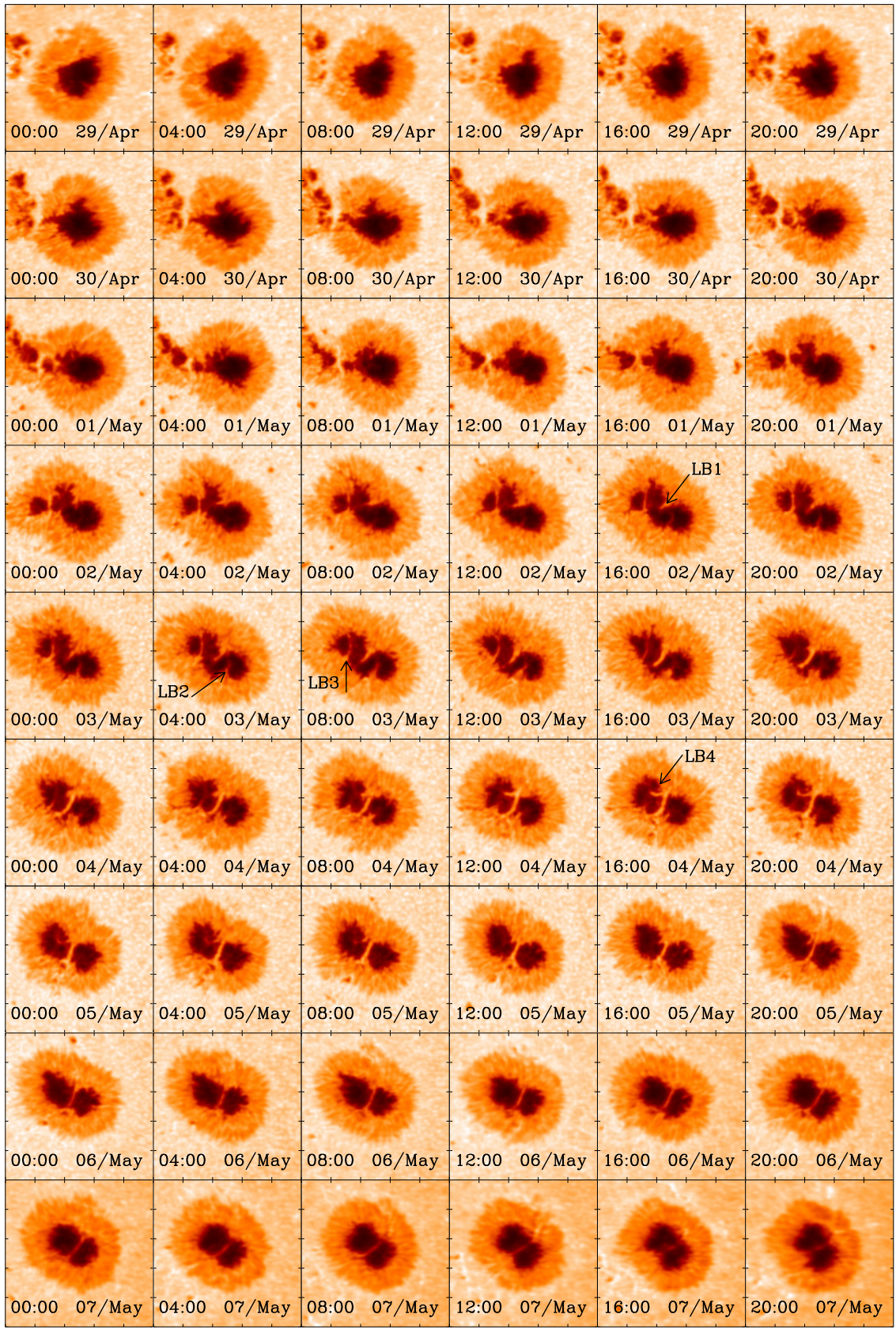}
    \caption{Temporal evolution of the intensity maps of the leading sunspot of NOAA active region 12049 observed with the HMI instrument. There are four main LBs labelled LB1, LB2, LB3, and LB4, which evolve in different ways. The first three LBs are analysed in this work.}
    \label{Fig:hmiSunspots}
  \end{center}
\end{figure*}

The time coverage of the LBs with GRIS data is rather poor. A better time coverage of the spot was achieved by the HMI instrument (which continuously records the evolution of the visible surface) on board the Solar Dynamics Observatory (SDO; \citealt{pesnell2012}) space mission. In Figure \ref{Fig:hmiSunspots}, we can see the evolution of NOAA AR12049 and its LBs as it crossed the visible solar disc. From this sequence, it seems that each LB was formed as a consequence of different processes, and develops in a different way.
 
Light  Bridge 1 was formed due to a strong penumbral protrusion into the umbra. This protrusion starts to intrude into the umbra on May 2. This structure gradually grows, splitting the umbra into two parts. This process takes around two days, until May 4. This LB is the largest and widest of all the LBs hosted by this sunspot, and it is present until the end of the available time sequence, when the spot crosses the western limb of the Sun (see Figure \ref{Fig:hmiSunspots}).
 
In contrast, the other LBs that appear in the sunspot are quite thin and their lifetimes are shorter. LB2 was a weak intrusion of the penumbra into the umbra that took place to the west of LB1. This intrusion began on May 2 and reached the opposite side of the umbra at the end of the same day; it then started to retreat until its disappearance on May 4. In contrast to LB1, this one partially dissolved shortly after splitting both umbral cores. In this process, the northern part dilutes and forms an area of `bright' umbral core, while the southern part gets wider and brighter. On May 4, this remnant of LB2 seems to merge with the southern part of LB1.
 
Light Bridge 3 formed as two umbrae approached each other. On April 29, the leading sunspot of NOAA active region 12049 had a roundish shape with a full umbra (with no LBs). Some small pores were starting to approach it from the northeast. As the pores merged with the sunspot, it lost part of its penumbra, and a bright area appeared between the spot and pores (from April 30 until May 1). As this bright area became narrower, it eventually led to the birth of LB3, which continued to get narrower as the two umbrae got closer and finally merged on May 4 (starting from the northern part of the LB and later extending to the south). The sunspot developed a fourth LB, but since it was not observed with GRIS, we did not analyse it in this study. Table \ref{Tab:descripcionLBs} summarises the properties of the three LBs that we analyse in the coming sections.
 
\begin{table}[h]
  \caption{Summary of the formation and evolution of the LBs.}
  \label{Tab:descripcionLBs}
  \centering
    \begin{tabular}{c|c}
      \hline
      \hline
      Light Bridge & Formation \& Evolution \\
      \hline    
      \hline
      \multirow{2}{*}{LB1} & Strong and broad penumbral intrusion \\
          & that breaks the umbra into two parts \\
      \hline
      \multirow{3}{*}{LB2} & Weak and thin penumbral intrusion, \\
          & which, after splitting the umbra, retreats \\
          & until its disappearance \\
      \hline          
      \multirow{3}{*}{LB3} & Close approach of two umbra until the \\
          & LB disappears from the northern part and \\
          & finally from the south \\
      \hline
    \end{tabular}
\end{table}

\subsection{Data reduction}
After recording the Stokes parameters with GRIS, the software supplied by the instrument developers was applied (Collados et al., in prep.). With this dedicated software, the dark current was subtracted, the flat-field and the bad pixels were corrected, the data were demodulated, and the instrumental cross-talk was corrected. However, after applying these routines, certain features needed to be considered before proceeding with the data analysis. 

The first step was to correct the continuum intensity trends of the data. To do so, we used the preceding and the following flat-field images. We reduced them as though they were science products, and we obtained the intensity-averaged spectra of each flat-field map. These two averaged profiles were then fitted with a polynomial of order 13, avoiding spectral lines as these may worsen the fit of the continuum. We interpolated the intensity to be corrected from the times at which the flat fields were taken to the time when each slit position of the data was recorded. Finally, we corrected the data (the four Stokes parameters), dividing each observed Stokes profile (including $Q$, $U$, and $V$) by this interpolated intensity correction. 

In the next step, we performed the wavelength calibration of the data by fitting a second-order polynomial to the core of the same lines in both the Fourier transform spectrometer atlas \citep{livingston1991} and the flat-field averaged-intensity spectra. The linear fit between these two sets results in the wavelength calibration of the observed spectral grid. 

The third step was to remove some high-frequency fringes that persist even after the flat-fielding correction, mostly in the intensity data. To do that, we applied Fourier passband filtering in the 384--417 \ m\AA\ frequency range. 

While observing, there were sudden and short episodes of very bad seeing. For these slit positions, the accuracy of the demodulation scheme was worse. This is evident in the form of strong polarimetric interference fringes in those slit scanning steps, which are more evident in Stokes $Q$, $U$, and $V$. Since these slit scanning steps strongly affect their neighbouring ones during the deconvolution process (explained in following section), we substituted these columns with an interpolated column obtained from the previous and following scan scanning steps, but they are not analysed in what follows.

\subsection{Data deconvolution}
\label{deconvolution}
When data are recorded, the optical devices in the optical path modify the image coming from the Sun. This is usually known as the telescope point spread function (PSF). In addition, for ground-based observations, the Earth's atmosphere blurs the image coming from the Sun. This is usually known as the seeing PSF and mostly determines the spatial resolution of the observations (even when using adaptative optics). In order to remove these effects, we deconvolved each map by applying a regularised principal component analysis and using a static PSF whose performance over space-borne data and methodology is described in \cite{quintero2015}.

To calculate the PSF of the optical system (for both the telescope and the Earth's atmosphere), we followed Collados's procedure (GREGOR internal report) developed with the spectropolarimetric data taken in the 1.56 $\mu$m range during the transit of Mercury on 2016 May 9. In order to get the PSF, an ideal Mercury image was convolved by a 2D Gaussian given by 

\begin{equation}
    \phi (r)=\alpha G_{1}(r,\sigma_{1})+(1-\alpha)G_{2}(r,\sigma_{2})  \, , \\
 \label{Eq:psf_profile}
\end{equation}where $G_{1}$ and $G_{2}$ are two Gaussians with the standard deviation given by $\sigma_{1}$ and $\sigma_{2}$, respectively. $G_{1}$ corresponds to the spatial resolution of the telescope (including the seeing), and $G_{2}$ accounts for the instrumental stray light. $\alpha$ is a constant and defines the relative weight of each Gaussian. The parameters $\alpha$, $\sigma_{1}$, and $\sigma_{2}$ were obtained following a non-linear fit to obtain
$\alpha = 0.710$, $\sigma_{1} = 1\farcs638$, and $\sigma_{2} = 8\farcs274$. In order to calculate the PSF profile of our observations, we used the same $\alpha$ and $\sigma_{2}$ as obtained for the Mercury transit. Regarding to obtain the $\sigma_{1}$ value, we needed to know the seeing value of our observation day. However, during the observations of the sunspot under study, there were communication problems between the GREGOR Adaptative Optics system and the instrument computer. For this reason, the data header did not record the seeing value ($r_{0}$) of our observed scans. That is why, in order to use an $r_{0}$ close to the real one, we deconvolved each scan using different seeing values to calculate the PSF and visually choose the one that did not produce any unusual artefact. Figure \ref{Fig:deconvolvedTests} shows the images of the same scan deconvolved using different seeing values to calculate the PSF profile. The number that appears in each map corresponds to the $r_{0}$ value (at 1.56 $\mu m$ of wavelength) used for the deconvolution. The deconvolved maps with $r_{0}<55.04\ $ cm seem to suffer an over-correction; thus, in order to be conservative, the $r_{0}=70.77\ $ cm ($r_{0}=18.06\ $ cm at 500 nm of wavelength) was eventually selected as the seeing value to be used to calculate the PSF shape. Once the $r_{0}$ value of the observations was estimated, and assuming that the first Gaussian ($G_{1}$) gave the angular resolution, we were able to calculate $\sigma_{1}$ following Equations \ref{Eq:angularResolution} and \ref{Eq:sigma1}:

\begin{figure}
  \centering
    \includegraphics[width=0.5\textwidth]{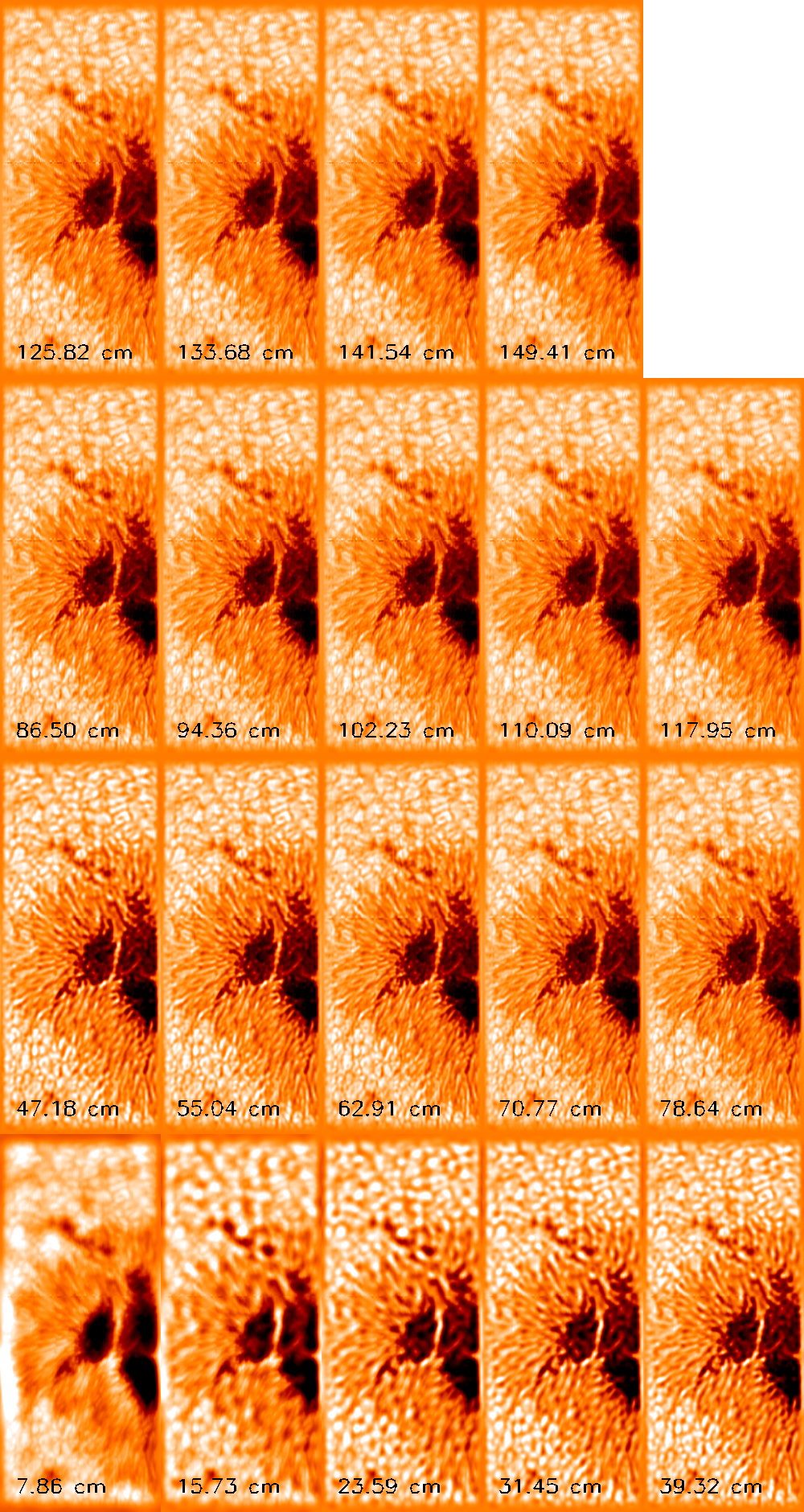}
  \caption{Deconvolved monochromatic images of the same scan for different values of seeing Gaussian (given by the lower number in each panel).}
  \label{Fig:deconvolvedTests}
\end{figure}

\begin{equation}
  \begin{array}{l}
    \theta\approx\dfrac{\lambda_{\rm obs}}{r_{0}'} \ 206265 \ \approx FWHM_{1} \, , \\
  \end{array}
  \label{Eq:angularResolution}
\end{equation}
where $\theta$ is the angular resolution and $206265$ is the unit conversion to give the result in seconds of arc. Bearing in mind that $FWHM_{1} = 2\sqrt{2ln2} \ \sigma_{1}$, we can calculate $\sigma_{1}$ as

\begin{equation}
  \begin{array}{l}
    \sigma_{1} = \dfrac{\lambda_{\rm obs}}{r_{0}'} \ \ 206265 \ \ \dfrac{1}{2\sqrt{2ln2}}  \, , \\
  \end{array}
  \label{Eq:sigma1}
\end{equation}where $\lambda_{\rm obs}$ is the central wavelength of this study ($1.5648\ \mu m$), and $r_{0}'$ is the seeing value for our spectral range ($70.77\ $ cm). This way, we obtained $\sigma_{1} = 0\farcs193672$. 

\begin{figure}
  \centering
    \subfigure{\includegraphics[width=0.4\textwidth]{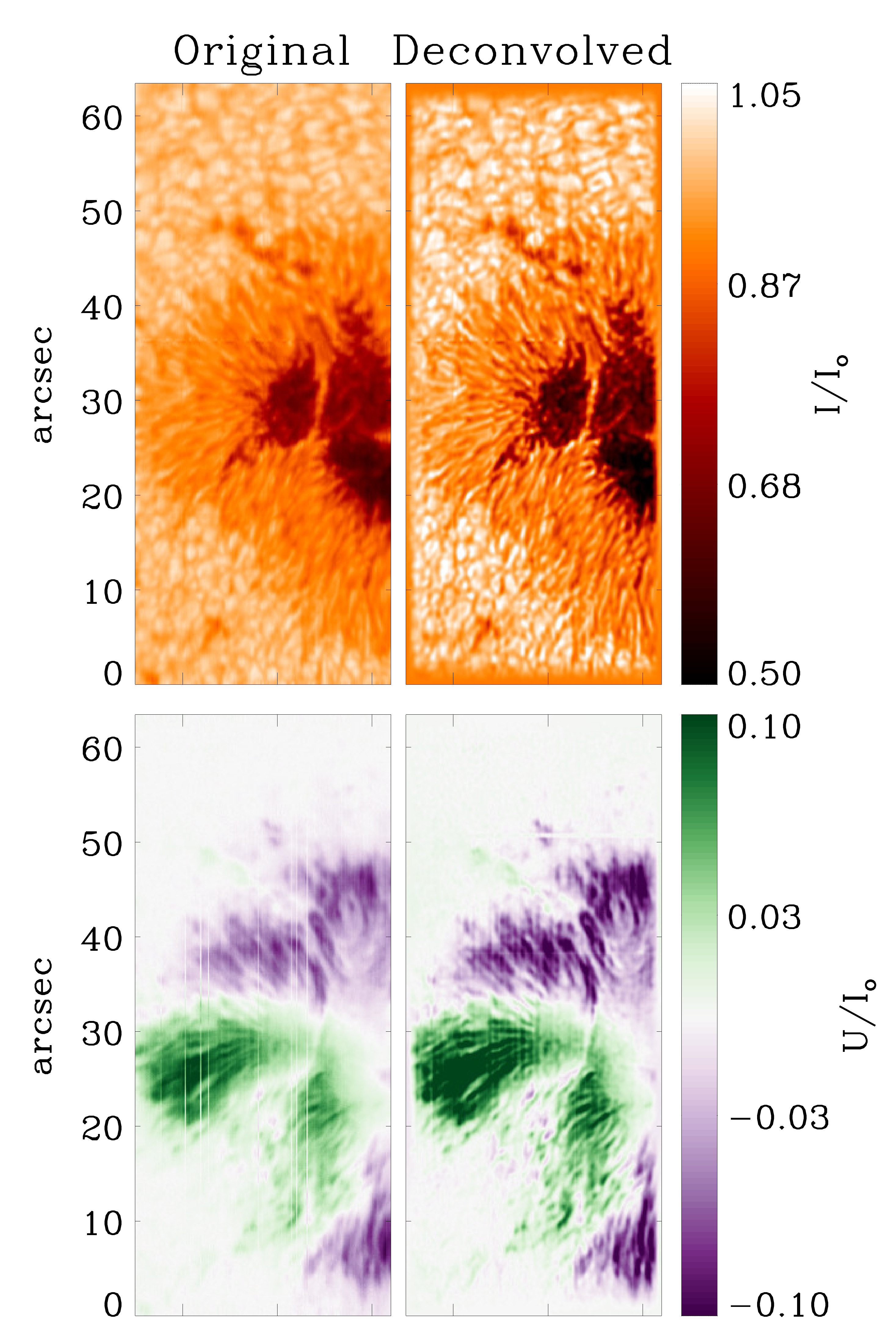}}
    \subfigure{\includegraphics[width=0.4\textwidth]{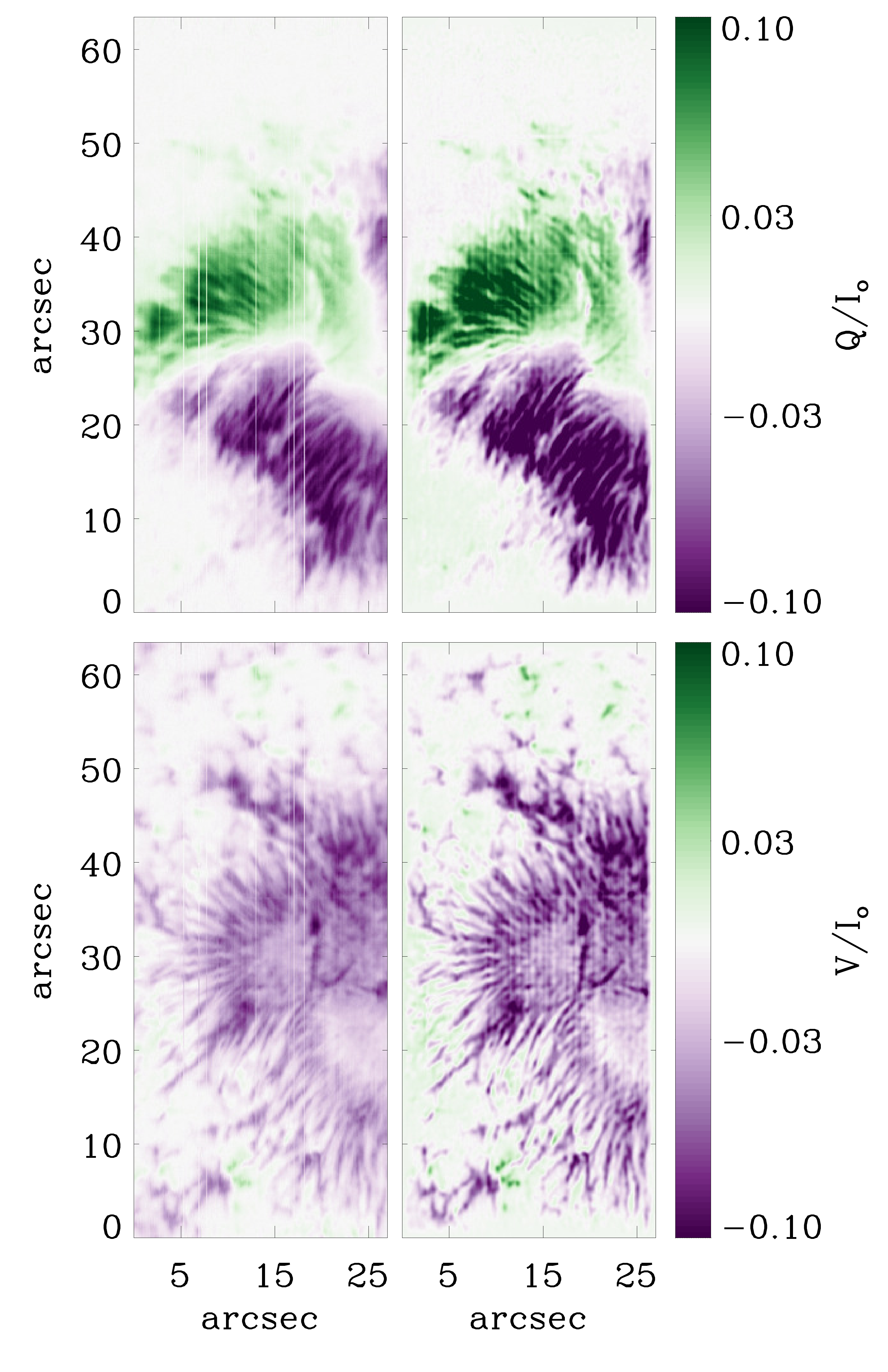}}
  \caption{Stokes parameter deconvolution. Each pair of panels is a monochromatic map of the Stokes parameter. From top to bottom: Stokes I at the continuum, Stokes Q and Stokes U both at the core of the line, and finally the Stokes V at the red wing. On the left of each pair, the interpolated original map is shown, and on the right is the same map after the deconvolution.}
  \label{Fig:deconvolvedMaps}
\end{figure}

Once $\alpha$, $\sigma_{1}$, and $\sigma_{2}$ variables were calculated, the observed spectropolarimetric maps were deconvolved using a principal component analysis regularisation \citep{ruiz2013} using a PSF given by Eq. \ref{Eq:sigma1}. Figure \ref{Fig:deconvolvedMaps} shows the comparison of the Stokes parameters before and after applying the deconvolution. 

\subsection{Inversion}
\label{sec:inversion}
When the maps are ready to use, in order to infer the magnetic and thermodynamic properties of the solar atmosphere, where the observed spectral lines were formed, we used the Stokes inversion based on response functions (SIR; \citealt{ruiz1992}) code. This code assumes local thermodynamic equilibrium (LTE) to solve the radiative transfer equation for polarised light in order to compute the synthetic Stokes profiles. SIR computes perturbations in the physical quantities at specific locations across the optical depth grid called nodes, and then carries out an interpolation to yield values at all grid points. Finally, the code compares the synthetic with the observed Stokes profiles, modifying an initial atmospheric model until the differences between the observed and synthetic profiles are minimised. 

The inversion code returns the optical depth stratifications of temperature ($T$), magnetic field strength ($|\vec{B}|$), magnetic field inclination with respect to the line of sight ($\gamma_{\rm los}$), magnetic field azimuth on the plane of the sky ($\phi_{\rm los}$),\footnote{The inclination is measured with respect to the line-of-sight, the $0^\circ$ azimuth reference is in the north solar direction, and the azimuth values increase in an anti-clockwise direction.} line-of-sight velocity ($v_{\rm los}$), micro-turbulence velocity ($v_{\rm mic}$), electronic pressure ($P_{\rm e}$), gas pressure ($P_{\rm g}$), and density ($\rho$). The last two quantities are derived from the temperature and electronic pressure stratifications after assuming hydrostatic equilibrium. Moreover, the macro-turbulence velocity ($v_{\rm mac}$), the filling factor ($ff$) of the magnetic component, and the stray light fraction ($sl$) might be provided by the code.

In the observed spectral range, there are some spectral lines that are sensitive to the magnetic field, and here we use two neutral iron spectral lines, 15648.515 and 15662.018 \AA, with respective Land\'e factors of 3.0 and 1.5. We avoided the other spectral lines in the observed spectral window either because the atomic parameters were not well characterised (i.e. the 15631.948 \AA\ spectral line) or because of contamination by molecular blends that clearly appear in lower temperature regions, mostly in umbra pixels (i.e. the 15645.016 and 15652.881 \AA\ spectral lines). The atomic parameters of the spectral lines inverted simultaneously are listed in Tab. \ref{Tab:atomicParameters}.

\begin{table*}
  \caption{Characteristics of the spectral lines inverted in this study to obtain the atmospheric parameters of the solar atmosphere where these lines are formed (\citealt{borrero2013}, \citealt{bloomfield2007}).}
  \label{Tab:atomicParameters} 
  \centering
\begin{tabular}{ccccccccccc}
\hline
\hline
Element & Ion & $\lambda\ (\AA)$ & Exc. Pot. (eV) & log(gf) & Term & $\alpha$ & $\sigma (a_{o}^{2}$) \\ 
\hline
\hline
Fe & I & 15648.515 & 5.426 & -0.669 & $^{7}D_{1} - ^{7}D_{1}^{0}$ & 0.229 & 975 \\
Fe & I & 15662.018 & 5.830 & -0.190 & $^{5}F_{5} - ^{5}F_{4}^{0}$ & 0.240 & 1197 \\
\hline
    \end{tabular}
    \tablefoot{The first and second columns are the atomic element and the ionisation state of the atom in the inverted spectral lines. The third column corresponds to the air wavelength of the transition in angstroms, the next column is the excitation potential of the lower level, the fifth column is the oscillation strength, and the lower and upper levels are listed in the sixth column. The seventh and eighth columns are the Barklem velocity exponential and Barklem cross-section (in units of the Bohr radius squared), respectively. The last two values were calculated following the ABO theory (\citealt{barklem1998}; \citealt{anstee1995}) and  can be obtained from the tabulated tables in \cite{anstee1995}.}
\end{table*}

We followed different inversion strategies depending on the inverted area: light bridges (LB), umbra (UM), penumbra (PE), and quiet Sun (QS).

\begin{itemize}
    \item[$\ast$] The UM and PE pixels were inverted with one magnetic atmospheric component by using five nodes for the $T$, one node for $v_{\rm mic}$, three nodes for $|\vec{B}|$, and two nodes for $\gamma$, $\phi$, and $v_{\rm los}$. The atmospheric guess model was the cool umbral model by \cite{collados1994}.

    \item[$\ast$] For the QS pixels, we followed a slightly different strategy: all the parameters but $T$ (seven nodes) were constant with optical depth and the atmosphere guess model was the Harvard-Smithsonian reference atmosphere model (HSRA; \citealt{gingerich1971}).

    \item[$\ast$] Finally, the LB pixels were inverted with one magnetic atmosphere and one non-magnetic atmosphere. The nodes used for the inversion of the magnetic component were the same as for the UM and PE pixels. The atmospheric guess model of the magnetic component was again the cool umbral model. The non-magnetic component was initialised using the atmosphere obtained in the inversion of the average the Stokes $I$ profiles coming from the surrounding QS areas. In addition, we only inverted the $v_{\rm los}$ of the non-magnetic component (constant with optical depth) and the $ff$. In other words, we used a magnetic component and a non-magnetic component with free $v_{\rm los}$ and filling factor. The $T$ was not inverted since the information about this parameter was not enough as we checked from inversion of these pixels with more complex strategies (allowing the non-magnetic $T$ to vary). This way, the $T$ stratification used was the inferred one in the inversion of the Stokes $I$ of the surrounding QS areas.
\end{itemize}

A list of the free parameters used in the strategies followed for the different areas can be found in Tab. \ref{Tab:nodes}. Following \cite{pastor2018}, the inversion process was repeated 50 times in each pixel with random initial values of the free parameters ($|\vec{B}|$, $\gamma$, $\phi$, $v_{\rm los}$, and $v_{\rm mic}$) in a given range (see Tab. 3 of \citealt{pastor2018}).

\begin{table*}
  \caption{Free parameters used in the strategies followed for the different regions.}
  \label{Tab:nodes}
  \centering
    \begin{tabular}{ccccccccccccc}
      \hline
      \hline
Region & $T^{1}$ & $|\vec{B}|^{1}$ & $\gamma^{1}$ & $\phi^{1}$ & $v_{\rm los}^{1}$ & $v_{\rm mic}^{1}$ & $T^{2}$ & $|\vec{B}|^{2}$ & $\gamma^{2}$ & $\phi^{2}$ & $v_{\rm los}^{2}$ & $v_{\rm mic}^{2}$ \\ 
      \hline
      \hline
LB & 5 & 3 & 2 & 2 & 2 & 1 & - & - & - & - & 1 & - \\
UM & 5 & 3 & 2 & 2 & 2 & 1 & - & - & - & - & - & - \\
PE & 5 & 3 & 2 & 2 & 2 & 1 & - & - & - & - & - & - \\
QS & 7 & 1 & 1 & 1 & 1 & 1 & - & - & - & - & - & - \\
      \hline
    \end{tabular}
    \tablefoot{The values represent the number of nodes used in the inversion process for each parameter and the superscripts 1 and 2 represent the two components used for the inversions.}
\end{table*}

A final step is required in order to obtain ready-to-use inferred parameters. This step involves the resolution of the so-called $180^{\circ}$ ambiguity present in the line-of-sight magnetic field azimuth ($\phi_{\rm los}$). This ambiguity comes from the fact that a given $\phi_{\rm los}$ leads to exactly the same spectral profiles as $\phi_{\rm los} + 180^{\circ}$. This ambiguity involves two possible magnetic field topologies, both in the line-of-sight and local reference frame, for each pixel. It is thus mandatory to properly resolve this ambiguity for the whole observed area in order to study the magnetic field topology. Here, we handled this step using the technique described by \cite{georgoulis2005} and the supplied software. This code returns the disambiguated values of azimuth, which together with the inclination of the magnetic field can be transformed to the local solar reference frame. Hereafter, the inclination ($\gamma$) is the angle between the magnetic field vector and the local solar vertical, the azimuth ($\phi$) is on the tangent plane to the solar surface, the $0^\circ$ azimuth reference is in the north solar direction, and the azimuth values increases anti-clockwise.
We focused our study on $T$, $|\vec{B}|$, $\gamma_{\rm los}$, and $v_{\rm los}$ (from magnetic and non-magnetic atmospheres).

\subsection{Velocity calibration}
After inferring the line-of-sight velocity ($v_{\rm los}$), it is necessary to calibrate it with respect to a reference zero velocity. There are two different ways to do this: absolute and relative calibrations. The absolute calibration can be done using either telluric lines or a laser-based calibration, but we have neither of them. Regarding relative calibrations, there are two methods that are commonly used in the bibliography: setting the average QS velocities or the mean umbral velocities to zero.

\begin{figure}[h]
  \begin{center}
    \includegraphics[width=0.5\textwidth]{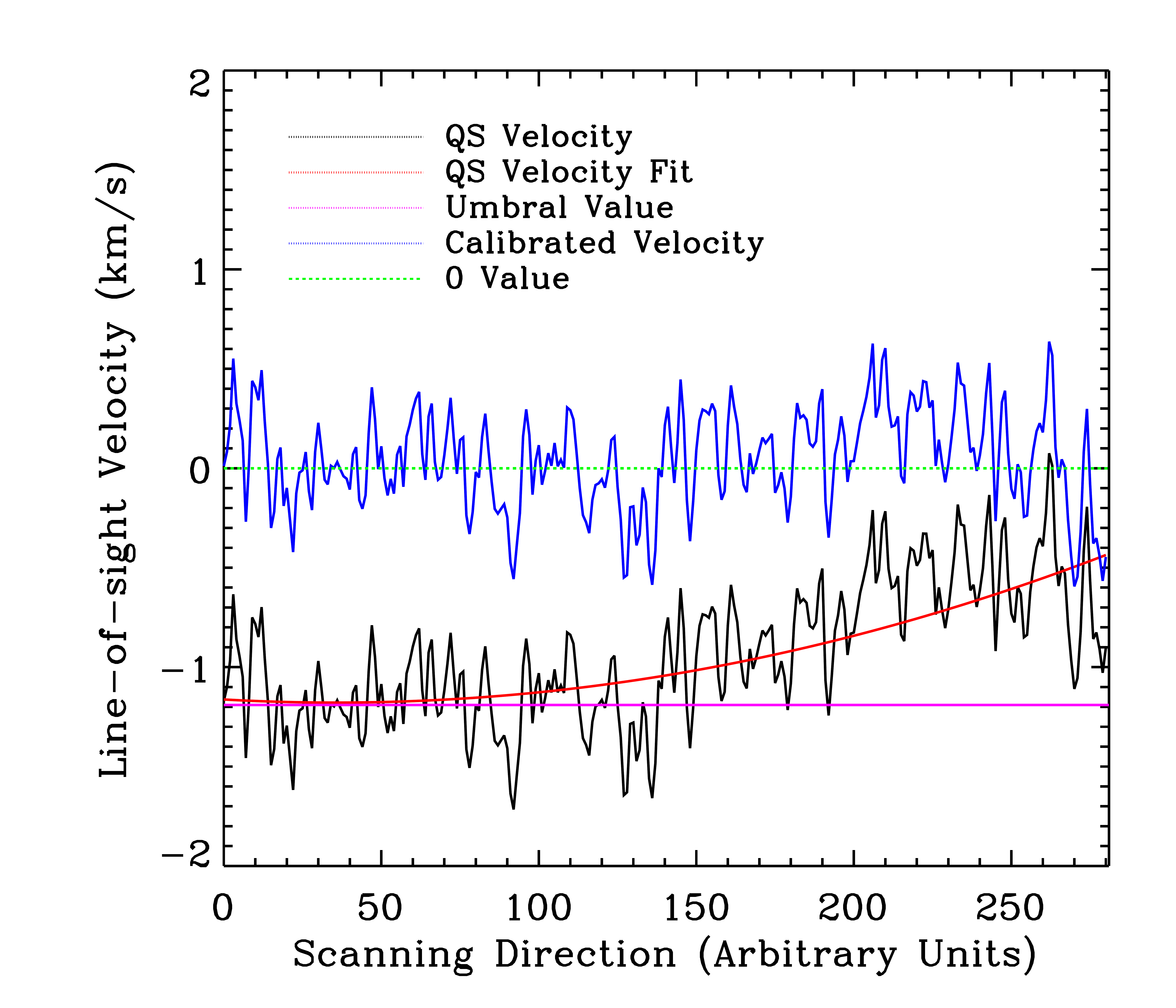}
    \caption{Velocity calibration: The black line represents the average $v_{\rm los}$ of the QS area of Scan 3\_05 along part of the slit used to correct the trend produced by an instrumentation problem. The red line is a second-degree polynomial that fits the black line. The pink line corresponds to the value obtained by averaging the velocity of the UM after correcting the instrumental trend. Finally, the blue line corresponds to the final result of the averaged QS $v_{\rm los}$ after applying the previous corrections. The green line marks the 0 value.}
    \label{Fig:velocityCalibration}
  \end{center}
\end{figure}

We decided to use the relative calibration that sets the umbral average velocity to 0. After applying this correction, we realised that there was still a residual velocity offset with time, thus\ along the scanning direction. This effect was seen most clearly in the QS average velocity along the scanning direction (see black line of Figure \ref{Fig:velocityCalibration}). This residual velocity was most probably due to an instrumental problem (private communication with Dr. Collados). To solve it, we identified a sufficiently large area of QS in the field of view present in all scanning positions. We then averaged along the slit the velocities associated with these QS pixels. In this way, assuming that the QS does not vary its average velocity with time, we obtained the velocity variation as the scan was recorded. Finally, we fitted these variations (one for each map) with a second-order polynomial (black and red lines of Figure \ref{Fig:velocityCalibration}, respectively) and used it to correct the velocity trend. After this trend correction, we applied the umbral relative calibration by calculating the average umbral $v_{\rm los}$ (pink line in Figure \ref{Fig:velocityCalibration}). These two corrections (the trend produced by the instrumentation and umbral calibration) were applied to all the optical depths, leading to ready-to-use velocities. 

\subsection{Response Functions}
\begin{figure}
  \centering
    \subfigure{\includegraphics[width=0.5\textwidth]{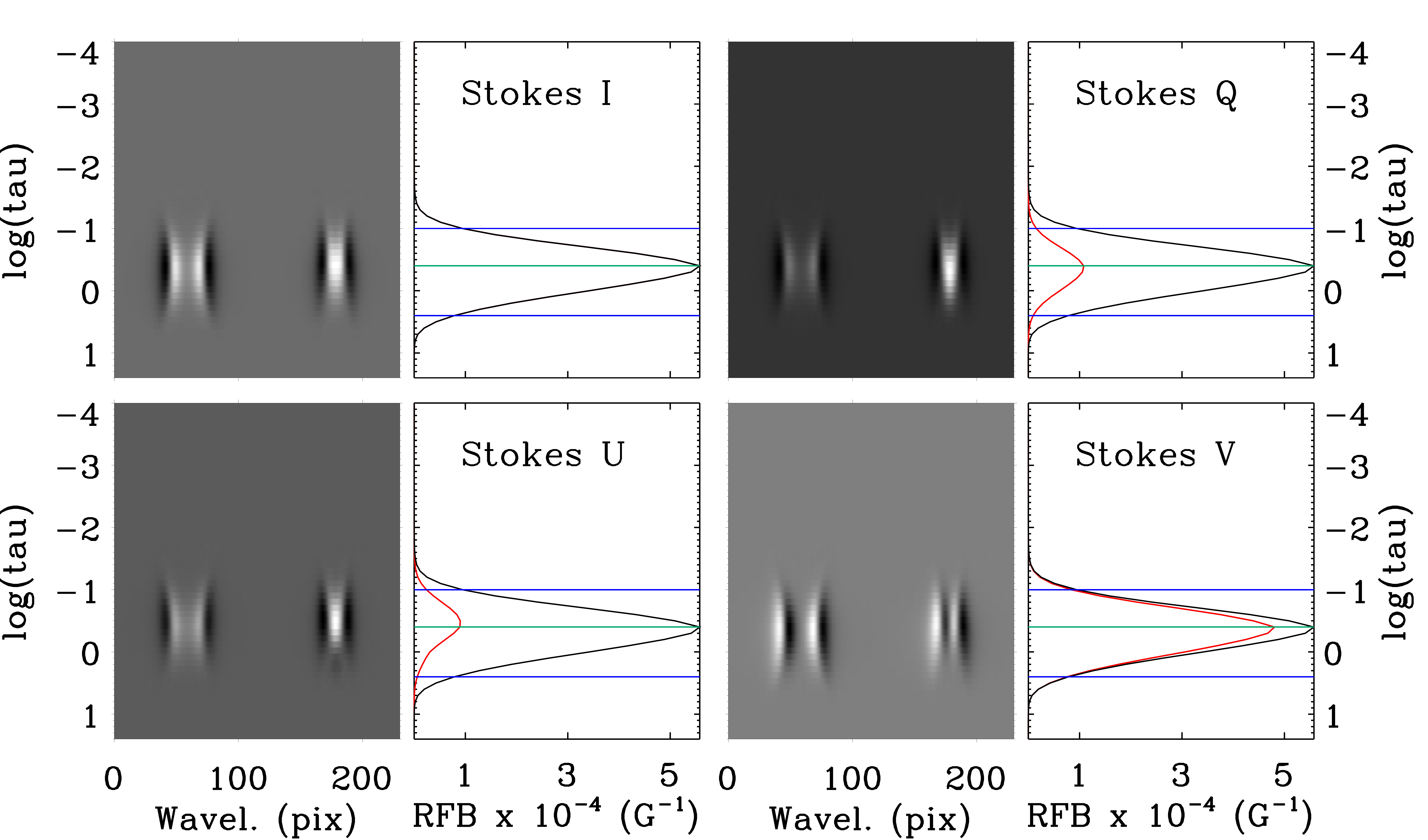}}
    \subfigure{\includegraphics[width=0.5\textwidth]{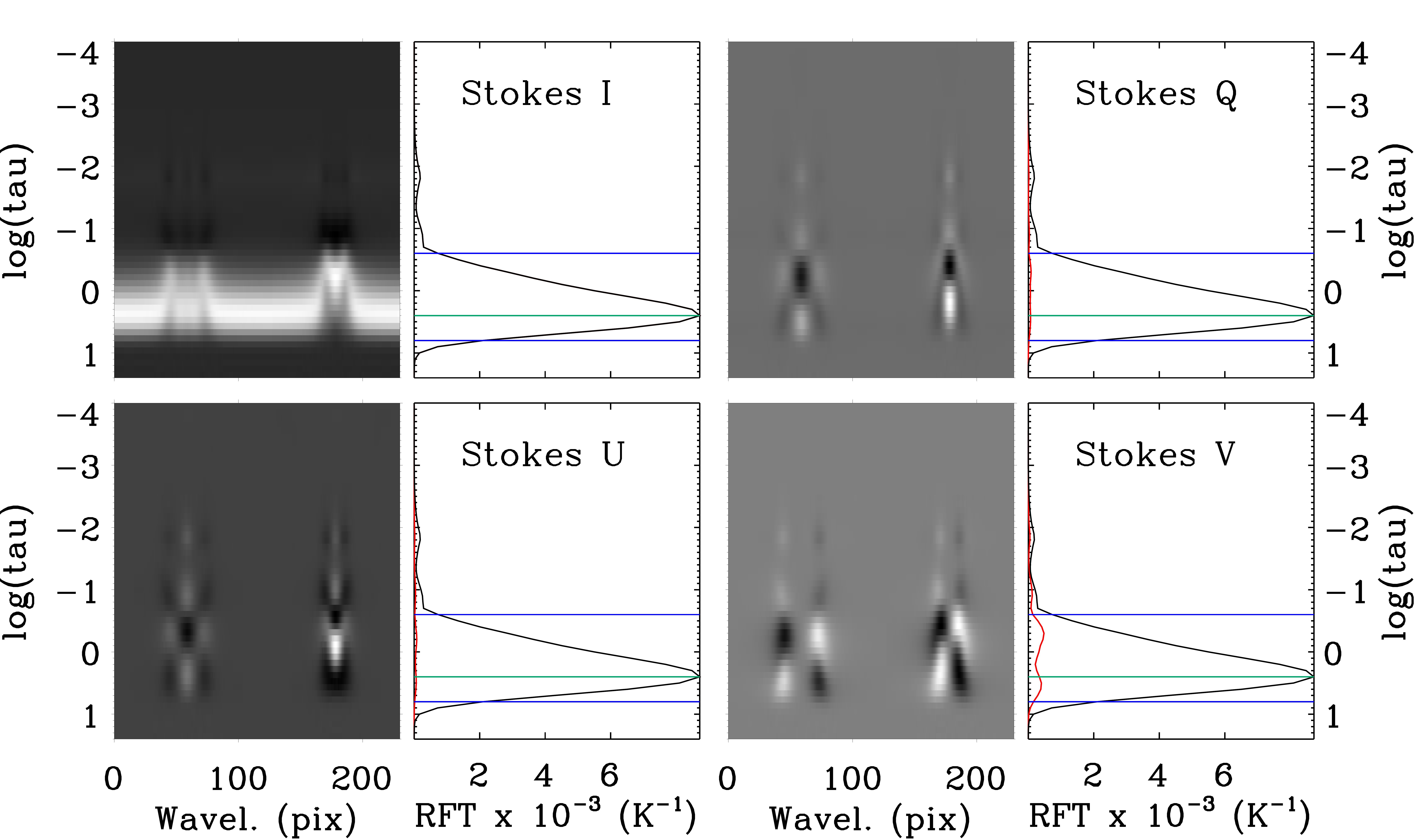}}
  \caption{Response functions of the Stokes parameters of the spectral lines used in the inversion of the magnetic field strength (top eight panels) and of the temperature (bottom eight panels). Each map was normalised to its total absolute value. The left line corresponds to the 15648.515~\AA\ spectral line and the right one to 15662.018~\AA. The grey scale images show the RFs of the four Stokes parameters, and the plots on the side depict the integral of each RF along the wavelength direction. The graphs depict the integral of each RF in the wavelength direction. The horizontal blue lines mark the optical depths for which the area of the integrated RF between these blue lines adds up to 95\% of the total, and the green horizontal line corresponds to the optical depth where the integrated RF is maximum.}
  \label{Fig:responseFunction}
\end{figure}

A crucial point is to know the optical depths where the Stokes profiles are sensitive to the physical parameters of the atmosphere, that is,\ where (and by how much) Stokes profiles change as one physical parameter is changed at a given optical depth. This is interesting because the inferred stratifications are not reliable for the whole grid of optical depths, but only for some ranges of it. In order to know at which optical depths the observed lines are sensitive to perturbations of the physical atmospheric parameters, we calculated the response functions (RFs) of each atmospheric parameter analysed in this work for a set of profiles representative of PE, UM, and LB. An example of the RFs is represented in Figure \ref{Fig:responseFunction}.

One of the analyses in the next section compares the behaviour of the atmospheric parameters in the UM, PE, and LBs. However, the spectral lines are not sensitive at the same optical depths for these regions. In order to study the parameters in a common set of optical depths for the three areas, we took four representative pixels from the UM and from the LB. We then calculated the optical depth ranges where each integrated RF contributes 95\% of the total area (this contribution is marked with blue horizontal lines in Figure \ref{Fig:responseFunction}). Finally, we took the optical depths that are common to all regions, the LB, the PE, and the UM. In this case, the observed infrared spectral lines show sensitivity mainly between $\log(\tau) = -0.7$ and $\log(\tau) = 0.4$ for the magnetic field vector and $v_{\rm los}$, and the values for the $T$ are between $\log(\tau) = -0.5$ and $\log(\tau) = 0.8$. Hereinafter, for the representation of the physical parameters, we use the values shown in Tab. \ref{Tab:valuesResponseFunction}.
 
\begin{table}[h]
  \caption{Optical depth ranges used for the analysis of the atmospheric parameters of the LBs.}
  \label{Tab:valuesResponseFunction}
  \centering
    \begin{tabular}{c|c|c}
      \hline
      \hline
            & \multirow{2}{*}{Sensitivity Range} & Differences      \\ 
            &                                   & Upper Layer - Lower Layer    \\ 
      \hline
      \hline
      $|\vec{B}|$& (-0.7 , 0.4) & (-0.7 , -0.4) - (0.1 , 0.4) \\ 
      $\gamma$   & (-0.7 , 0.4) & (-0.7 , -0.4) - (0.1 , 0.4) \\ 
      $v_{los}$  & (-0.7 , 0.4) & (-0.7 , -0.4) - (0.1 , 0.4) \\ 
      $T$        & (-0.5 , 0.8) & (-0.5 , -0.2) - (0.5 , 0.8) \\ 
      \hline
    \end{tabular}
    \tablefoot{For the figures in which we compare the parameters between different layers, we compared the variation of the various physical parameters considered in two layers at different averaged optical depths. The optical depths used for these averages are shown in the `Upper Layer' and `Lower Layer' columns. For the figures in which we study the temporal evolution of the atmospheric parameters, we used the values of the `Sensitivity Range' column.}
\end{table}

\subsection{Inversion products}
\begin{figure*}
  \begin{center}
    \includegraphics[width=1.0\textwidth]{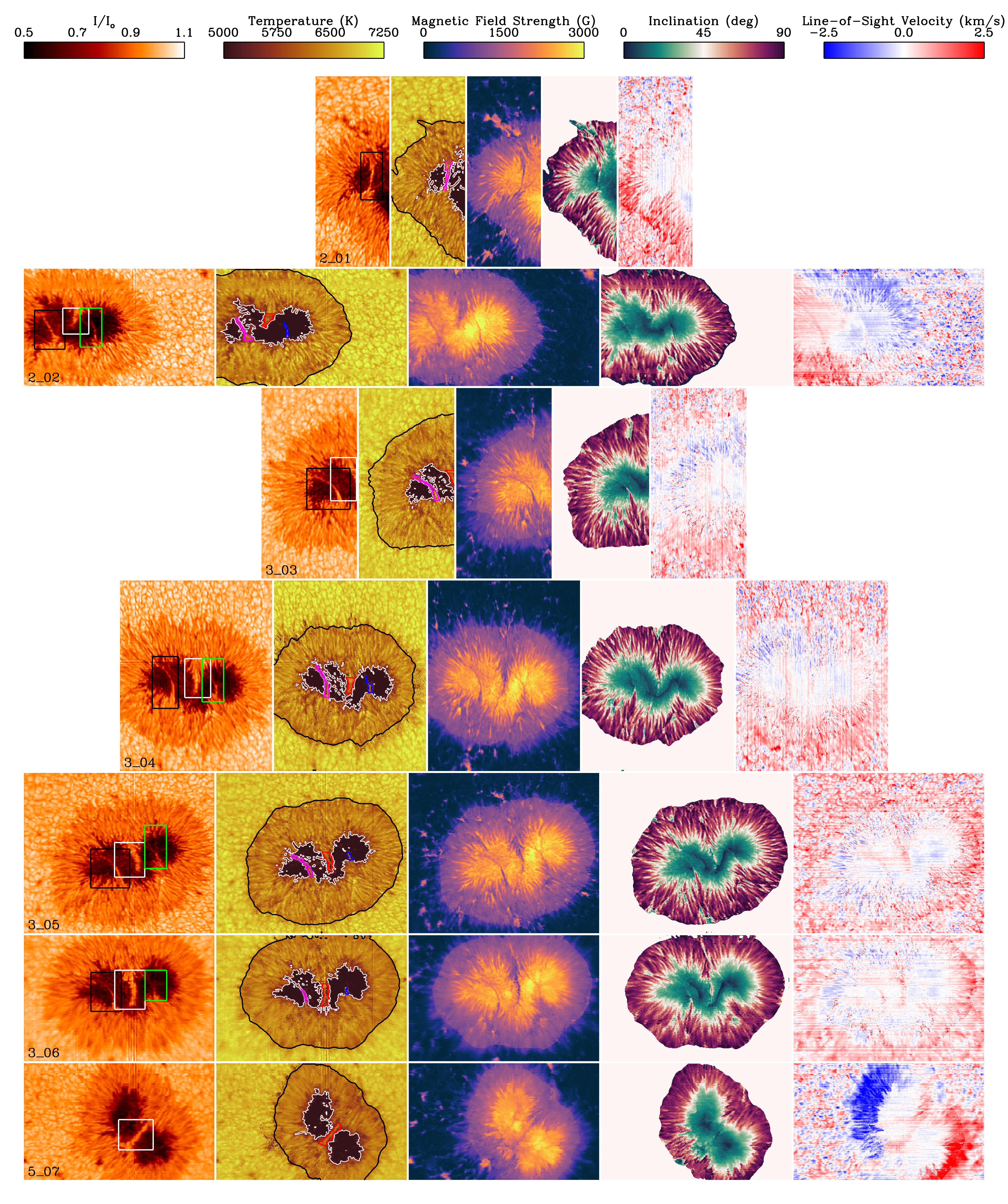}
    \caption{Maps of the atmospheric parameters obtained after the Stokes parameter inversions for the magnetic atmosphere averaged in optical depth between the values shown in Tab. \ref{Tab:valuesResponseFunction} (Sensitivity Range column). Columns (from left to right): normalised observed continuum intensity with respect to the QS, temperature, magnetic field strength, inclination in the solar reference frame, and velocity in the line-of-sight reference frame. The rectangles drawn on the intensity maps mark the regions used in the next section to analyse the different LBs. The white square indicates LB1, the green one is the area taken for LB2, and the black one corresponds to LB3. The contours represent the different areas taken for analysis of the results. The black, white, red, blue and pink contours mark the PE, UM, LB1, LB2 and LB3, respectively. The labels indicate the observation day and the number of the scan of the third column of Tab. \ref{Tab:observationsInfo} (observationDay\_scan).}
    \label{Fig:parametrosAtmosfericos}
  \end{center}
\end{figure*}

\begin{figure}
  \begin{center}
    \includegraphics[width=0.5\textwidth]{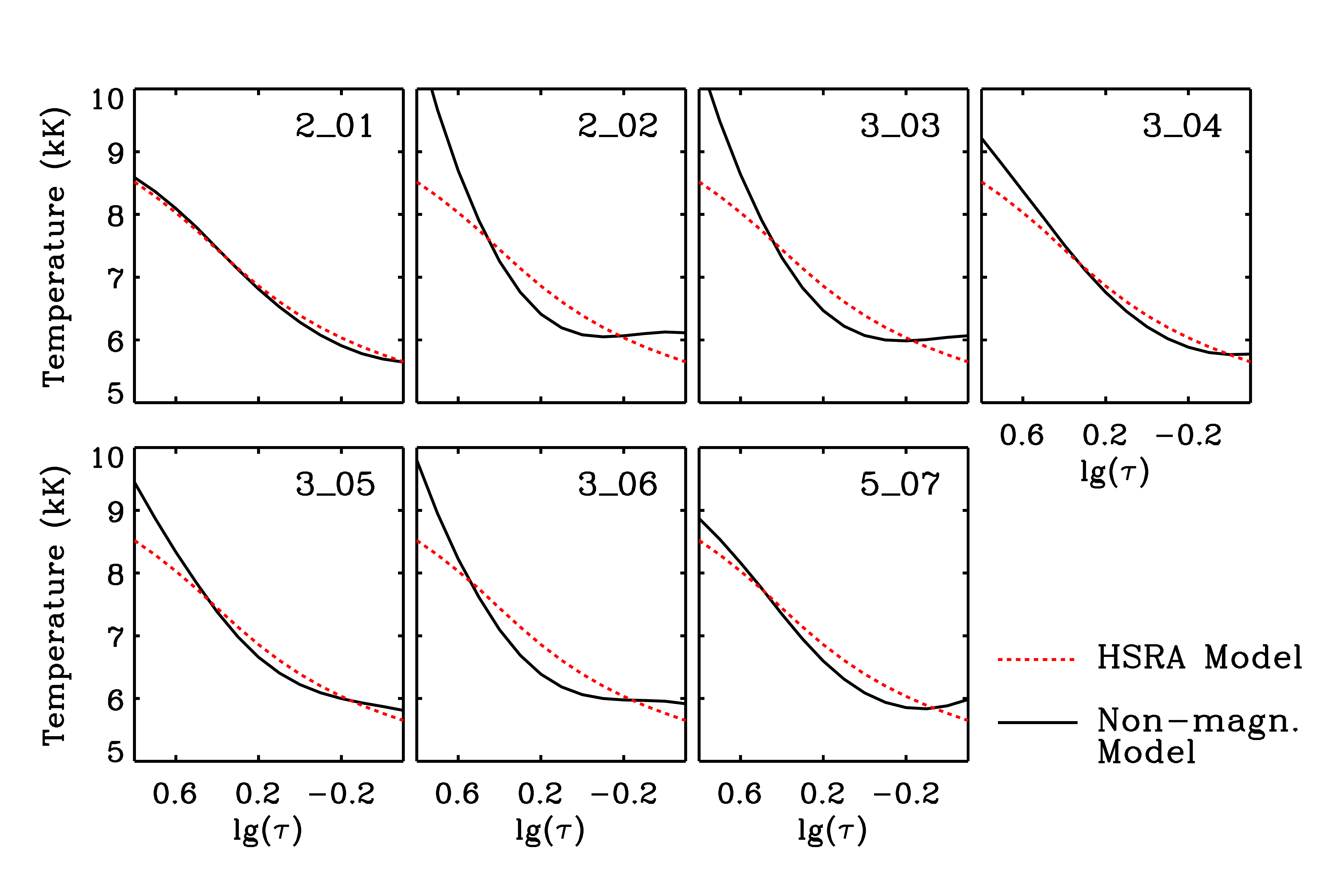}
    \caption{Temperature stratifications. The solid dark line corresponds to the non-magnetic component for each observed map represented between the optical depth values of the left column of Tab. \ref{Tab:valuesResponseFunction}. The red dotted lines correspond to the temperature stratification of the HSRA model.}
    \label{Fig:estratificacion_temp_non_mag}
  \end{center}
\end{figure}

Maps of the normalised observed continuum intensity with respect to the QS, $I/I_{\rm o}$, temperature, $T$, magnetic field strength, $|\vec{B}|$, magnetic field inclination, $\gamma$ (in the local reference frame), and line-of-sight velocity of the magnetic component, $v_{\rm los}^{m}$ (after velocity calibration), are shown in Figure \ref{Fig:parametrosAtmosfericos}. All these maps, except $I/I_{\rm o}$, are the results of the inversions of Stokes profiles averaged in optical depth between the values shown in Tab. \ref{Tab:valuesResponseFunction} (Sensitivity Range column) for the \textbf{magnetic atmosphere}, that is,\ the line-of-sight velocity of the non-magnetic component, $v_{\rm los}^{nm}$, of the LB areas is not represented in this figure. The temperature stratification of the non-magnetic atmosphere for each map, together with the temperature stratification of the HSRA model, are shown in Figure \ref{Fig:estratificacion_temp_non_mag} with black and red lines, respectively. The filling factor maps, $ff$, and $v_{\rm los}^{nm}$  are analysed in detail and shown in Section \ref{subsec:comparison_mag_nonmag}.

At this point, we have our data ready for analysis, which is presented in the next section.

\section{Analysis and results}
The main focus of this work is the study of the temporal evolution of the LB atmospheric parameters compared with that of the UM, PE, and QS.

\subsection{Magnetic and thermal parameters: $|\vec{B}|$, $\gamma$, and $T$}
\label{results_bit}

In order to better characterise these LBs, we performed three different analyses. First, in Section \ref{subsec:temporalEvolutionLB}, we study the temporal evolution of $|\vec{B}|$, $\gamma$, and $T$ (where $T=ff*T_{\rm mag} + (1-ff)*(T_{no-mag})$, $ff$ is the filling factor, and $T_{\rm mag}$ and $T_{\rm no-mag}$ are the temperature of the magnetic and non-magnetic component, respectively). To do so, we compared their global behaviour with that of the UM, PE, and QS. We averaged the physical properties
in each LB both horizontally and along the line of sight. This analysis distinguishes neither the spatial properties of the LBs nor the optical depth variations, but it allows a preliminary insight into the general properties of these LBs. In order to characterise the former, we present a second analysis in Section \ref{subsec:spatialDistributionLB}, where we focus on the temporal evolution and spatial distribution of the inferred parameters averaged in optical depth to analyse how the LBs change with time and space. Finally, in order to characterise the optical depth variation, we compared the difference of the LB parameters between the upper and lower layers with that of the UM and the PE in Section \ref{subsec:heightDifferenceLB}.

\subsubsection{Temporal evolution of averaged  parameters} \label{subsec:temporalEvolutionLB}
\begin{figure*}
  \centering
  \includegraphics[width=1\textwidth]{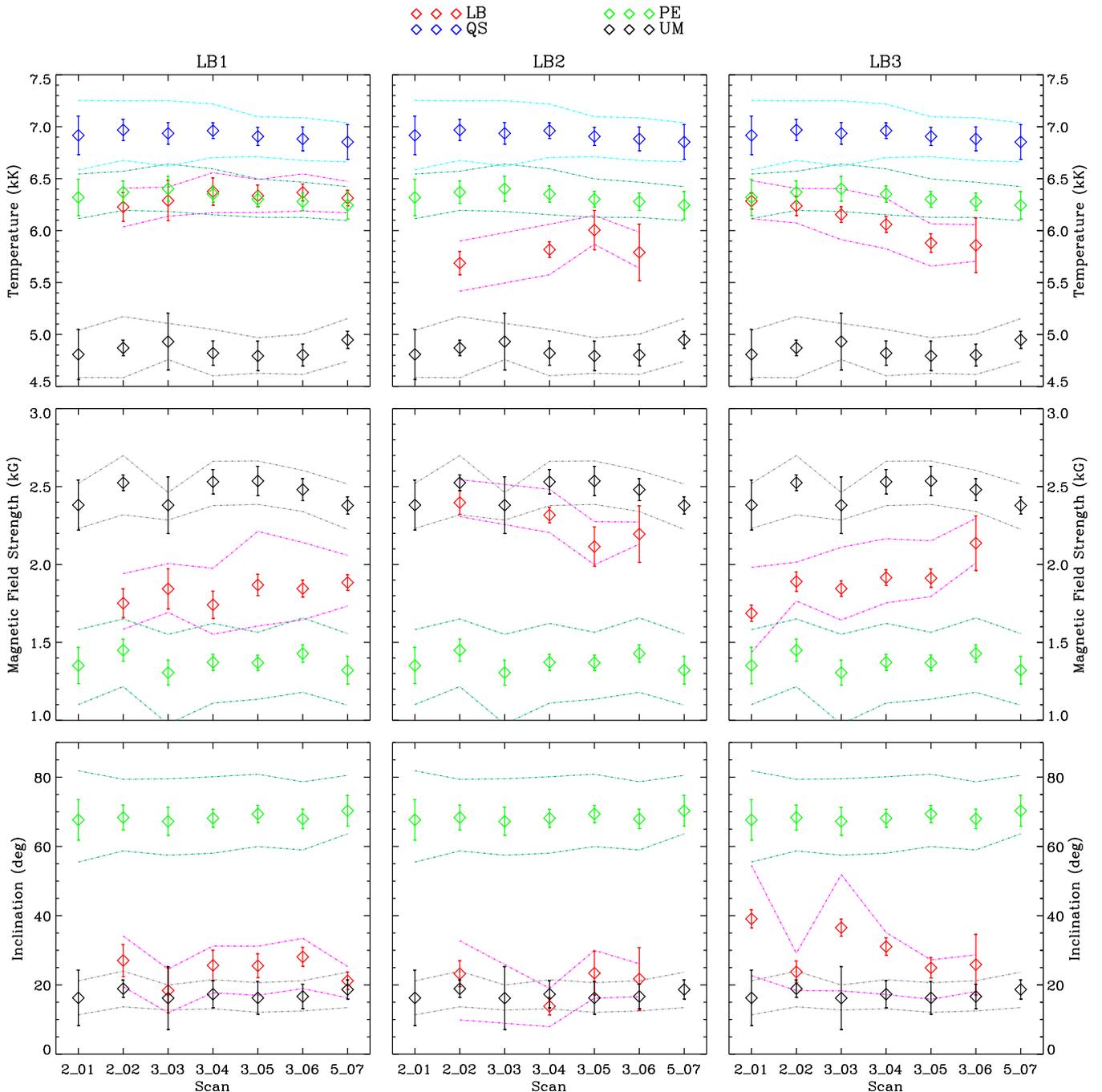}
  \caption{From top to bottom: Temporal evolution of temperature, inclination, and magnetic field strength, respectively, averaged over the optical depth values of the left column of Tab. \ref{Tab:valuesResponseFunction} and over the LB, UM, PE, and QS pixels. From left to right: LB1, LB2, and LB3, respectively. Black and grey lines are for UM values, red and pink for LB pixels, and light and dark green for the PE, and, in the temperature panel, light and dark} blue for the QS pixels. The diamonds correspond to the values and the dash-dotted lines indicate the 25th and 75th percentiles. The $x$-axis represents the scan, of which the format is day\_scan-number.
  \label{Fig:evol_temporal_promedio}
\end{figure*}

We computed the spatial and optical depth averages of $|\vec{B}|$, $\gamma$ (in local reference frame), and $T$ for the LB (hereinafter, $|\vec{B}|_{\rm lb}$, $\gamma_{\rm lb}$, and $T_{\rm lb}$), where the magnetic parameters are applicable to the pixel portion of the corresponding magnetic component, and which is detailed later in Section \ref{subsec:ff}. Furthermore, the temperature is a weighted sum of the magnetic and non-magnetic components, UM ($|\vec{B}|_{\rm um}$, $T_{\rm um}$, and $\gamma_{\rm um}$), and PE ($|\vec{B}|_{\rm pe}$, $T_{\rm pe}$, and $\gamma_{\rm pe}$). We additionally included the QS temperature ($T_{\rm qs}$) in order to have a standard reference. The pixels taken for each region are indicated in Figure \ref{Fig:parametrosAtmosfericos} with black and white contours on the temperature maps. These calculations are shown in Figure \ref{Fig:evol_temporal_promedio}. 

For reference, we first considered the case for the UM, PE, and QS. $|\vec{B}|$, $\gamma$, and $T$ hardly vary with time during the observed scans and for any of the regions of interest. On average, the PE and the UM are $\approx$600~K and $\approx$2000~K cooler than the QS, respectively. For the magnetic field, the PE has values of $\approx$1000~G lower than the UM and its magnetic field lines are $\approx$50~deg more inclined than that of the UM, which is $\approx$20~deg with respect to the line-of-sight.

It might surprise the reader to see that the magnetic field lines of the UM are on average more inclined than one would expect for this structure, where close-to-vertical fields are expected. This is because the inclination values of the UM have a broad distribution from 0 to 35~deg, and only in the centremost part of the UM is the magnetic field inclination really close to vertical fields. When moving from the very centre, the inclination increases, reaching values up to $\approx$30~deg. That is why, when averaging the whole UM inclination, we get values close to 20~deg.

Analysing the temporal evolution of the different atmospheric parameters $T$, $|\vec{B}|$, and $\gamma$ of the three LBs, we found that LB1 exhibits very little time variation, whereas LB2 and LB3 do, LB2 heats and decreases its magnetic field strength, while LB3 cools and increases its magnetic field strength, which becomes more vertical. The temporal evolution of the parameters shows a variety of behaviours that could be related to the evolution process of the LB, as is seen in the HMI image (Figure \ref{Fig:hmiSunspots}). The properties of LB3 behave as if they were changing from a penumbra-like structure into an umbra-like one. This could be due, as is seen from HMI data, to the fact that this LB forms by the coalescence of two UMs and finally disappears. Additionally, there are some differences between LB1 and LB2. The temperature of LB1 is closer to the penumbral temperature than that of LB2. Also, the magnetic field strength of LB2 is closer to the UM magnetic fields than that of LB1, so LB1 essentially looks more like the PE and LB2 behaves more like the UM. Table \ref{Tab:valuesParametrosAtmosfericos} shows the quantitative differences between the atmospheric parameters of each LB and the PE, UM, and QS (when applicable) together with their standard deviation as a measurement of the dispersion.

\begin{table}
  \caption{Comparison of LB results with those of the PE, UM, and QS.}
  \label{Tab:valuesParametrosAtmosfericos}
  \centering
  \begin{tabular}{c||c|c|c|c}
   \hline
   \hline
          & \multirow{2}{*}{Region}    & $T_{lb} - T_{reg}$ & $|\vec{B}|_{lb} - |\vec{B}|_{reg}$ & $\gamma_{lb} - \gamma_{reg}$ \\
          &     &  (K) &  (G) &  (deg) \\
   \hline    
   \hline    
          & UM     & $ 1460\pm110$         & $ -650\pm120$                         & $  7\pm4$                       \\      
   LB1    & PE  & $   -7\pm 97$         & $  450\pm100$                         & $-44\pm4$                       \\      
          & QS & $ -600\pm 80$         & -                                     & -                               \\      
   \hline    
          & UM     & $ 1000\pm160$         & $ -260\pm120$                         & $  3\pm5$                       \\      
   LB2    & PE  & $ -500\pm160$         & $  850\pm110$                         & $-48\pm4$                       \\      
          & QS & $-1100\pm160$         & -                                     & -                               \\      
   \hline    
          & UM     & $ 1240\pm160$         & $ -570\pm120$                         & $ 13\pm7$                       \\      
   LB3    & PE  & $ -260\pm150$         & $  520\pm120$                         & $-38\pm7$                       \\      
          & QS & $ -850\pm160$         & -                                     & -                               \\      
   \hline
  \end{tabular}
  \tablefoot{The first column is the LB number, the second is the region for the comparison, and the last three columns correspond to the results for $T$, $|\vec{B}|$, and $\gamma$, respectively. The last three columns were\ calculated as the average of the differences between the LB parameters and the region parameters. The standard deviation of these averages is also shown after each difference value.}
\end{table}

\begin{figure}[h]
  \begin{center}
    \includegraphics[width=0.5\textwidth]{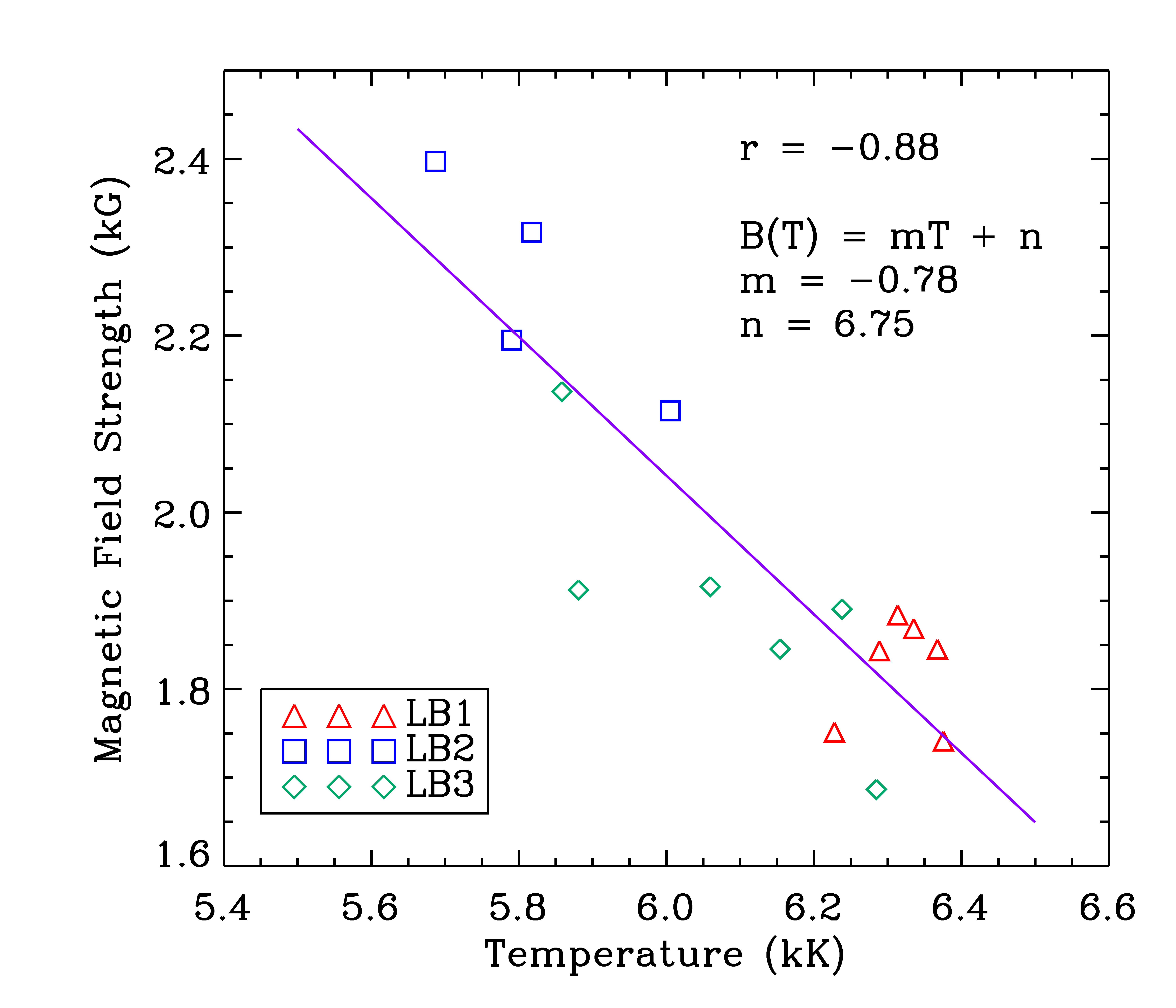}
    \caption{Relation between the magnetic field strength and the temperature for the three LBs. The red triangles, blue squares, and green diamonds correspond to the values for LB1, LB2, and LB3, respectively. The purple line represents the linear fit of the values whose equation is written on the right of the graph. The $r$ value corresponds to the Pearson correlation coefficient.}
    \label{Fig:correlation_bb_tt}
  \end{center}
\end{figure}

Analysing the global behaviour of the three LBs, it seems that there might be an anti-correlation between the temperature and the magnetic field strength, meaning the cooler the LB, the greater the magnetic field strength. Figure \ref{Fig:correlation_bb_tt} shows this anti-correlation between $T_{\rm lb}$ and $|\vec{B}|_{\rm lb}$. The pink line represents the linear fit, with linear and independent coefficients $-0.78~G/kK$ and $6.75~G$, respectively, and the Pearson correlation coefficient is $-0.88$.

\subsubsection{Spatial distribution of the inferred parameters}
\label{subsec:spatialDistributionLB}
\begin{figure}[h]
  \begin{center}
    \includegraphics[width=1\textwidth,angle=90]{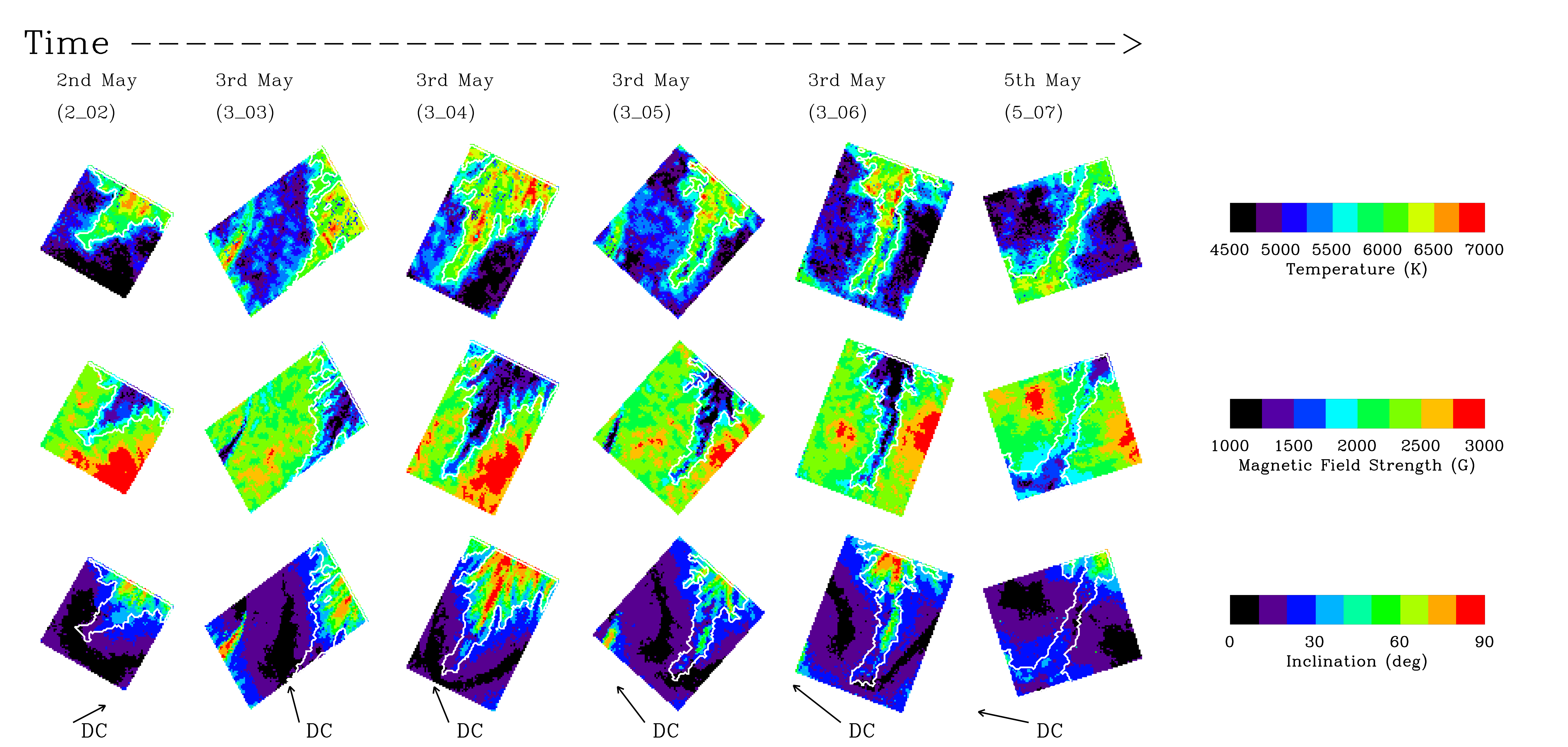}
    \caption{Temporal evolution maps of the temperature, inclination in the local reference frame, and the magnetic field strength of LB1. The parameters are averaged over the optical depths shown in the left column of Tab. \ref{Tab:valuesResponseFunction}. The date and the hour of each observation are written at the top of the figure. Arrows indicate the disc centre direction.}
    \label{Fig:resultsLB1}
  \end{center}
\end{figure}

\begin{figure}[h]
  \begin{center}
    \includegraphics[width=0.75\textwidth,angle=90]{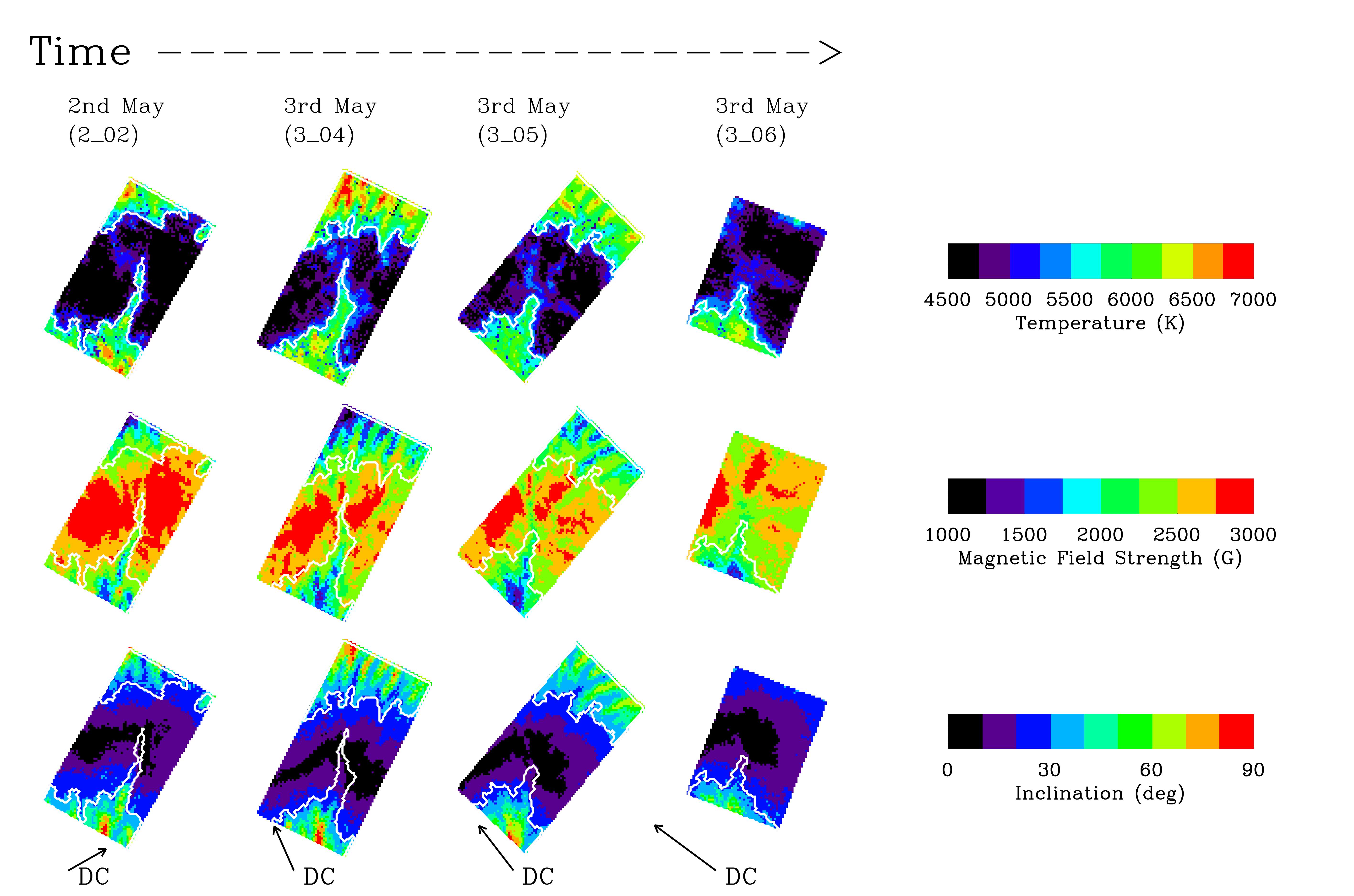}
    \caption{Same as Figure \ref{Fig:resultsLB1} for LB2.}
    \label{Fig:resultsLB2}
  \end{center}
\end{figure}

\begin{figure}[h]
  \begin{center}
    \includegraphics[width=1\textwidth,angle=90]{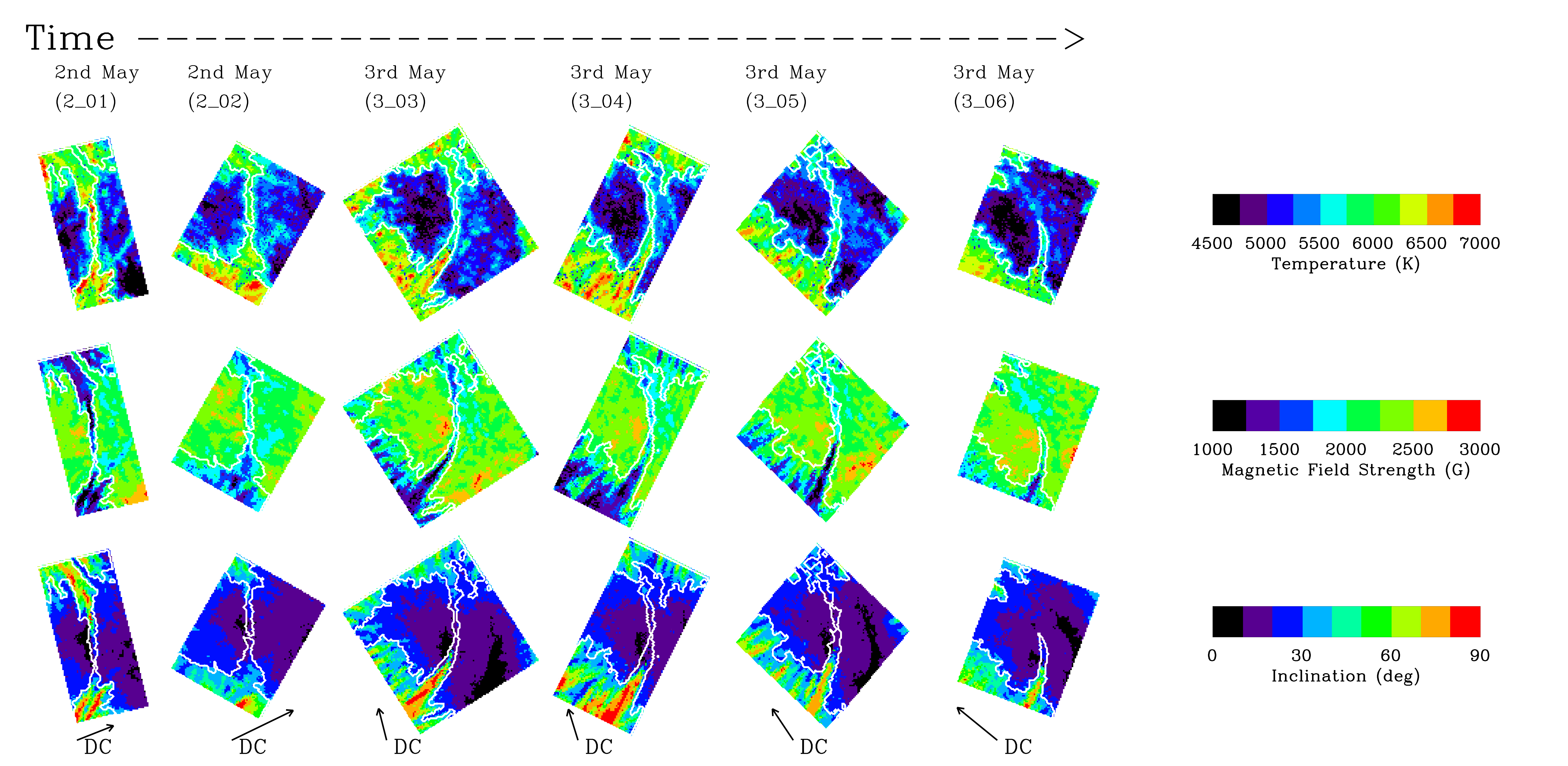}
    \caption{Same as Figure \ref{Fig:resultsLB1} for LB3.}
    \label{Fig:resultsLB3}
  \end{center}
\end{figure}

The previous analysis does not consider spatial properties of the physical parameters in the LBs. In order to analyse this spatial behaviour, Figures \ref{Fig:resultsLB1}, \ref{Fig:resultsLB2}, and \ref{Fig:resultsLB3} present the values of the atmospheric parameters as a function of spatial location averaged over the optical depth listed in the left column of Tab. \ref{Tab:valuesResponseFunction}. Each row of the figures represents the analysed atmospheric parameter ($T$, $|\vec{B}|$, and $\gamma$), and the different columns are the various observations of each LB (the observation date and label are written at the top of the figure).

The temporal evolution and the spatial distribution of LB1 (Fig. \ref{Fig:resultsLB1}) show that the LB area increases from a tiny intrusion into the UM (Scan 2\_02) until it breaks the UM into two different umbral cores (Scan 5\_07). Some general properties observed in Fig. \ref{Fig:evol_temporal_promedio} are also seen here, such as the fact that $T_{\rm lb}$, $|\vec{B}|_{\rm lb}$, and $\gamma_{\rm lb}$ lie between those for the PE and the UM. Morphologically, the temperature in the LB shows a patchy structure where some small structures are cooler than their surroundings (blue areas of the Scans 3\_05 and 3\_06). This cooler region could be related to the dark lane as it is more or less centred along the LB axis, but we lack good enough spatial resolution to assess this with confidence. Also, some scans (3\_03 and 3\_04) show a sudden heating of the axis of the LB as compared to the rest of the LB but the scarce temporal cadence does not allow us to further address this point. Regarding the magnetic field strength, values of the $|\vec{B}|_{\rm lb}$ are lower than those of the surrounding UM. It is worth noting that the UM also presents a strong spatial variation as the magnetic field is weaker to the east of LB1 than to the west. For most of the area, the magnetic field strength values are in the range $\approx$1500--2000~G. The spatial distribution of this parameter shows that the magnetic field in the middle of the LB is weaker than at the borders. Also, the magnetic field lines of LB1 show a broad inclination distribution from $\approx$10--60~deg, while it is very vertical with values between $\approx$0--30~deg in the surrounding UM. In Scan 5\_07, when the LB is totally formed, the behaviour of the inclination is different, probably because the heliocentric angle for this scan (see Tab. \ref{Tab:observationsInfo}) is higher than for the rest observed maps and this produces that the region of the atmosphere to which the spectral lines are sensitive may be different (higher in the atmosphere) for this particular scan, showing a different behaviour of inclination angle. All the magnetic LB structure is almost vertical with values very similar to those of the UM. An interesting feature of the inclination in contrast to magnetic field strength and temperature (which seem to be homogeneous along the whole LB) is that the magnetic lines are more vertical at the tip of LB1 (the end of the LB inside the UM) and more inclined in the rest of the body. This property stays constant throughout the time sequence: when a LB is forming, $T$ and $|\vec{B}|$ change their values before $\gamma$ does.

The spatial evolution of LB2 (Figure \ref{Fig:resultsLB2}) shows how the LB retreats until it almost disappears. A similar behaviour to that of LB1 (Figure \ref{Fig:resultsLB1}) is visible for this LB, but in this case the magnetic field strength of LB2 is higher than that of LB1 ($|\vec{B}|_{lb2}$ $\approx$2000~G, $|\vec{B}|_{\rm lb1}$ $\approx$1500--2000~G). The temperature values for this LB are very similar to those for LB1. The magnetic field lines of this LB are very vertical with values very similar to that of the surrounding UM (between $\approx$0--30~deg), being more vertical at the tip of the LB and more inclined in the rest of the LB. In contrast to LB1, LB2 is disappearing, so it seems that $\gamma$ precedes $T$ and $|\vec{B}|$ in its withdrawal.

Light Bridge 3 (Figure \ref{Fig:resultsLB3}) is formed when two UMs approach each other and the remnant granulation gets trapped in between. In this LB, $T$ and $|\vec{B}|$ behave much like those of LB1 and LB2 ($|\vec{B}|_{\rm lb3} < |\vec{B}|_{\rm um}$ and $T_{\rm lb3} > T_{\rm um}$), except for the evident penumbral intrusion suffered by the LB in Scans 3\_03, 3\_04, and 3\_05. This event is characterised by a sudden change of $|\vec{B}|$ and $\gamma$, changing from 1750~G to 1500~G and 30~deg to 80~deg, respectively. This sudden change in LB properties is mentioned in Section \ref{subsec:temporalEvolutionLB}, but now, with the spatial information, we speculate that this event is caused by a sudden intrusion of penumbral filament into the umbral area through the LB spine.

\subsubsection{Optical depth difference characterisation}
\label{subsec:heightDifferenceLB}
\begin{figure}[h]
  \begin{center}
    \includegraphics[width=0.5\textwidth]{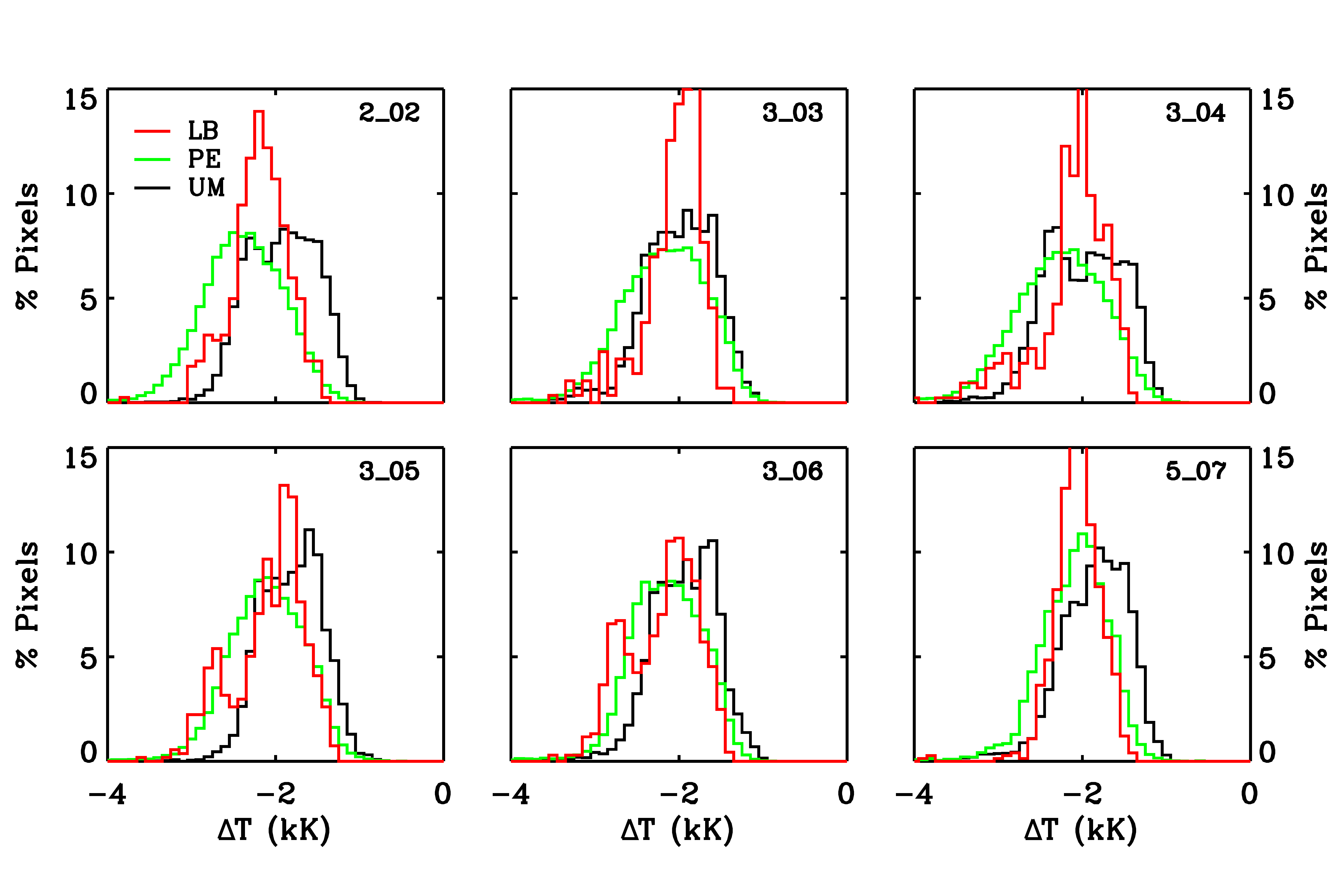}
    \caption{Distribution of the difference between the average temperature in the upper and lower layers of LB1 ($\Delta T = T_{\rm upper} - T_{\rm lower}$). The red line corresponds to the LB distribution. The penumbral is shown in green and the umbral distribution in black.}
    \label{Fig:diferencias_tt_lb1}
  \end{center}
\end{figure}

\begin{figure}[h]
  \begin{center}
    \includegraphics[width=0.5\textwidth]{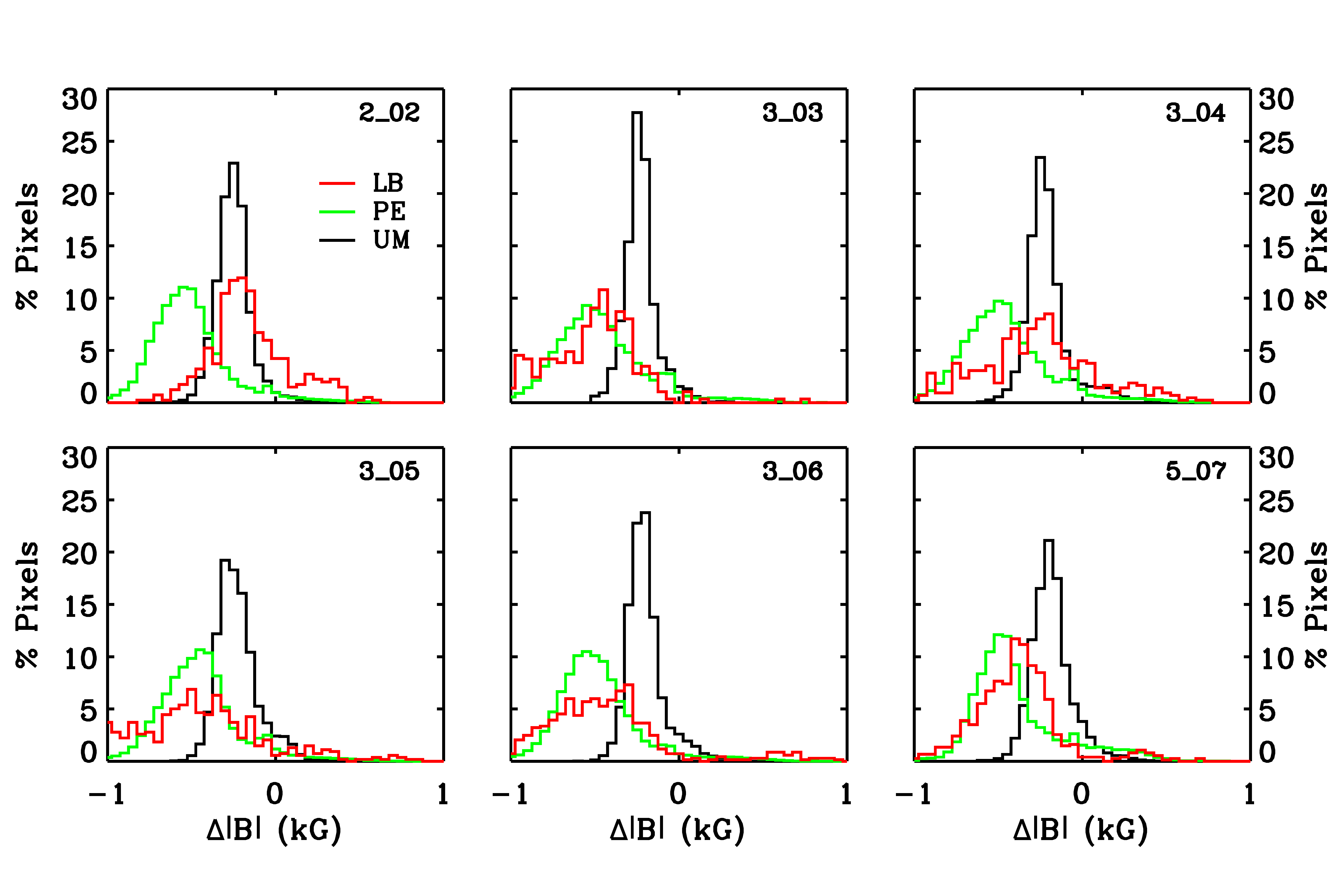}
    \caption{Same as Figure \ref{Fig:diferencias_tt_lb1}, but for the magnetic field strength of LB1 ($\Delta |\vec{B}| = |\vec{B}|_{\rm upper} - |\vec{B}|_{\rm lower}$).}
    \label{Fig:diferencias_bb_lb1}
  \end{center}
\end{figure}

\begin{figure}[h]
  \begin{center}
    \includegraphics[width=0.5\textwidth]{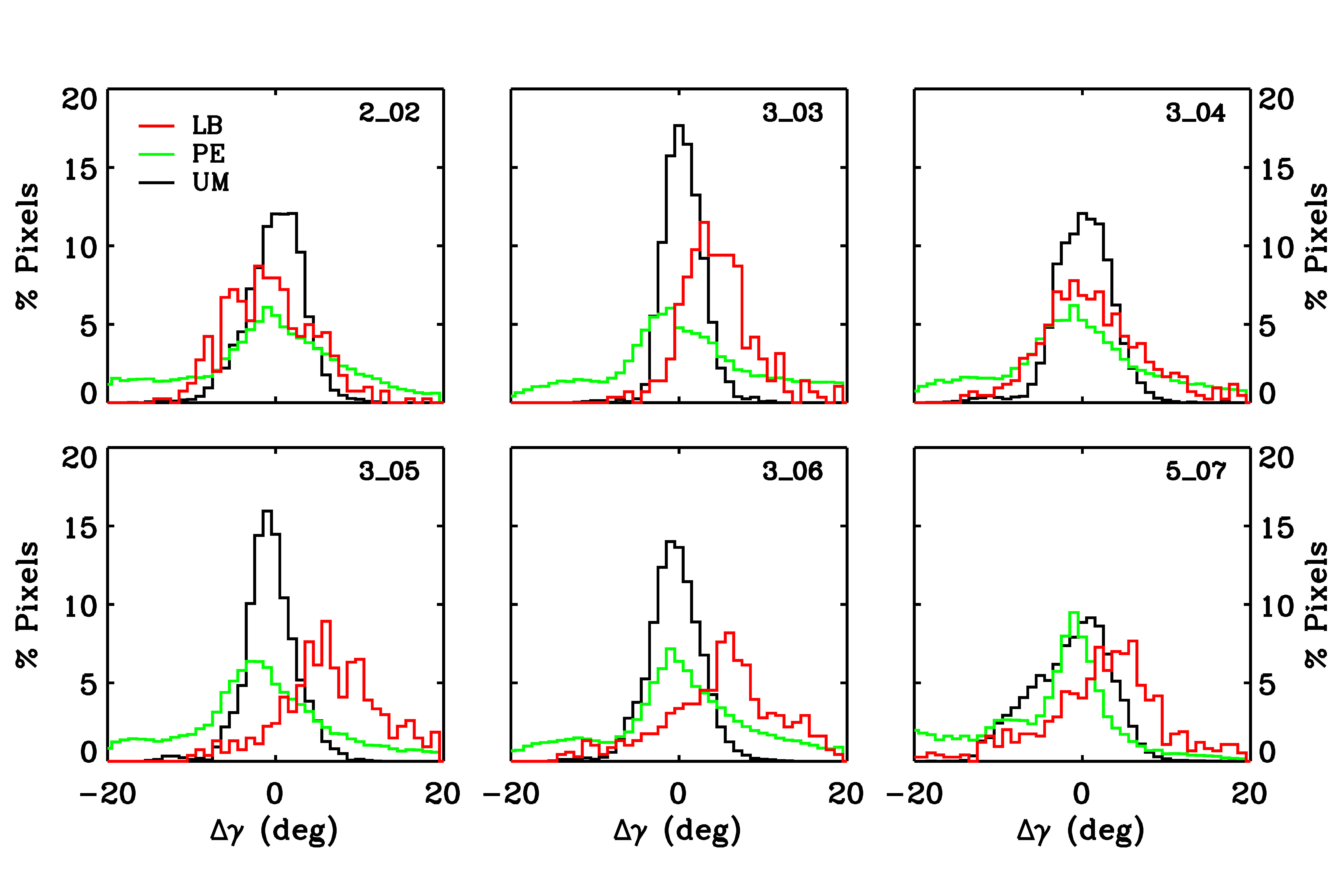}
    \caption{The same as Figure \ref{Fig:diferencias_tt_lb1}, but for the magnetic field inclination of LB1 ($\Delta \gamma = \gamma_{\rm upper} - \gamma_{\rm lower}$).}
    \label{Fig:diferencias_ii_lb1}
  \end{center}
\end{figure}

Another way to study the behaviour of the LBs is to compare the variation in optical depth of the atmospheric stratification for these structures with those seen for the PE and the UM. To do this, we calculated the difference between the average values of various atmospheric parameters at higher optical depths and the average at lower optical depths (the optical depth values for which the various averages were made are shown in Tab. \ref{Tab:valuesResponseFunction}, Differences column) to obtain a rough sense of their vertical gradients. We did this calculation for $T$, $|\vec{B}|$, and $\gamma$ for the penumbral, umbral, and LB regions. Figures \ref{Fig:diferencias_tt_lb1} to \ref{Fig:diferencias_ii_lb1} show these differences and their probability density function (PDF) distribution for LB1. We used it as a representative of all LBs as they are very similar (see Appendix \ref{appA} for the results of LB2 and LB3). Having a broad distribution means that there is a large range of gradients (steep and shallow), so the difference between the parameter value in the higher and lower layers can be large. When the distribution is narrow, the whole set of pixels behave similarly. When it is centred on the value '0', this means that, on average, the analysed parameter does not change with height. If the distribution is centred in the positive (negative) values, we see that the values in the higher layers are, on average, larger (smaller) than in the lower ones.

The penumbral area (green line) cools faster with optical depth than the UM, hence the gradient of temperature for the penumbral area is lower than for the umbral one, as is seen in the histograms of Figure \ref{Fig:diferencias_tt_lb1}. The temperature in the lower layers for both regions is higher than in the upper layers, and their distributions have a similar widths ($\Delta T\approx$-1000~K), however, the average difference for the UM is <$\Delta T>\approx$1900~K and for the PE is <$\Delta T>\approx$-2200~K. As for the magnetic field strength difference (Figure \ref{Fig:diferencias_bb_lb1}), the umbral distribution has a width of $\Delta |\vec{B}|\approx$200~G and its average is <$\Delta |\vec{B}|>\approx-200$~G (the negative sign means that at higher layers the magnetic field strength is weaker), while the penumbral distribution is broader ($\Delta |\vec{B}|\approx$400~G) and the mean value is <$\Delta |\vec{B}|>\approx-450$~G. A similar behaviour is seen for the magnetic field inclination difference distribution (see Figure \ref{Fig:diferencias_ii_lb1}). For the umbral region, the histogram is narrower (its width is $\Delta \gamma\approx$6~deg) than for the penumbral area (its width is $\Delta \gamma\approx$19~deg), and the average umbral distribution is <$\Delta \gamma>\approx$-0.25~deg and <$\Delta \gamma>\approx$-1.5~deg for the UM and PE, respectively (i.e.\ the orientation of the magnetic field lines hardly changes with optical depth for the UM and PE).

\begin{figure}[h]
  \begin{center}
    \includegraphics[width=0.5\textwidth]{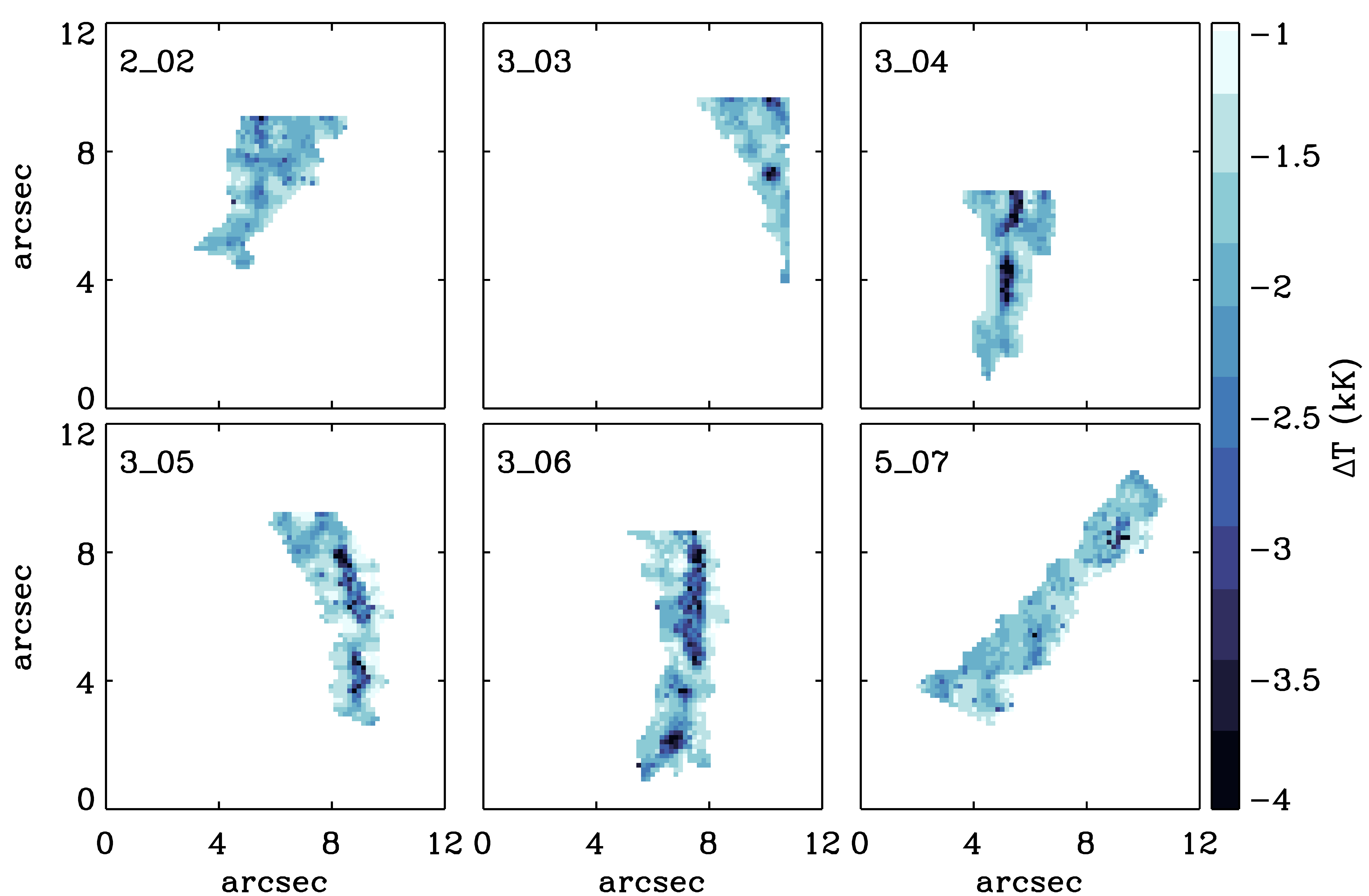}
    \caption{Difference maps between the average temperature in the upper and lower layers of LB1.}
    \label{Fig:mapa_dif_tt_lb1}
  \end{center}
\end{figure}

Focusing on the LBs, the distribution of the differences in temperature (see Figures \ref{Fig:diferencias_tt_lb1} and the figures in Appendices \ref{appA}, \ref{Fig:diferencias_tt_lb2}, and \ref{Fig:diferencias_tt_lb3}, the last two figures for LB2 and LB3) shares similarities with the UM. The distributions are broad with a width of $\Delta T \approx$ 800~K and the average is <$\Delta T>\approx -1900$~K (the negative sign means that the plasma is cooler at higher layers). Yet it seems that the differences of the LB are closer to the minimum of the umbral distribution (around $-2000$~K), meaning the gradient of the temperature for the LB pixels is higher than for the umbral area. Additionally, some scans (3\_04, 3\_05 and 3\_06 for LB1, 3\_05 and 3\_06 for LB2, and 2\_01, 3\_03 and 3\_06 for LB3) show a bi-modal behaviour with an additional peak around $\Delta T \approx-3000$~K. Figures \ref{Fig:mapa_dif_tt_lb1}, \ref{Fig:mapa_dif_tt_lb2}, and \ref{Fig:mapa_dif_tt_lb3} show the temperature difference maps for the different scans of the LBs. They show two behaviours: the borders of the LBs cool less than some areas in the middle of the structure, which cool faster with optical depth. This behaviour could be related to the dark lane reported by other authors (\citealt{sobotka1994}, \citealt{bharti2007}, \citealt{louis2008}, \citealt{rouppe2010}, see Section \ref{intro} for more details), which we are unable to see at this spatial resolution.

The global behaviour of the magnetic field strength for all the scans of the LBs (Fig. \ref{Fig:diferencias_bb_lb1} and figures from Appendix \ref{appA}, \ref{Fig:diferencias_bb_lb2}, and \ref{Fig:diferencias_bb_lb3}) shows a trend towards negative values, that is,\ the magnetic field strength of deeper layers is stronger than that in the upper ones. The width of the distributions is $\Delta |\vec{B}|\approx$500~G, similar to that of the PE. 

The distribution of the magnetic field inclination (Fig. \ref{Fig:diferencias_ii_lb1} and figures from Appendices \ref{appA}, \ref{Fig:diferencias_ii_lb2}, and \ref{Fig:diferencias_ii_lb3}) is broad (its width is $\Delta \gamma\approx$11~deg), and the average is <$\Delta \gamma>\approx$3~deg. For this parameter, it seems that the LB behaves differently with respect to both the PE and the UM as its distribution peaks to positive (in most of them) $\Delta \gamma$ (more horizontal above), and its shape is broader than that for the UM and narrower than that inferred for the PE.

\subsection{Filling factor and dynamic parameters:  $ff$, $v_{\rm los}^{m}$, and $v_{\rm los}^{\rm nm}$}
\label{results5_vlos}
Other important parameters to analyse are the filling factor ($ff$) and the line-of-sight velocity ($v_{\rm los}$). Because of the strategy followed in the Stokes parameter inversion for the LBs, we have a line-of-sight velocity of the magnetic component, $v_{\rm los}^{m}$, and a line-of-sight velocity of the non-magnetic atmosphere $v_{\rm los}^{\rm nm}$. First, we analysed the behaviour of the $ff$ values, and then we considered two different analyses: 1) a comparison of the $v_{\rm los}^{\rm m}$ with $v_{\rm los}^{\rm nm}$ , and 2) a characterisation of the variation in the optical depth of $v_{\rm los}^{\rm m}$ ($v_{\rm los}^{\rm nm}$ is constant with optical depth). The latter is also done for the PE and UM.

\subsubsection{Filling factor}
\label{subsec:ff}
In order to infer the atmospheric parameters of the observed spectral profiles, we inverted the Stokes parameters using an inversion strategy (see Section \ref{sec:inversion}) with two atmospheric components: one magnetic and one non-magnetic, weighted by the $ff$. This way, the $ff$ goes with the magnetic component, while $(1-$ff$)$ goes with the non-magnetic.

\begin{figure}[h]
  \begin{center}
    \includegraphics[width=0.5\textwidth]{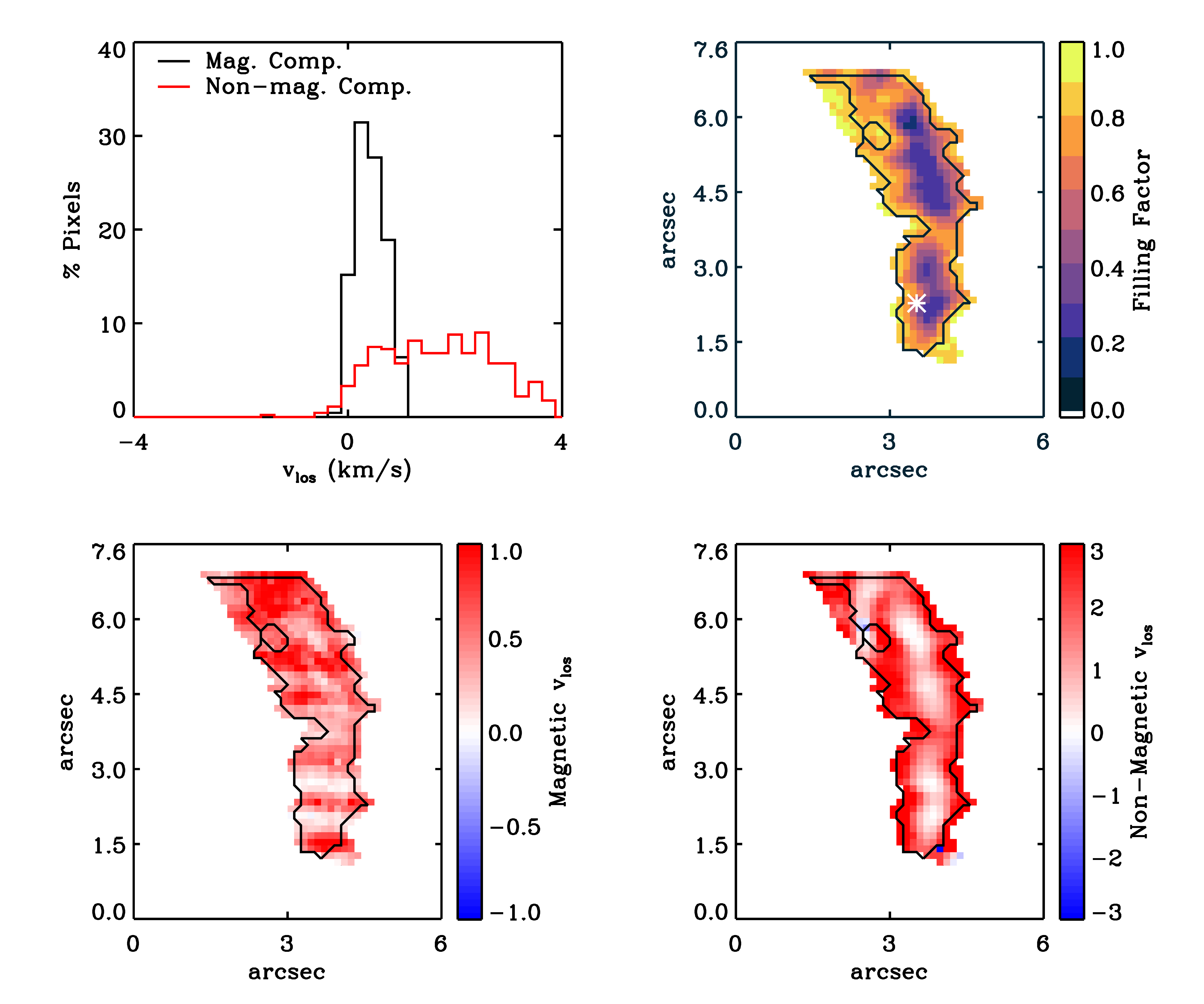}
    \caption{Comparison between magnetic and non-magnetic depth-averaged velocities of LB1 in Scan 3\_05. The histogram shows the distributions of $v_{\rm los}^{\rm m}$ (black line) and $v_{\rm los}^{\rm nm}$ (red line). The upper right panel corresponds to $ff,$ and the white asterisk indicates the pixel taken to analyse the Stokes $I$ and $V$ profiles in Fig. \ref{Fig:comparacion_perfiles}. The lower panels represent $v_{\rm los}^{\rm m}$ on the left and $v_{\rm los}^{\rm nm}$ on the right. The black contours of the maps enclose the area with $ff$ below 85\%.}
    \label{Fig:03may_012_comparacion_vv_lb2}
  \end{center}
\end{figure}

Table \ref{Tab:ff_median} shows the median of the $ff$ for each scan and LB together with two percentiles ($14^{th}$ and $86^{th}$, sub- and super- index, respectively). For LB1, the thickest one, there are clearly two different behaviours (see i.e. top-right panel of Figures \ref{Fig:03may_012_comparacion_vv_lb2} and \ref{appA_lb1}): the core is highly non-magnetic, while for the outer parts of the LB the magnetic $ff$ increases. For LB2 and LB3, this behaviour is not that clear, which may be related to the fact that these LBs are thinner (figures from Appendices \ref{appA_lb2} and \ref{appA_lb3}) and they are mainly magnetic. 

\begin{table*}[h]
  \caption{Median of the filling factor for each scan and LB.}
  \label{Tab:ff_median}
  \centering
  \begin{tabular}{c|ccccccc}
    \hline
    \multirow{2}{*}{LB} & \multicolumn{7}{c}{Scan}                             \\ 
        & 2\_01 & 2\_02 & 3\_03 & 3\_04 & 3\_05 & 3\_06 & 5\_07 \\ 
    \hline
    \hline
    \\
    LB1 & -     & $0.84^{0.91}_{0.68}$  & $0.86^{0.92}_{0.72}$  & $0.82^{0.90}_{0.59}$  & $0.71^{0.85}_{0.39}$  & $0.61^{0.80}_{0.33}$  & $0.62^{0.77}_{0.44}$  \\ 
    \\
    LB2 & -     & $0.94^{0.97}_{0.89}$  & -     & $0.93^{0.96}_{0.84}$  & $0.86^{0.92}_{0.77}$  & $0.88^{0.91}_{0.78}$  & -     \\ 
    \\
    LB3 & $0.88^{0.93}_{0.75}$  & $0.81^{0.87}_{0.73}$  & $0.89^{0.94}_{0.77}$  & $0.88^{0.94}_{0.76}$  & $0.91^{0.96}_{0.78}$  & $0.81^{0.89}_{0.71}$  & -     \\ 
    \\
    \hline
  \end{tabular}
\end{table*}

\subsubsection{Comparison between $v_{\rm los}^{m}$ and $v_{\rm los}^{\rm nm}$} 
\label{subsec:comparison_mag_nonmag}
We now focus on the behaviour of the velocities of the magnetic ($v_{\rm los}^{\rm m}$) and non-magnetic ($v_{\rm los}^{\rm nm}$) components on the LB. To do so, we compared both velocities by analysing their velocity distributions averaged between the optical depths shown in Tab. \ref{Tab:valuesResponseFunction} (Sensitivity Range column). This is not relevant for the $v_{\rm los}^{\rm nm}$ case as it is considered to be constant with optical depth. For the non-magnetic component, we avoided those LB pixels with $ff$ above 75\% because, in most of these cases, this component was used to fit non-spectral line features (see Section \ref{sec:inversion}). Figure \ref{Fig:03may_012_comparacion_vv_lb2} shows this comparison for LB1 in Scan 3\_05 (the figures for the other scans and LBs are in Appendix \ref{appB}, from \ref{Fig:02may_017_comparacion_vv_lb2} to \ref{Fig:03may_027_comparacion_vv_lb1}, as similar results are observed for most of them).\footnote{Except for the LB1 in Scan 2\_02 shown in Figure \ref{Fig:02may_017_comparacion_vv_lb2}. A possible explanation for this diffference is given in Appendix \ref{appB}.} The top-left panel of the figure shows the distribution of $v_{\rm los}^{\rm m}$ and $v_{\rm los}^{\rm nm}$. The magnetic distribution (black line) is narrow with a width of $\Delta v_{\rm los}^{\rm m}\approx$0.5~km/s centred close to 0~km/s, so\ the magnetic component for LB1 is essentially at rest, while the distribution for $v_{\rm los}^{\rm nm}$ (red line) is broader (its width is $\Delta v_{\rm los}^{\rm nm}\approx$2.5~km/s) with values up to 4~km/s and with an average value of 2~km/s. This result is the same for all the scans and LBs except for the one shown in \ref{Fig:02may_017_comparacion_vv_lb2}. A possible explanation for a different result can be seen in \ref{appA_lb1}.

The top-right panel of Figure \ref{Fig:03may_012_comparacion_vv_lb2} shows the $ff$ map. This parameter shows a clear spatial distribution, which is smaller close to the centre of the LB (less than $\approx$15\%). Closely related to this spatial picture is the $v_{\rm los}^{\rm nm}$ map (bottom right panel of Figure \ref{Fig:03may_012_comparacion_vv_lb2}). The spine has areas of zero velocities or even upflows that are surrounded by (strong) downflows. The areas where the rest/upflows are present coincide with the area with smaller $ff$ values. This means that in those areas the LBs are mainly non-magnetic and show a convective-like pattern. In clear contrast, the magnetic component shows an average downflow character. A similar behaviour is seen in all the other scans and LBs except Scan 2\_02 (see Appendix \ref{appA} for the results and for a discussion of this exception).

\subsubsection{$v_{\rm los}^{\rm m}$ optical depth difference characterisation}
\label{subsec:height_difference_magnetic}
\begin{figure}[h]
  \begin{center}
    \includegraphics[width=0.5\textwidth]{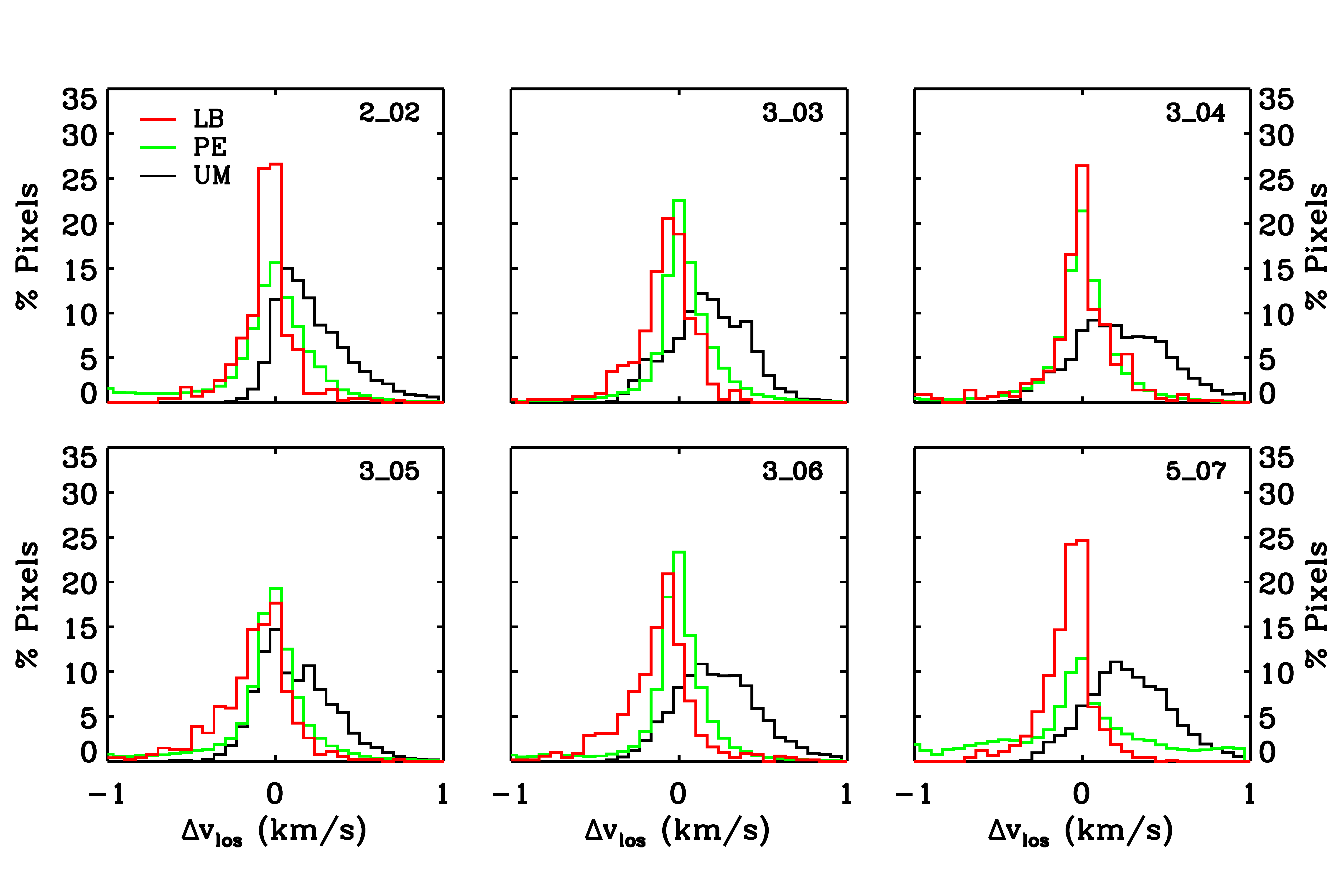}
    \caption{Same as Figure \ref{Fig:diferencias_tt_lb1}, but for the line-of-sight velocity of LB1 of the magnetic component ($\Delta v_{\rm los} = v_{\rm los}^{\rm upper} - v_{\rm los}^{\rm lower}$).} 
    \label{Fig:diferencias_vv_mag_lb1}
  \end{center}
\end{figure}

\begin{figure}[h]
  \begin{center}
    \hspace*{0.5in}
    \includegraphics[width=0.5\textwidth]{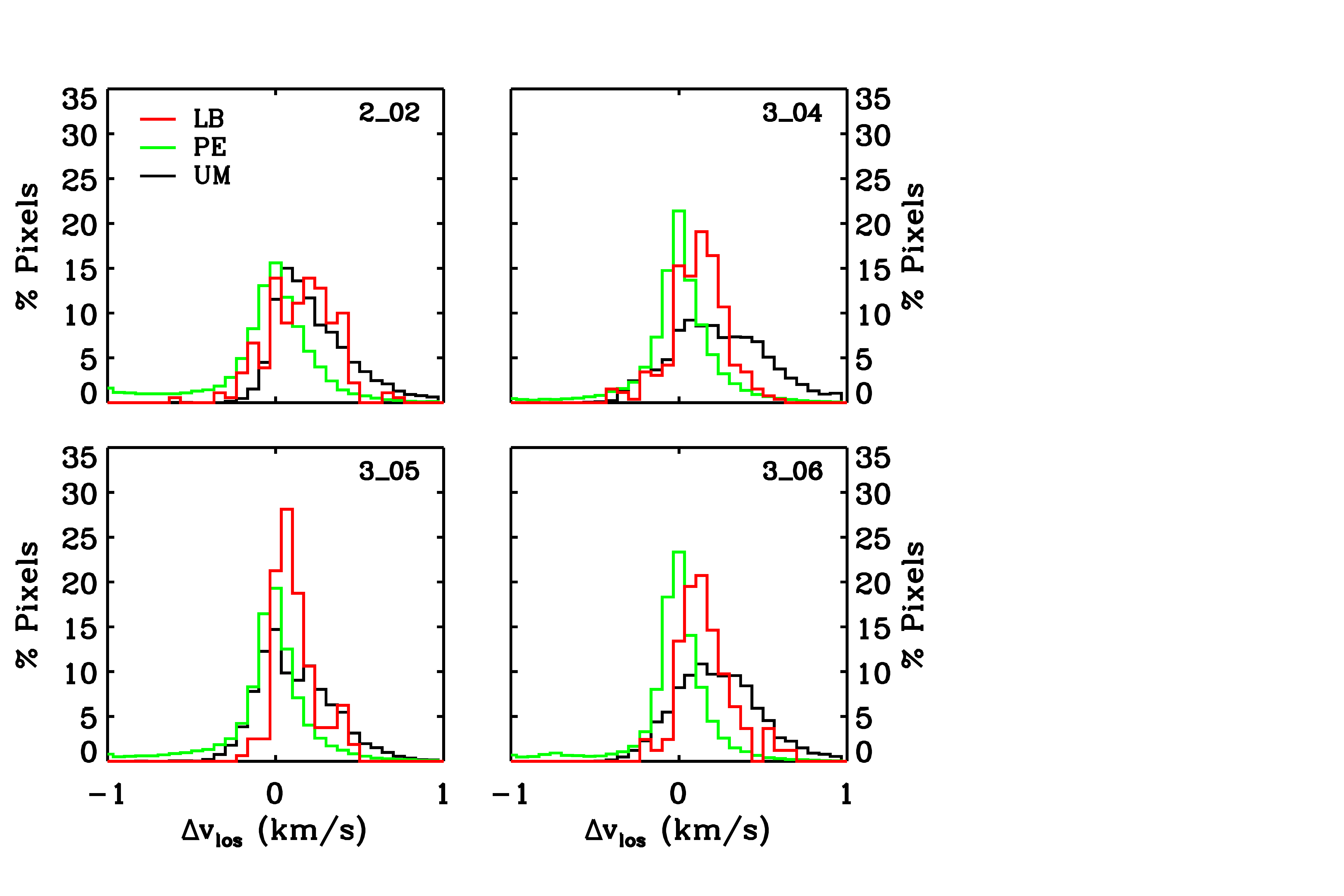}
    \caption{Same as Figure \ref{Fig:diferencias_vv_mag_lb1}, but for LB2.}
    \label{Fig:diferencias_vv_mag_lb2}
  \end{center}
\end{figure}

\begin{figure}[h]
  \begin{center}
    \includegraphics[width=0.5\textwidth]{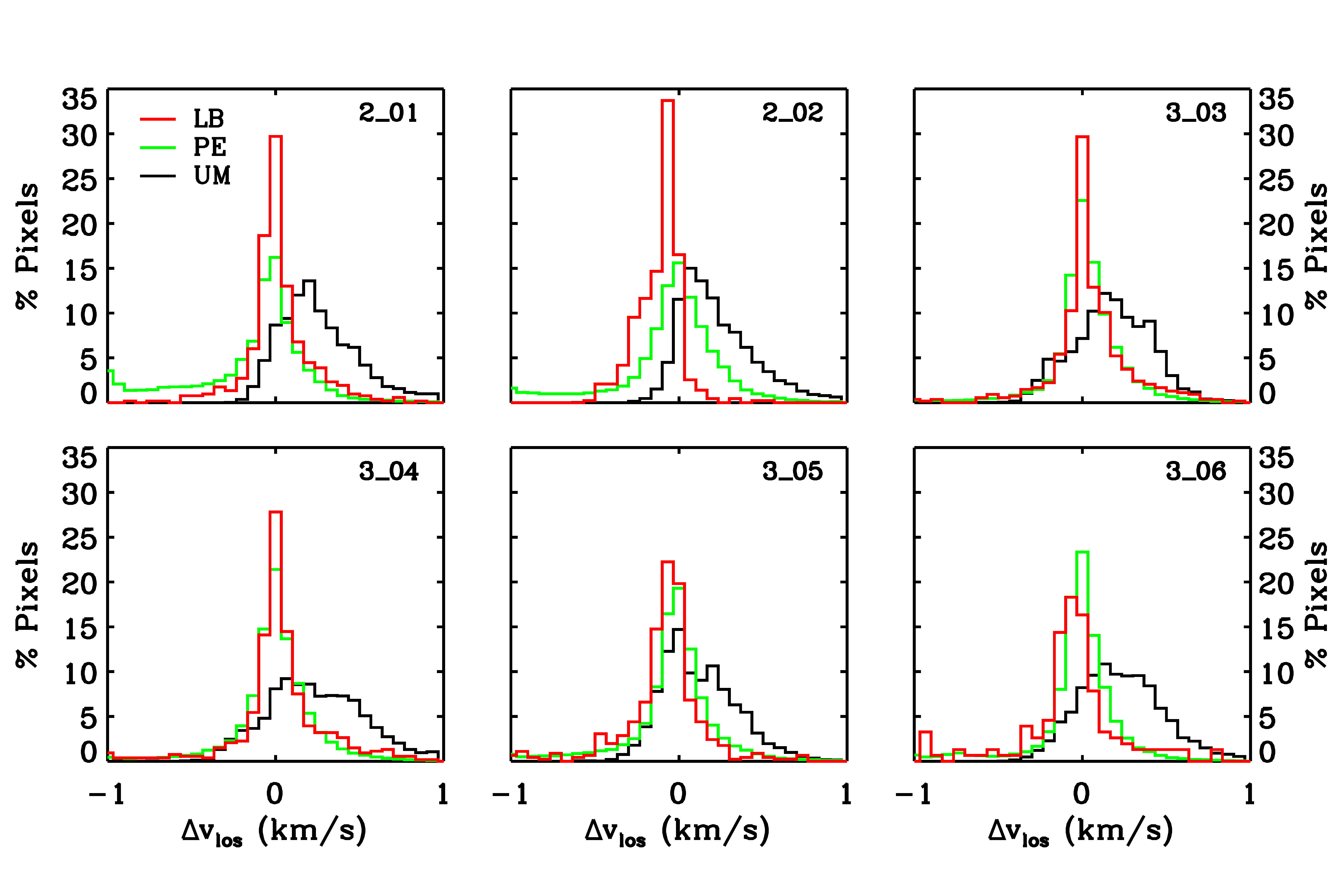}
    \caption{Same as Figure \ref{Fig:diferencias_vv_mag_lb1}, but for LB3.}
    \label{Fig:diferencias_vv_mag_lb3}
  \end{center}
\end{figure}

Following the same method used in Section \ref{results_bit} for $T$, $|\vec{B}|$, and $\gamma$, we calculated the differences of $v_{\rm los}^{\rm m}$ between the averaged upper layers and the averaged lower layers (the optical depth values to average are shown in the Tab. \ref{Tab:valuesResponseFunction}, Differences
column). Figures \ref{Fig:diferencias_vv_mag_lb1}, \ref{Fig:diferencias_vv_mag_lb2}, and \ref{Fig:diferencias_vv_mag_lb3} show the histograms of this analysis for LB1, LB2, and LB3, respectively. 

The distribution of the penumbral velocity differences for all the scans (i.e. Figure \ref{Fig:diferencias_vv_mag_lb1}, in green) is narrow and the mean is <$v_{\rm los}^{\rm m}>\approx -0.05$~km/s. Also, the distributions have long tails reaching up to $\pm$1~km/s . This difference for the UM is characterised by positive values and an asymmetric distribution. The mean value of this distribution is around 0.5~km/s, with values of 0.15~km/s of averaged velocities in the upper layers and -0.25~km/s in the lower layers, meaning\ there are weak average downflows in higher layers and weak average upflows in lower ones. On average, both distributions have a width of $\approx$0.7~km/s.

For LBs, there are two different behaviours, as seen from the distributions of the $v_{\rm los}^{\rm m}$ differences between upper and lower layers. LB1 and LB3 are very similar with average values of $\approx$-0.15~km/s and their width being $\approx$0.5~km/s. This is similar to the penumbral case, but the distributions of the differences for the LBs are slightly shifted to negative values. As for LB2, the distribution width is $\approx$0.4~km/s, and its values tend to be positive with an average of $\approx$0.25~km/s. The difference between the behaviours of LB1 and LB3 and that of LB2 could be because LB2 is very thin and there could be contamination of the surrounding umbra, as LB2's behaviour is more similar to the umbral one.

\section{Discussion and conclusions}
\label{discussion}
Spectropolarimetric data observed with GRIS have been used to study the long-term evolution of the thermodynamic and magnetic parameters of three LBs formed in the same sunspot. The spectropolarimetric data were inverted in order to infer the atmospheric parameters and study the temporal evolution (spanning a few days), spatial distribution, and optical depth variation of different atmospheric parameters in a sunspot with 3 LBs: the temperature ($T$), the magnetic field strength ($|\vec{B}|$), the local magnetic field inclination ($\gamma$), and the line-of-sight velocity ($v_{\rm los}$) in three LBs hosted by the same sunspot. Also, the relevance of the magnetic field was studied through the analysis of the relative filling factor ($ff$). Using the continuum intensity provided by the HMI instrument on board the SDO satellite, we find that each LB was formed in a different way. LB1 was formed by a strong intrusion of the PE that grows until it breaks the UM into two parts, LB2 formed by a weak intrusion of the PE that does not break the UM completely and then retreats until it disappears, and LB3 was formed by the remnant granulation as two UMs approach each other and (eventually) merge. Data from the GRIS instrument allowed us to study the topology and thermodynamics of these LBs in more detail. 

In general, the three LBs are characterised by similar atmospheric structures and behaviours. In order to explain the observed spectral profiles of LBs, we need two different atmospheric components in the inversion process: one magnetic and one non-magnetic. The magnetic component occupies a significant fraction of LB pixels, reaching up to 80\% at the borders, while in the axis of the LB it occupies between 10 and 60\%. Regarding the physical parameters, these structures are hotter than UM and cooler than PE and the QS. Their magnetic field strength is lower than that of the UM and higher than the magnetic field of the PE. Furthermore, the magnetic field lines of LBs are more inclined than the magnetic lines of the UM and more vertical than those of the PE.

Analysing the temporal evolution of the different atmospheric parameters, we find that LB1 exhibits very little time variation, whereas LB2 and LB3 exhibit much more. LB2 heats and decreases its magnetic field strength, while LB3 cools and increases its magnetic field strength, which becomes more vertical. Furthermore, it looks as if the properties of LB3 are changing from a penumbra-like structure into an umbra-like one. This could be due, as is seen from HMI data (Figure \ref{Fig:hmiSunspots}), to the fact that this LB forms by the coalescence of two UMs and finally disappears. Additionally, there are some differences between LB1 and LB2. The temperature of LB1 is closer to the penumbral temperature than that of LB2. Also, the magnetic field strength of LB2 is closer to the UM magnetic fields than that of LB1, so LB1 essentially looks more like the PE and LB2 behaves more like the UM. 

This general picture is in agreement with the bibliography (see Tab. \ref{Tab:resultadosOtrosAutores} for a summary of quantitative measurements of LBs in the literature). We find average temperature values of 6000~K, 5600~K, and 5900~K for LB1, LB2, and LB3, respectively. These values are slightly hotter than those found by \cite{jurcak2006} and \cite{lagg2014}. This could be due to the fact that we used observations in the near infrared, while they observed in the visible, so we probed deeper (hotter) into the atmosphere. We note that the temperature value in \cite{falco2016} is an upper boundary (for the biggest granules of the LB).

\begin{table*}
  \caption{Comparison of the results presented in this chapter with previous studies.}
  \label{Tab:resultadosOtrosAutores}
  \centering
  \resizebox{\textwidth}{!}{
    \begin{tabular}{c|c|c||c|c|c||c|c|c||c}
    \hline
    \hline
    \multirow{2}{*}{Paper} & \multirow{2}{*}{Spectral Lines (\AA)} & \multirow{2}{*}{Region} & \multicolumn{3}{c||}{$|\vec{B}|$ (G)} & \multicolumn{3}{c||}{$\gamma$ (deg)} & $T$ (K) \\ \cline{4-10} 
                                        &                           &                        & MIN  & MAX  & $<\ >$ & MIN & MAX & $<\ >$ & $<\ >$  \\ 
    \hline 
    \hline
    \multirow{2}{*}{\cite{kneer1973}}   & \multirow{2}{*}{6150}               & $LB$                   &      &      & 1900   &     &     &        &         \\ 
                       &                    & $Umbra$                &      &      & 2700   &     &     &        &         \\ 
    \hline
    \multirow{2}{*}{\cite{ruedi1995}}   & \multirow{2}{*}{15648.5 \& 15652.9} & $LB_{sc}$              & 2900 & 3200 &        &     &     & 40     &         \\
                       &                    & $LB_{wc}$              & 1800 & 1900 &        & 60  & 70  &        &         \\ 
    \hline
    \multirow{2}{*}{\cite{leka1997}}    & \multirow{2}{*}{6301.5 \& 6302.5}   & $15 LBs$               & 1350 & 2378 & 1729.2 & 9   & 38  & 22.8   &         \\ 
                       &                    & $Umbra$                & 1844 & 2936 & 2343.4 & 8   & 20  & 12.2   &         \\ 
    \hline
    \multirow{15}{*}{\cite{jurcak2006}}  & \multirow{15}{*}{6301.5 \& 6302.5}   & $LB1_{0km}^{narrow}$   &      &      & 1000   &     &     & 32     & 5670    \\ 
                       &                    & $LB1_{100km}^{narrow}$ &      &      & 1560   &     &     & 26     & 5250    \\ 
                       &                    & $LB1_{200km}^{narrow}$ &      &      & 1700   &     &     & 25     & 4890    \\ 
                       &                    & $LB1_{0km}^{broad}$    &      &      & 705    &     &     & 44     & 5990    \\ 
                       &                    & $LB1_{100km}^{broad}$  &      &      & 860    &     &     & 39     & 5310    \\ 
                       &                    & $LB1_{200km}^{broad}$  &      &      & 1070   &     &     & 35     & 5125    \\ 
                       &                    & $LB2_{0km}$            &      &      & 1475   &     &     & 45     & 5490    \\ 
                       &                    & $LB2_{100km}$          &      &      & 1615   &     &     & 41     & 5020    \\ 
                       &                    & $LB2_{2000km}$         &      &      & 1800   &     &     & 41     & 4680    \\ 
                       &                    & $Umbra1_{200km}$       &      &      & 2200   &     &     & 14     & 4785    \\ 
                       &                    & $Umbra1_{100km}$       &      &      & 2240   &     &     & 7      & 4280    \\ 
                       &                    & $Umbra1_{0km}$         &      &      & 2150   &     &     & 8      & 4100    \\ 
                       &                    & $Umbra2_{200km}$       &      &      & 2640   &     &     & 12     & 4440    \\ 
                       &                    & $Umbra2_{100km}$       &      &      & 2600   &     &     & 11     & 4000    \\ 
                       &                    & $Umbra2_{0km}$         &      &      & 2620   &     &     & 16     & 3740    \\ 
    \hline
    \multirow{2}{*}{\cite{shimizu2011}} & \multirow{2}{*}{6301.5 \& 6302.5}   & $LB$                   & 1000 & 1600 &        & 110 & 150 &        &         \\ 
                       &                    & $Umbra$                & 1900 & 2000 &        & 170 & 170 &        &         \\ 
    \hline
    \cite{rezaei2012}  & 6173               & $LB$                   & 1400 & 1600 &        & 37  & 50  &        &         \\ 
    \hline
    \multirow{4}{*}{\cite{louis2012}}   & \multirow{4}{*}{6301.5 \& 6302.5}   & $LB1$                    &      &      & 1720   &     &     & 158    &         \\ 
                       &                    & $LB2$                    &      &      & 1800   &     &     & 159    &         \\ 
                       &                    & $LB3$                    &      &      & 1620   &     &     & 135    &         \\ 
                       &                    & $LB_{RF}$              &      &      & 1280   &     &     & 150    &         \\ 
    \hline
    \multirow{3}{*}{\cite{lagg2014}}    & \multirow{6}{*}{6301.5 \& 6302.5}   & $LB_{log(\tau)=0}^{Int}$     &      &      & 170    &     &     & 108    & 6590    \\
                       &                    & $LB_{log(\tau)=-0.8}^{Int}$  &      &      & 60     &     &     & 108    & 5330    \\ 
                       &                    & $LB_{log(\tau)=-2}^{Int}$    &      &      & 280    &     &     & 120    & 4810    \\ 
                       &                    & $LB_{log(\tau)=0}^{Bou}$     &      &      & 320    &     &     & 95     & 6290    \\                        
                       &                    & $LB_{log(\tau)=-0.8}^{Bou}$  &      &      & 1320   &     &     & 129    & 5550    \\ 
                       &                    & $LB_{log(\tau)=-2}^{Bou}$    &      &      & 1400   &     &     & 142    & 4980    \\                        
    \hline
    \cite{louis2014}   & 6301.5 \& 6302.5   & $LB$                   & 800  & 1700 &        & 140 & 180 &        &         \\ 
    \hline
    \multirow{2}{*}{\cite{louis2015}}   & \multirow{2}{*}{6301.5 \& 6302.5}   & $LB$                   &      &      & 1688   &     &     & 156    &         \\ 
                       &                    & $Umbra$                &      &      & 2611   &     &     & 164    &         \\ 
    \hline
    \multirow{2}{*}{\cite{falco2016}}   & \multirow{2}{*}{6301.5 \& 6302.5}   & $LB_{N}$               & 700  & 1000 &        &     &     &        & 6400*   \\ 
                       &                    & $LB_{S}$               &      &      & 1500   &     &     &        &         \\ 
    \hline
    \multirow{2}{*}{\cite{felipe2016}}  & \multirow{2}{*}{10827 \& 10839}     & $LB_{log(\tau)=0.3}$   &      &      & 30     &     &     & 76     &         \\ 
                       &                    & $LB_{log(\tau)=-2.2}$  &      &      & 1300   &     &     & 150    &         \\ 
    \hline
    \cite{okamoto2018} & 6301.5 \& 6302.5   & $LB$                   &      & 6250 &        &     &     &        &         \\
    \hline
    \multirow{3}{*}{\cite{guglielmino2019}} & \multirow{3}{*}{6301.5 \& 6302.5} & $UF$        &1500 &2000 & 2240   &   &        & 78 &     \\
                                            &                                   & $Umbra$     &     &     & 2900   &     &     & 147 &     \\ 
                                            &                                   & $Penumbra$  &     &     & 1900   &     &     & 113 &     \\ 
    \hline
    \multirow{4}{*}{This work}        & \multirow{4}{*}{15648.5 \& 15662.0} & $LB1$                  &      &      & 1800   &     &     & 25     & 6300    \\ 
                       &                    & $LB2$                  &      &      & 2300   &     &     & 20     & 5800    \\ 
                       &                    & $LB3$                  &      &      & 1900   &     &     & 30     & 6000    \\ 
                       &                    & $Umbra$                &      &      & 2500   &     &     & 17     & 4800    \\ 
    \hline
    \end{tabular}
    }
    \tablefoot{The first column is the paper reference, the second is the spectral lines used for the study, the third is the region analysed, the next six columns correspond to the minimum, maximum, and mean of the magnetic field strength and inclination, and the last column is the average temperature. When the authors do not study a parameter, the cell is empty. The meaning of the subindices are: \textit{sc} and \textit{wc} from \citealt{ruedi1995} are strong and weak atmospheric components, respectively; \textit{FR} from \citealt{louis2012} means LB in formation; \textit{N} and \textit{S} from \citealt{falco2016} means the north and south area of the LB; and * from \citealt{falco2016} means that the temperature was calculated in the biggest granule of the northern part of the LB. The superindices from \citealt{lagg2014} mean that the values are calculated in the interior and in the boundary of a granule in a LB.}
\end{table*}

On average, the magnetic field strengths found for LB1, LB2, and LB3 are 1800~G, 2300~G, and 1900~G, respectively. These values are similar to those found by \cite{kneer1973}, \cite{ruedi1995}, \cite{leka1997}, \cite{jurcak2006}, \cite{rezaei2012}, \cite{shimizu2011}, \cite{louis2012}, \cite{louis2014}, \cite{louis2015}, and \cite{falco2016}. Our data suggest that there might be an anti-correlation between the temperature and the magnetic field strength: the cooler the LBs, the stronger its magnetic field.

The magnetic field lines of the LBs are more horizontal than those of the UM. The averaged values for LB1, LB2, and LB3 are 25, 20, and 30~deg, respectively. These values are in agreement with those reported by \cite{leka1997}, \cite{jurcak2006}, \cite{louis2012}, \cite{louis2014}, \cite{louis2015}, and \cite{guglielmino2019}. As before, despite the similar global behaviour, LB3 seems to show an evolution with time of the average inclination. Initially, the magnetic field lines are quite inclined ($\approx$40~deg), and, as time evolves, they shift to an average of $\approx$25~deg. In addition, on May 3, LB3 suffers an abrupt change to more inclined fields, which is in agreement with the HMI sequence (a protrusion of the PE into the UM is seen in the southern part of the LB in the panel `20:00 02/May/2014').

From these results, although the three LBs show characteristic properties of these structures, each one has its own peculiarities. LB1 does not change its atmospheric parameters with time, while LB2 and LB3 do. LB1 is more similar to the PE (hotter and with weaker magnetic field strength), while LB2 is cooler and has a stronger magnetic field (more similar to the UM). LB3 cools and its magnetic field strength increases with time, going from values close to that of the PE to that of the UM. 

The spatial distribution of $T$, $\gamma,$ and $|\vec{B}|$ of the LBs is in general agreement with previous results of other authors ($|\vec{B}|_{\rm um}>|\vec{B}|_{\rm lb}>|\vec{B}|_{\rm pe}$, $T_{\rm pe}>T_{\rm lb}>T_{\rm um}$, and $\gamma_{\rm pe}>\gamma_{\rm lb}>\gamma_{\rm um}$). In addition, there are some features, such as the intrusion that LB3 suffers in Scans 3\_03, 3\_05, and 3\_06, where the physical properties are a bit different than for the rest of the LB (almost horizontal magnetic field lines, weak magnetic field strength, and hotter temperatures). Moreover, we found that the spatial extension of the LB-like physical properties is different for $T_{\rm lb}$ and $|\vec{B}|_{\rm lb}$ than for $\gamma_{\rm lb}$, as the former extend more than the latter. The inclination at the tip of LBs is always more vertical (similar to the umbral values) than the rest of the LB. This result could be consistent with the work done by \cite{lindner2020}, who studied the {Jur{\v c}{\'a}k} criterion \citep{jurcak2018}. This is probably a stability criterion for the UM (though this is still an open discussion, see \citealt{loptien2020}). For this particular sunspot, they found that the {Jur{\v c}{\'a}k} criterion applies to part of the LB, meaning\ it could be related to the part of the LB that has higher inclinations, while the tip would be ascribed to the fact that it is `under evolution/formation'. Another possibility is that the LB tip is associated with the emergence of a horizontal magnetic field \citep{toriumi2015b}, but we do not observe the convection pattern in the magnetic velocity so we would need more information to confirm if this is due to the horizontal emergence flux. Also, this study reveals another interesting feature: the biggest LB is wide enough to show a cool and weak patchy structure along the axis of the LB. This result could imply the presence of a dark lane whose existence cannot be determined as the spatial resolution is not high enough.

Making a differential analysis between two optical depths we found that the PE cools the most (2200~K) with optical depth, while the UM and LB are quite similar, with the latter cooling down slightly less than the UM. This result for the LB temperature is consistent with previous studies (see i.e. \citealt{jurcak2006} and \citealt{lagg2014} from Tab. \ref{Tab:resultadosOtrosAutores}). Moreover, the temperature of the LBs shows an additional peak ($\Delta T\approx$-3000~K) for some scans where the temperature changes faster with optical depth. These secondary peak values are located along the axis of the LB and could be related to the dark lane reported by other authors. The magnetic field strengths of the UM, PE, and LB are stronger in the lower layers than in the upper ones. The PE shows the steepest decrease in magnetic field, the UM the shallowest, and the LB magnetic field decreases between these two layers. Regarding the inclination of the magnetic field lines, the UM shows no average change between the two layers considered, while the PE and LBs do show an average variation of inclination with optical depth, with more inclined field lines in the upper layers for the LBs and with more vertical field lines in the upper layers for the PE. The results for the magnetic parameters are opposite to those presented by \cite{jurcak2006} and \cite{lagg2014}, as in our case $|\vec{B}|$ decreases and becomes more horizontal with optical depth. Whether this difference arises from the different spectral range used or because the LBs are intrinsically different from those studied by these authors is unclear. Ideally, LBs observed simultaneously in both wavelength ranges would be required to further address this point.

Analysing the filling factor values shown in Tab. \ref{Tab:ff_median}, we can see that the cores of LBs are highly non-magnetic, while the outer parts have a magnetic character. The fact that between 6\% and 29\%  (as inferred from the median value of the LBs analysed on this paper) of the LBs are non-magnetic is in agreement with \cite{leka1997}, who observed that often more than 20\% of the material in the LB is non-magnetic. An open question that remains from this work concerning the filling factor is the fact that the outer parts of the LBs are much more magnetic. This could be related to 1) a residual of stray light that could not be removed using the techniques applied to the data, or 2) the optical depths the observed spectral line is sensitive to are different for the core and the outer parts of the LB, that is, the physical properties probed by the spectral line are different in the core and the outer parts of the LBs. 

The analysis of $v_{\rm los}^{\rm m}$ shows that the UM is almost at rest, while the velocity difference of the penumbral material is positive with optical depth. We find that LB1 and LB3 behave similarly and hardly change with optical depth, while LB2 behaves like the PE with a positive velocity difference with optical depth (the velocity in the upper layers being 0.2~km/s greater than in the lower layers). $v_{\rm los}^{\rm nm}$ shows a common behaviour for all the LBs and scans (save for one, whose behaviour we cannot explain). We found that the velocity of the magnetic component is close to zero or shows small downflows, and the velocity of the non-magnetic component has a convective-like character. The spine of the LBs is at rest or even has small upflows and is surrounded by strong downflows. These results are in agreement with the bibliography (see Tab. \ref{Tab:resultadosVlos} for a summary of $v_{\rm los}$ measurements of LBs in the literature). Some authors, including \cite{rimmele2008}, \cite{giordano2008}, \cite{rouppe2010}, \cite{lagg2014}, \cite{toriumi2015a}, and \cite{guglielmino2017}, found that the spine of the LBs shows upflows, while the edges show downflows (the same behaviour as we see in the non-magnetic atmosphere). However, they do not distinguish between magnetic and non-magnetic components, so it is not easy to compare directly. Also, the $v_{\rm los}^{\rm m}$ values calculated in this paper are similar to the results obtained by \cite{ruedi1995} and \cite{shimizu2011}, who observed downflows in the whole LB.

\begin{table*}
  \caption{Comparison of the results of the $v_{\rm los}$ obtained in this work with previous studies.}
  \label{Tab:resultadosVlos}
  \centering
  \resizebox{\textwidth}{!}{
    \begin{tabular}{c|c|c|c|c}
    \hline
    \hline
    Paper & Spectral Lines & Upflows & Downflows & Upflows \& Downflows \\  
          & (\AA)          & (km/s)  & (km/s)    & (km/s)               \\  
    \hline
    \hline
        \cite{beckers1969} &  6173.3 \& 5576.1            & U         &          &   \\
    \hline
        \cite{ruedi1995}   &  15648.5 \& 15652.9          &           & 1.5 &   \\
    \hline
        \cite{leka1997} & 6301.5 \& 6302.5  &   &   &  U \& D \\    
    \hline
        \cite{rimmele1997} & 5576 \& 5691  &   &   & U \& D  \\    
    \hline
        \cite{hirzberger2002} & 5425  & 0.2 - 1.5   &   &   \\    
    \hline
        \cite{schleicher2003} & 15648.5 \& 15652.9  & 0.3   &   &   \\    
    \hline
        \multirow{2}{*}{\cite{bharti2007}} & \multirow{2}{*}{5576 \& 6302}  &   &   & U: 0.1 - 0.2 on the LB axis  \\  
          &  &  &    & D: 0.5 - 1.4 on the LB edges  \\  
    \hline
        \cite{katsukawa2007} & 6301.5 \& 6302.5  & 0.2  &   &   \\    
    \hline
        \cite{rimmele2008} & 5434  & U &   &   \\    
    \hline
        \multirow{2}{*}{\cite{giordano2008}} & \multirow{2}{*}{7090.4}  &   &   & U: 0.07 on the LB axis  \\  
          &  &  &    & D: 0.15 on the LB edges  \\    
    \hline
        \cite{louis2009} & 6301.5 \& 6302.5  &   & Supersonic  &   \\    
    \hline
        \multirow{2}{*}{\cite{rouppe2010}} & \multirow{2}{*}{6301.5 \& 6302.5}  &   &   & U: 0.5 - 1 on the LB axis  \\  
          &  &  &    & D: lower on the LB edges  \\    
    \hline
        \cite{shimizu2011} & 6301.5 \& 6302.5 &   & 0.5 - 2  &   \\    
    \hline
        \multirow{2}{*}{\cite{lagg2014}} & \multirow{2}{*}{6301.5 \& 6302.5}  &   &   & U: 2 on the LB axis  \\  
          &  &  &    & D: supersonic on the LB edges  \\
    \hline
        \multirow{2}{*}{\cite{toriumi2015a}} & \multirow{2}{*}{6301.5 \& 6302.5}  &   &   & U: 1 on the LB axis  \\  
          &  &  &    & D: 6 on the LB edges  \\   
    \hline
        \cite{guglielmino2019} & 6301.5 \& 6302.5  &   & D  &   \\ 
    \hline
        \multirow{2}{*}{This work} & \multirow{2}{*}{15648.5 \& 15662.0}  &   &   & U: on the LB axis  \\  
          &  &  &    & D: on the LB edges  \\   
    \hline
    \end{tabular}
  }
  \tablefoot{The first column is the reference paper, the second is the spectral lines used for the study, and the third, fourth, and fifth columns show if the LB shows upflows, downflows or both and their values, respectively. U means upflows and D downflows.}
\end{table*}

\begin{acknowledgements}
 ABGM would like to thank Matthias Rempel for the revision of the paper and his suggestions. The authors are grateful to the SDO/HMI team for their data. ABGM acknowledges Fundaci\'on La Caixa for the financial support received in the form of a Ph.D. contract. This work was supported by NASA Contract NAS5-02139 (HMI) to Stanford University. We thank the University Corporation for Atmospheric Research and their NCAR/HAO programs. This work was supported by the HAO Newkirk Fellowship. This project has received funding from the European Research Council (ERC) under the European Union's Horizon 2020 research and innovation programme (SUNMAG, grant agreement 759548). The Institute for Solar Physics is supported by a grant for research infrastructures of national importance from the Swedish Research Council (registration number 2017-00625). This material is based upon work supported by the National Center for Atmospheric Research, which is a major facility sponsored by the National Science Foundation under Cooperative Agreement No. 1852977. The authors gratefully acknowledge support from the Spanish Ministry of Economy and Competitivity through project AYA2014-60476-P (Solar Magnetometry in the Era of Large Solar Telescopes),as well as project PGC2018-102108-B-I00 and FEDER funds. This paper made use of the IAC Supercomputing facility HTCondor (http://research.cs.wisc.edu/htcondor/), partly financed by the Ministry of Economy and Competitiveness with FEDER funds, code IACA13-3E-2493. 
\end{acknowledgements}

\bibliographystyle{aa}
\bibliography{articulos} 


\clearpage

\begin{appendix} 

\section{Height difference characterisation of the magnetic and thermal parameters $|\vec{B}|$, $\gamma$, and $T$}
\label{appA}
\begin{figure}[h]
  \begin{center}
    \includegraphics[width=0.65\textwidth]{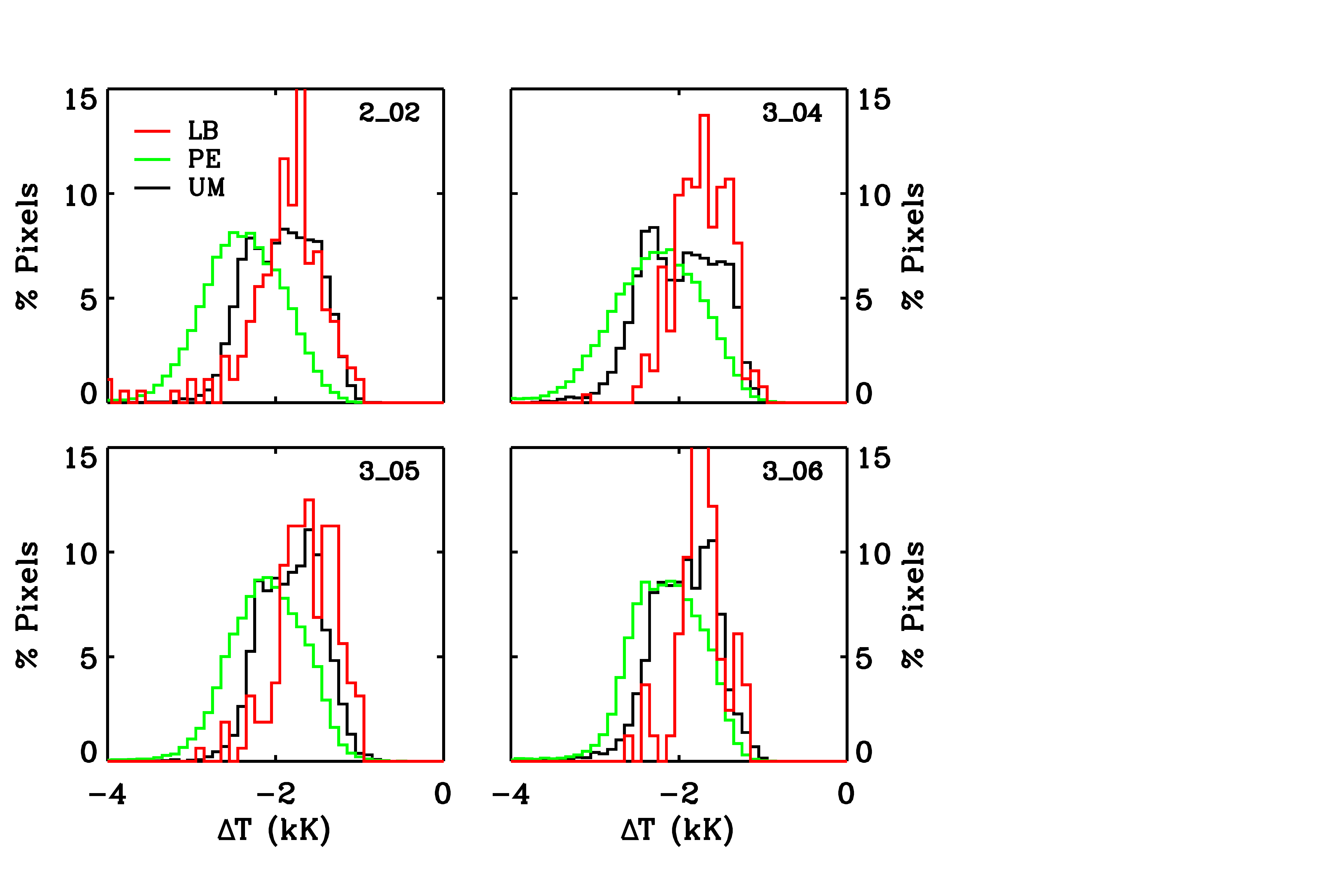}
    \caption{Same as Figure \ref{Fig:diferencias_tt_lb1}, but for LB2.}
    \label{Fig:diferencias_tt_lb2}
  \end{center}
\end{figure}

\begin{figure}[h]
  \begin{center}
    \includegraphics[width=0.5\textwidth]{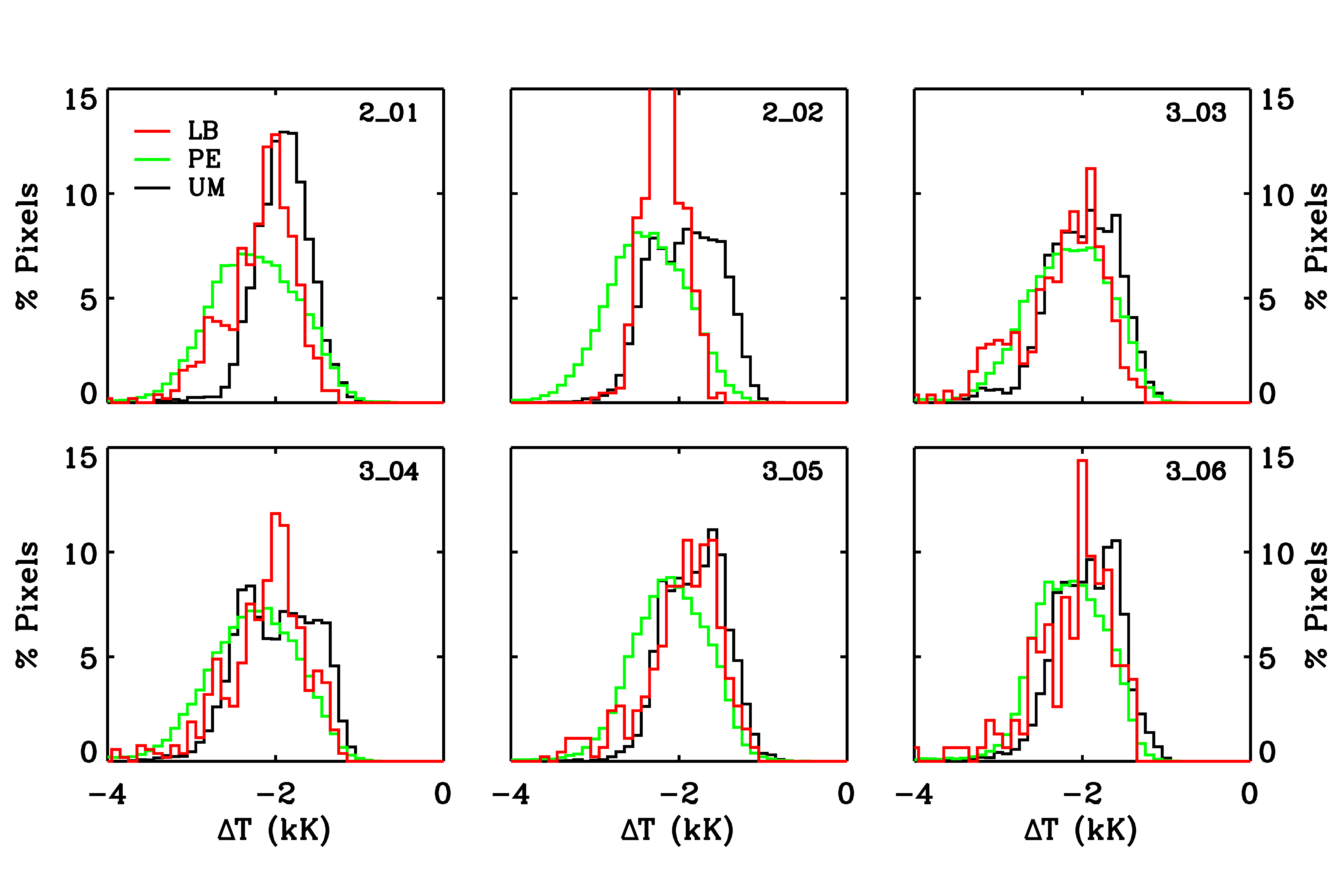}
    \caption{Same as Figure \ref{Fig:diferencias_tt_lb1}, but for LB3.}
    \label{Fig:diferencias_tt_lb3}
  \end{center}
\end{figure}

\begin{figure}[h]
  \begin{center}
    \includegraphics[width=0.5\textwidth]{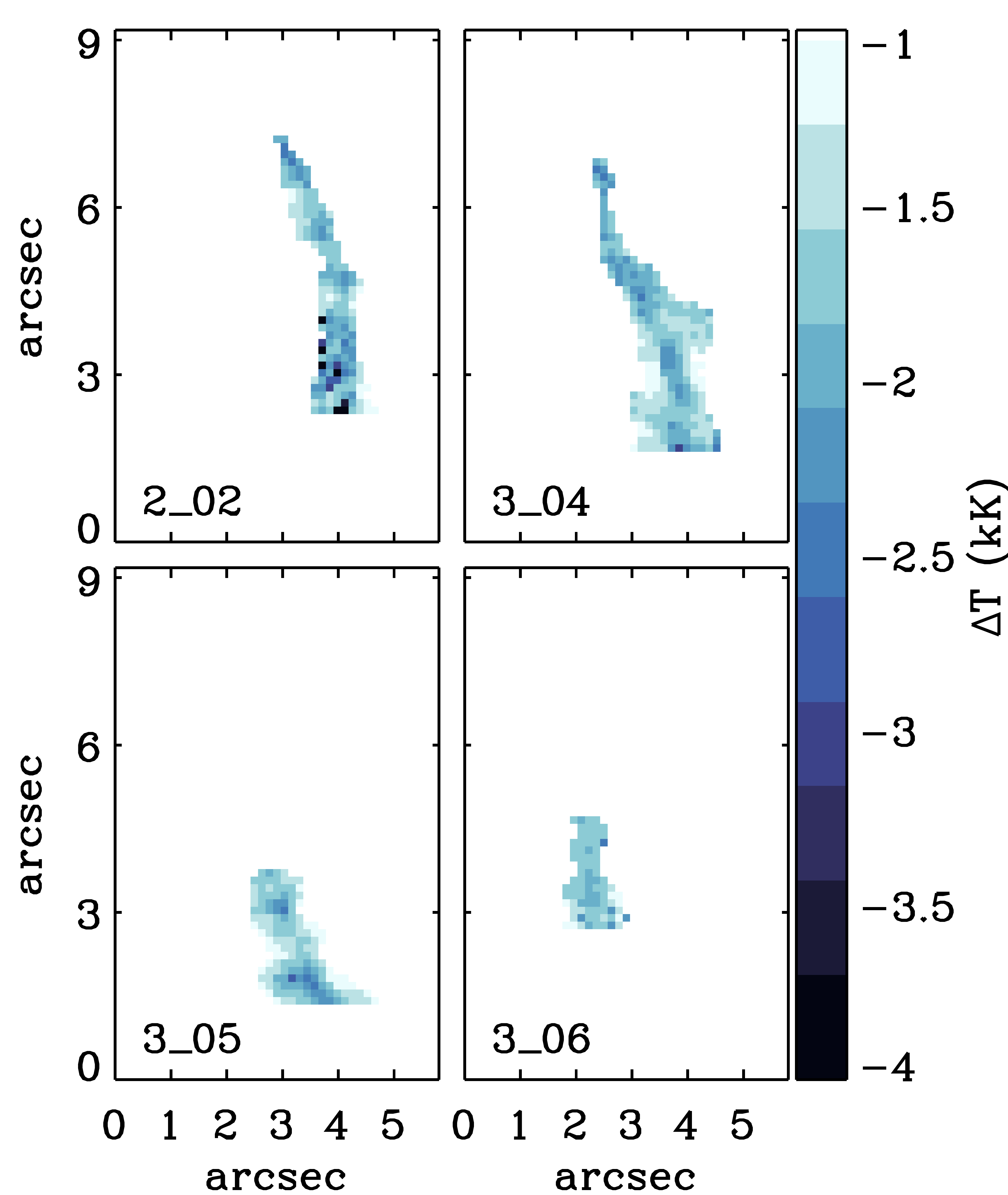}
    \caption{Same as Figure \ref{Fig:mapa_dif_tt_lb1}, but for LB2.}
    \label{Fig:mapa_dif_tt_lb2}
  \end{center}
\end{figure}

\begin{figure}[h]
  \begin{center}
    \includegraphics[width=0.5\textwidth]{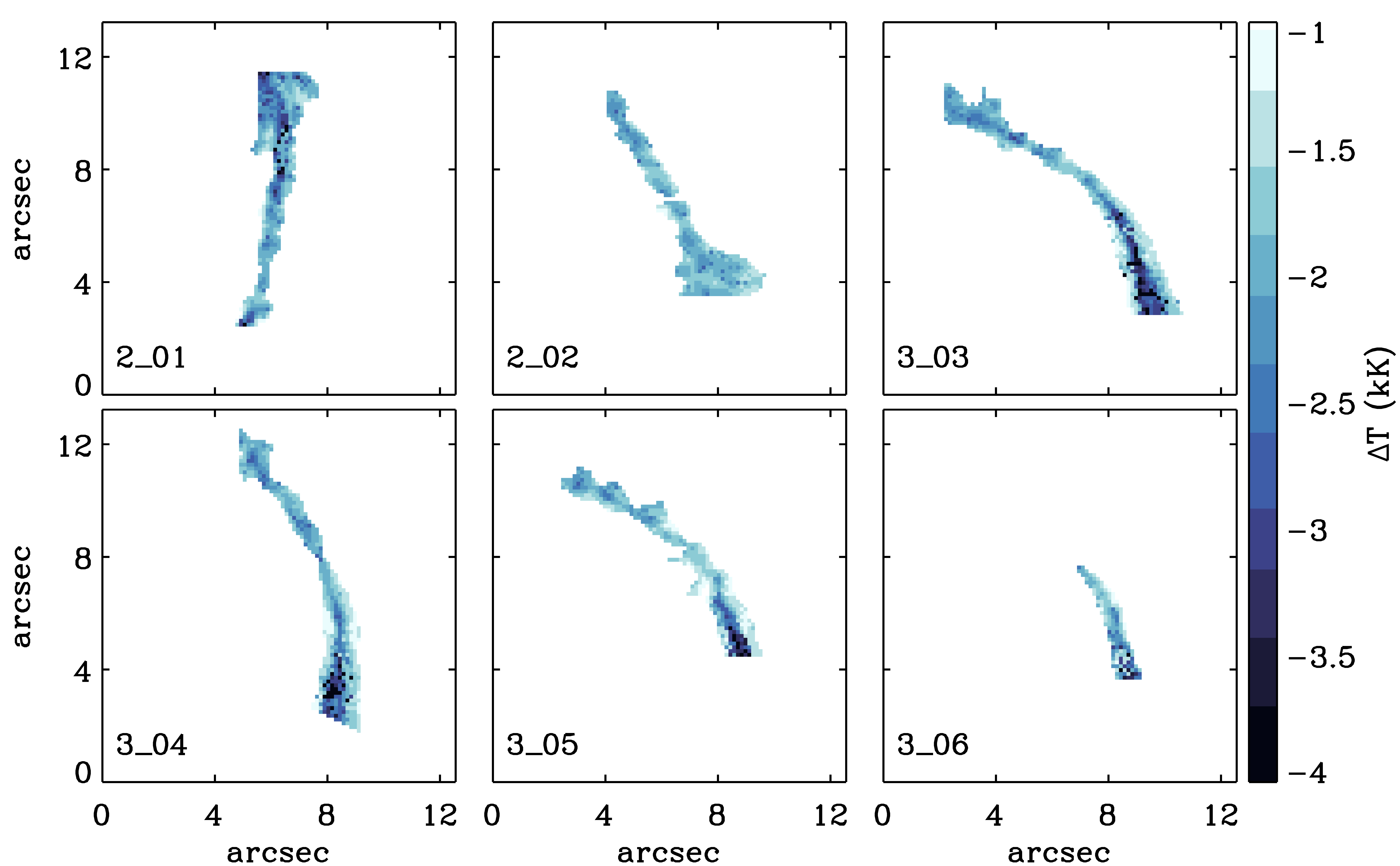}
    \caption{Same as Figure \ref{Fig:mapa_dif_tt_lb1}, but for LB3.}
    \label{Fig:mapa_dif_tt_lb3}
  \end{center}
\end{figure}

\begin{figure}[h]
  \begin{center}
    \includegraphics[width=0.65\textwidth]{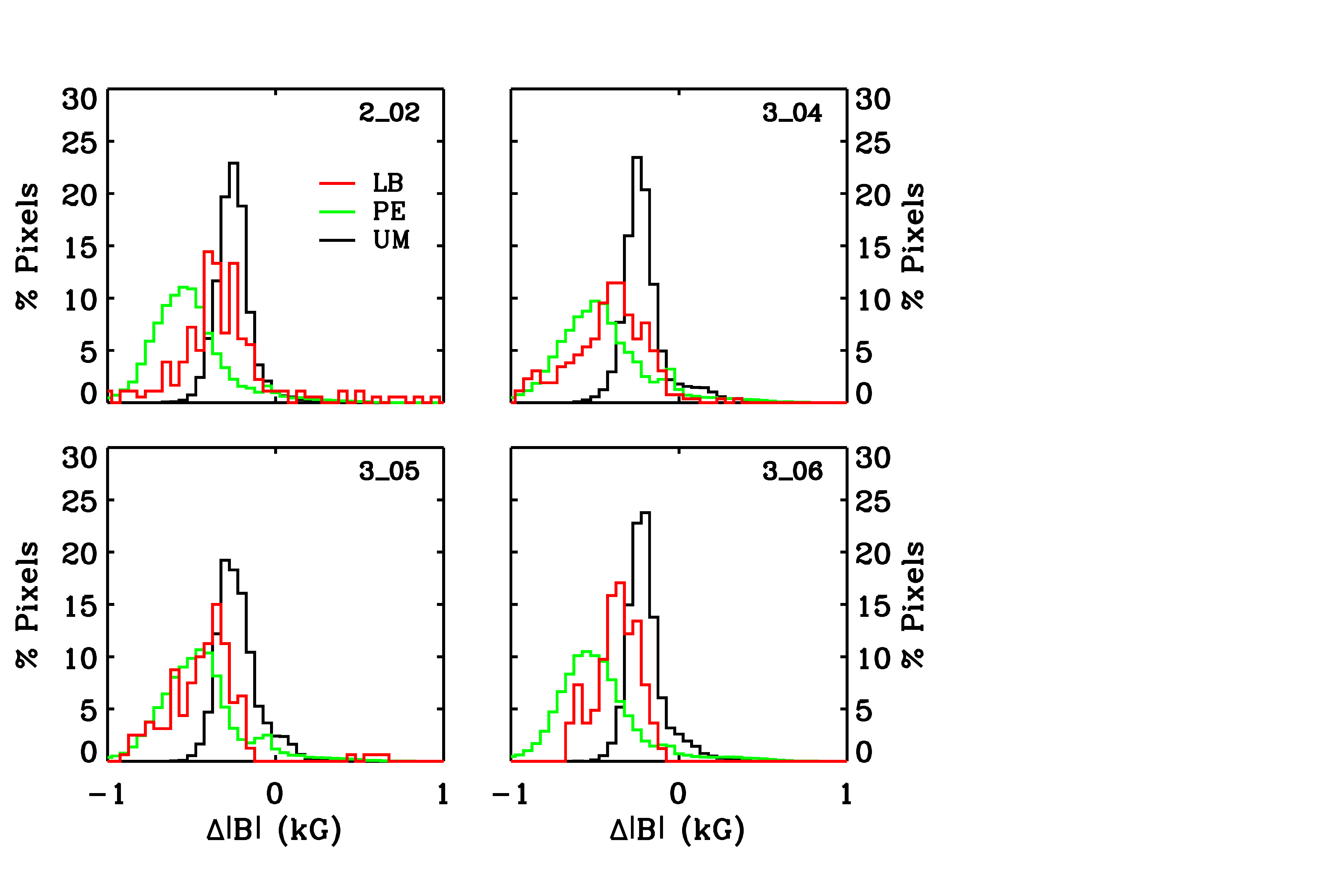}
    \caption{Same as Figure \ref{Fig:diferencias_bb_lb1}, but for LB2.}
    \label{Fig:diferencias_bb_lb2}
  \end{center}
\end{figure}

\begin{figure}[h]
  \begin{center}
    \includegraphics[width=0.5\textwidth]{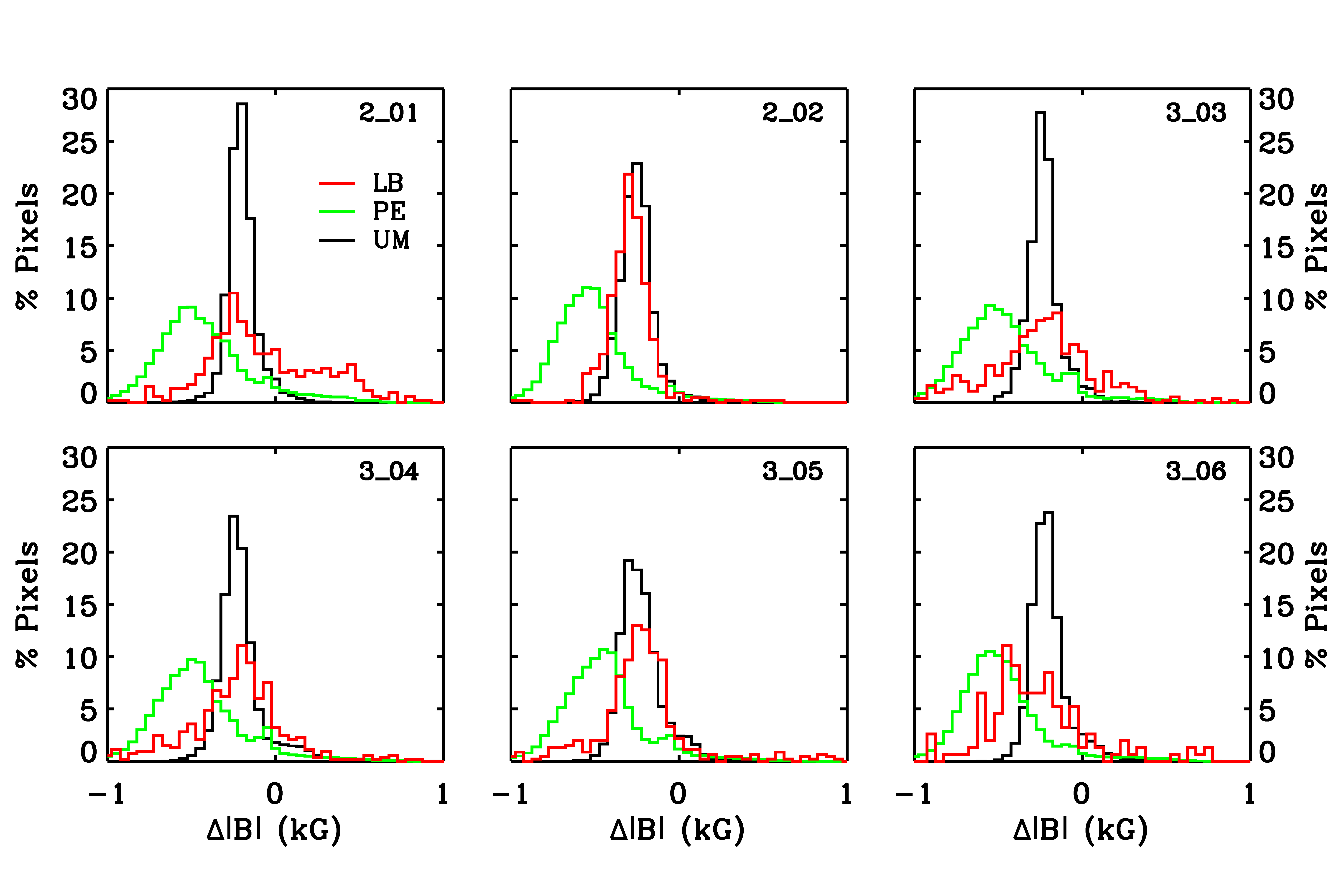}
    \caption{Same as Figure \ref{Fig:diferencias_bb_lb1}, but for LB3.}
    \label{Fig:diferencias_bb_lb3}
  \end{center}
\end{figure}

\begin{figure}[h]
  \begin{center}
    \includegraphics[width=0.65\textwidth]{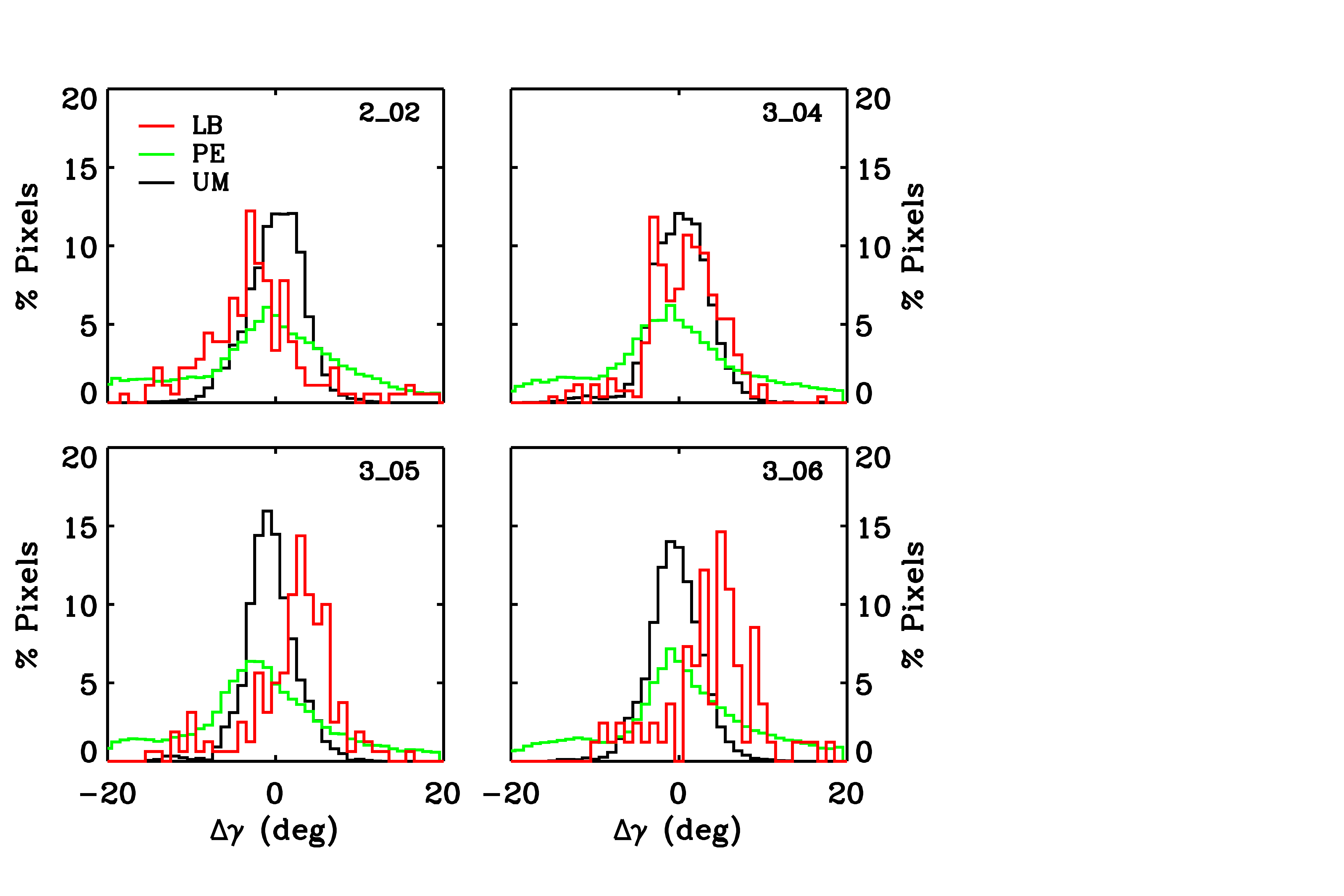}
    \caption{Same as Figure \ref{Fig:diferencias_ii_lb1}, but for LB2.}
    \label{Fig:diferencias_ii_lb2}
  \end{center}
\end{figure}

\begin{figure}[h]
  \begin{center}
    \includegraphics[width=0.5\textwidth]{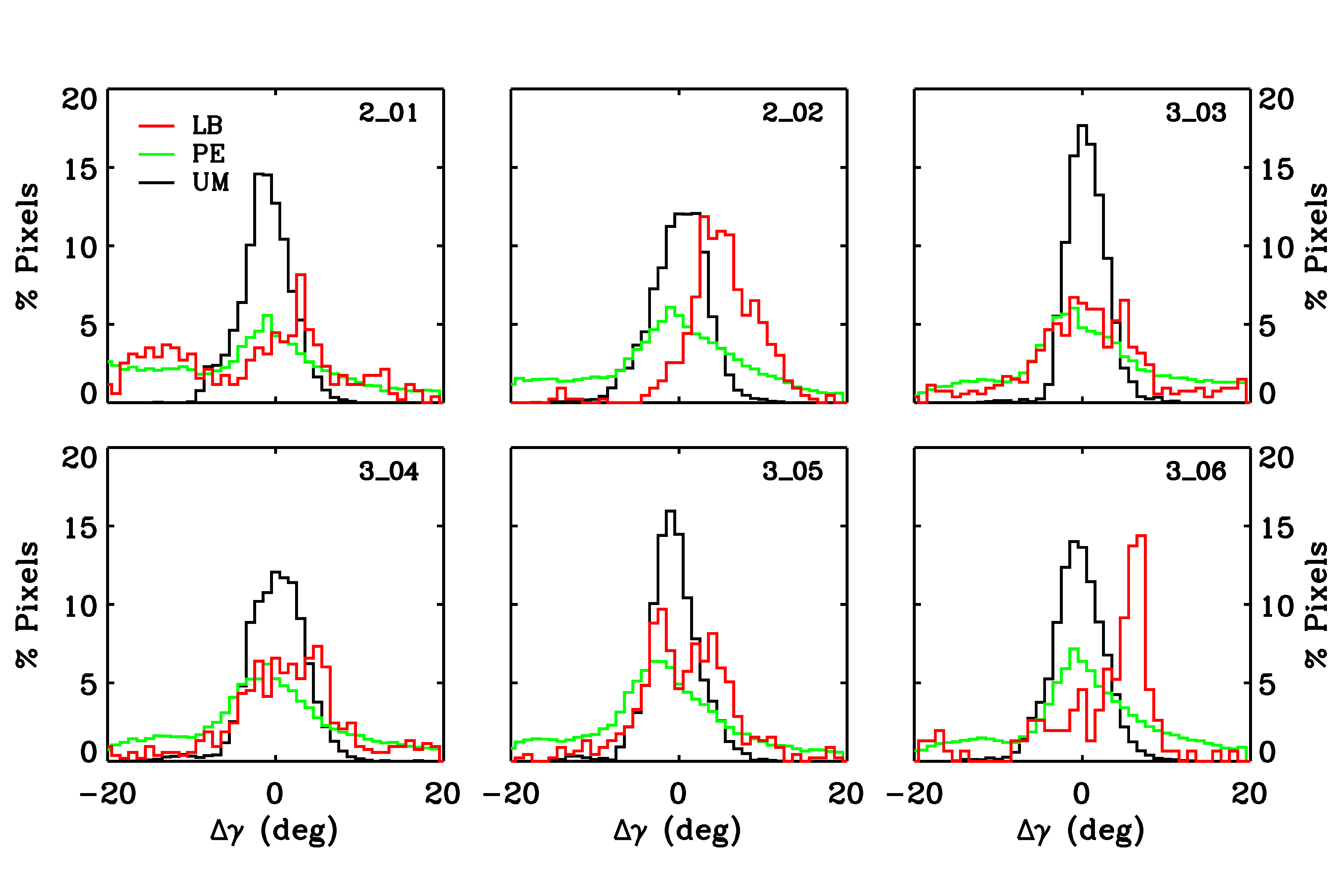}
    \caption{Same as Figure \ref{Fig:diferencias_ii_lb1}, but for LB3.}
    \label{Fig:diferencias_ii_lb3}
  \end{center}
\end{figure}

In this appendix, we present the results of the LBs described in Section \ref{results_bit}. All figures of this appendix show the height difference characterisation. Figures\ \ref{Fig:diferencias_tt_lb2}--\ref{Fig:mapa_dif_tt_lb3} are for the temperature analysis, Figures\ \ref{Fig:diferencias_bb_lb2} and \ref{Fig:diferencias_bb_lb3} are for the magnetic field strength analysis, and Figures\ \ref{Fig:diferencias_ii_lb2} and \ref{Fig:diferencias_ii_lb3} for the inclination analysis. The black line is for the pixels of the umbra, the green one is for the pixels from the penumbra, and red represents the distribution of the LB pixels. 

\subsection{Temperature}
\label{appA_tt}

\clearpage

\subsection{Magnetic field strength}
\label{appA_bb}

\subsection{Magnetic field inclination}
\label{appA_ii}

\clearpage


\section{Comparison between $v_{\rm los}^{m}$ and $v_{\rm los}^{\rm nm}$}
\label{appB}
In this appendix, we present the comparison between the line-of-sight velocities of the magnetic and non-magnetic components for the three LBs in the different scans. These plots were calculated following the procedures explained in Section \ref{subsec:comparison_mag_nonmag}.

\subsection{Light Bridge 1}
\label{appA_lb1}

\begin{figure}[h]
  \begin{center}
    \includegraphics[width=0.5\textwidth]{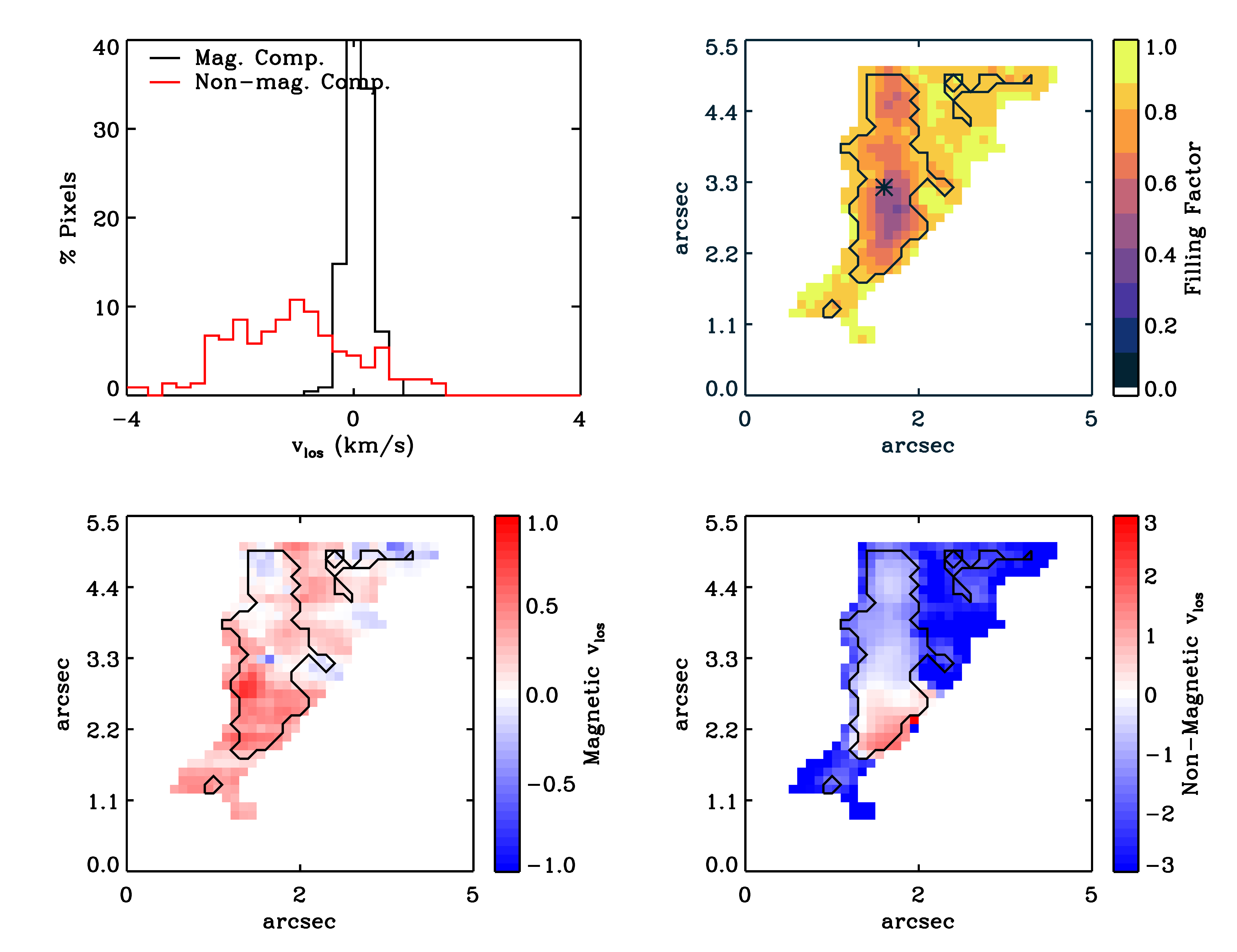}
    \caption{Same as Figure \ref{Fig:03may_012_comparacion_vv_lb2}, but for LB1 of Scan 2\_02.}
    \label{Fig:02may_017_comparacion_vv_lb2}
  \end{center}
\end{figure}

\begin{figure}[h] 
  \begin{center}
    \includegraphics[width=0.5\textwidth]{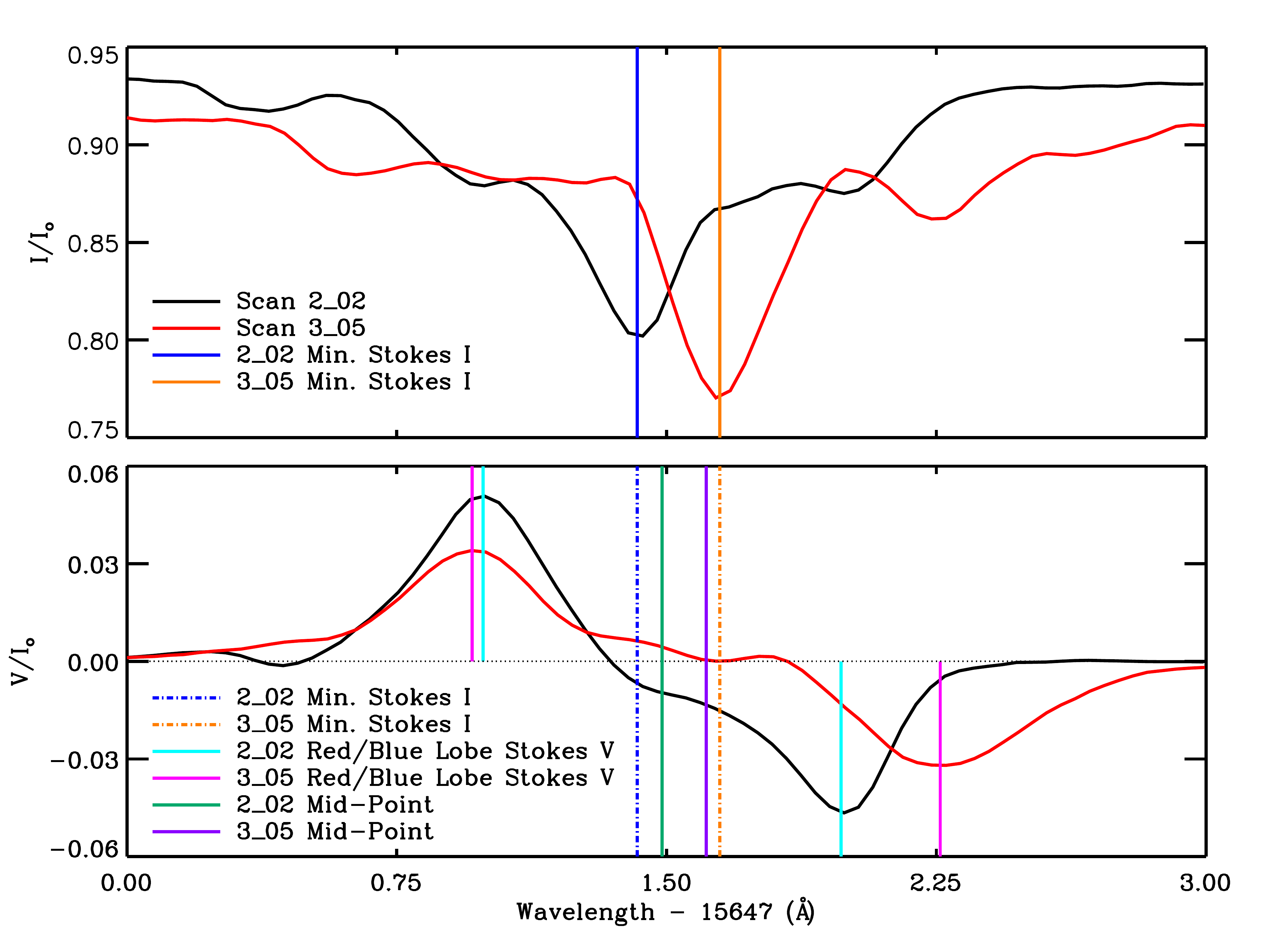}
    \caption{Comparison between Stokes $I$ and $V$ for the pixels marked by white asterisks in Figures \ref{Fig:03may_012_comparacion_vv_lb2} and \ref{Fig:02may_017_comparacion_vv_lb2}. The red and black lines correspond to the Stokes $I$ and $V$ profiles of Scans 3\_05 and 2\_02, respectively. The vertical solid lines of the top panel and the dashed lines of the bottom panel represent the core of the Stokes $I$ profiles. The vertical light blue and pink lines of the bottom panel mark the red and blue lobes of the Stokes $V$ profiles for Scans 2\_02 and 3\_05, respectively, and the green and purple lines correspond to the mid-point of Stokes $V$ profiles.}
    \label{Fig:comparacion_perfiles}
  \end{center}
\end{figure}

Figure \ref{Fig:02may_017_comparacion_vv_lb2} shows the comparison of $v_{\rm los}^{\rm m}$ and $v_{\rm los}^{\rm nm}$ for LB1 in Scan 2\_02. The histogram shows that the distribution of $v_{\rm los}^{\rm m}$ is similar to that of Figure \ref{Fig:03may_012_comparacion_vv_lb2}, narrow and around 0~km/s, while the distribution of $v_{\rm los}^{\rm nm}$ has negative values (instead of positive). In order to find out whether this change in behaviour is the result of the data reduction, deconvolution or inversion process, or whether there is an error in the data analysis, we compared the Stokes $I$ and $V$ profiles of two different pixels (white asterisks of Figures \ref{Fig:03may_012_comparacion_vv_lb2} and \ref{Fig:02may_017_comparacion_vv_lb2}). These pixels were selected because both atmospheric components have a similar contribution to the final profile ($ff\approx$50\%). In this way, we are quite sure that we can estimate both velocities separately and accurately enough. For these pixels (as representative pixels), we calculated the position of the minimum of the Stokes $I$ profiles and compared it with the mid-point of the Stokes $V$ profiles (see Figure \ref{Fig:comparacion_perfiles}). For Scan 3\_05, the minimum of the Stokes $I$ (orange line of top panel) is redshifted with respect to the mid-point of Stokes $V$ (purple line of bottom panel), while the minimum of the Stokes $I$ (dark blue line of top panel) of Scan 2\_02 is blueshifted with respect to the mid-point of Stokes $V$ (green line of bottom panel). Thus, the profiles are consistent with the results obtained with SIR. Now, the reason why the $v_{\rm los}^{\rm nm}$ distributions for the three LBs for this single scan are blueshifted is not clear and needs additional study. We can be sure that it is not a problem of the velocity calibration since the UM, PE, and QS behave consistently.

\begin{figure}[h] 
  \begin{center}
    \includegraphics[width=0.5\textwidth]{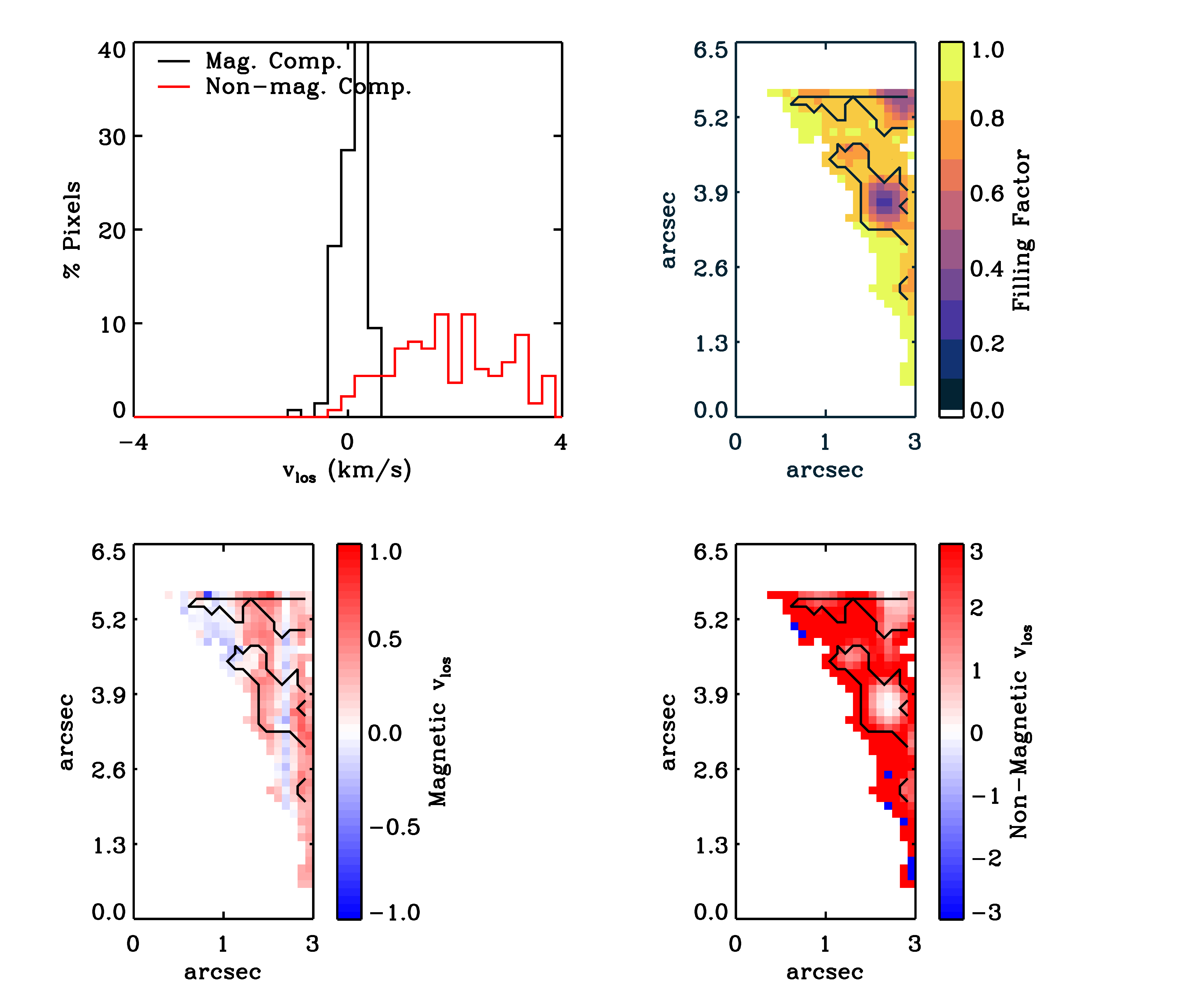}
    \caption{Same as Figure \ref{Fig:03may_012_comparacion_vv_lb2}, but for LB1 of Scan 3\_03.}
    \label{Fig:03may_004_comparacion_vv_lb2}
  \end{center}
\end{figure}

\begin{figure}[h] 
  \begin{center}
    \includegraphics[width=0.5\textwidth]{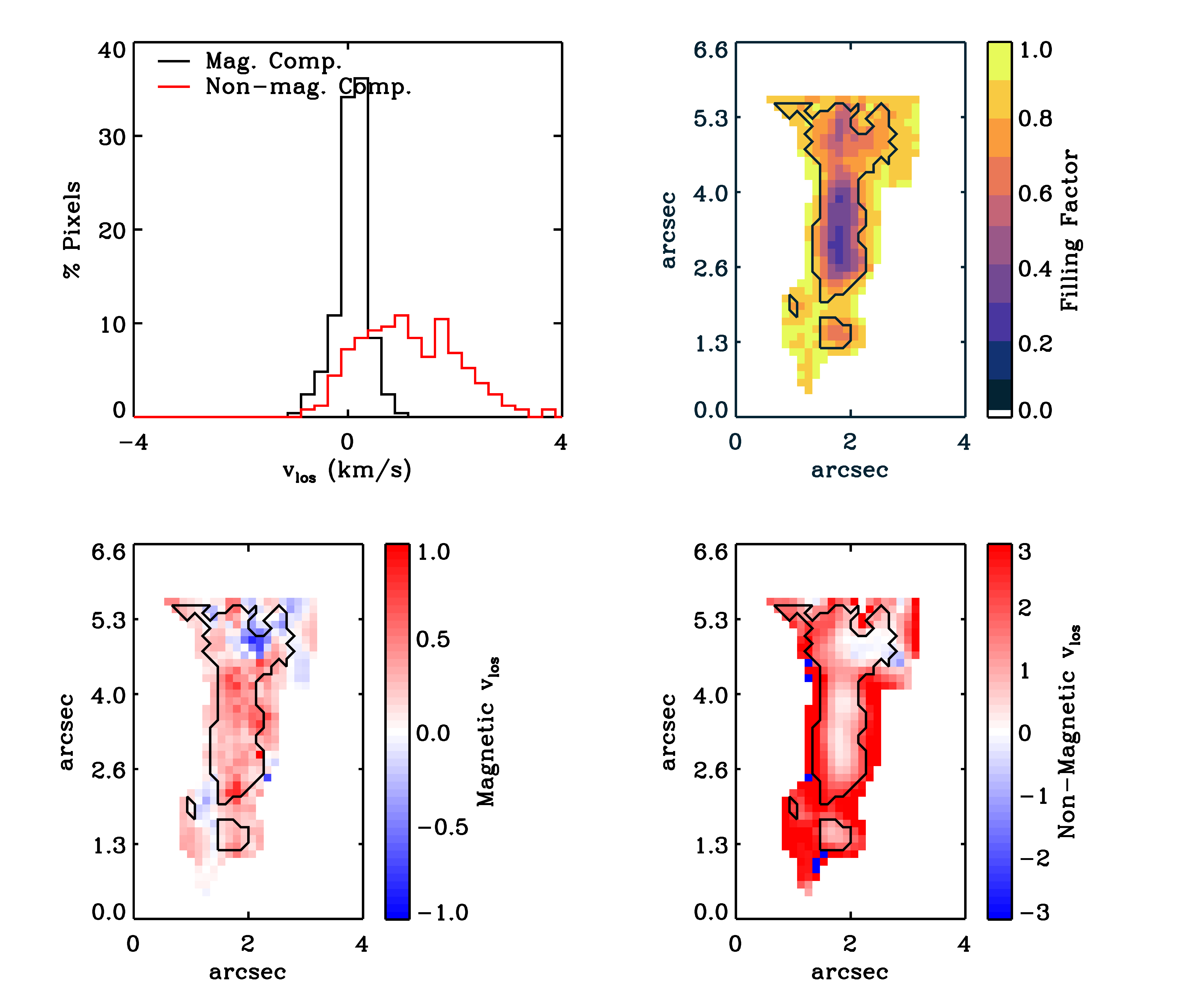}
    \caption{Same as Figure \ref{Fig:03may_012_comparacion_vv_lb2}, but for LB1 of Scan 3\_04.}
    \label{Fig:03may_007_comparacion_vv_lb2}
  \end{center}
\end{figure}

\begin{figure}[h] 
  \begin{center}
    \includegraphics[width=0.5\textwidth]{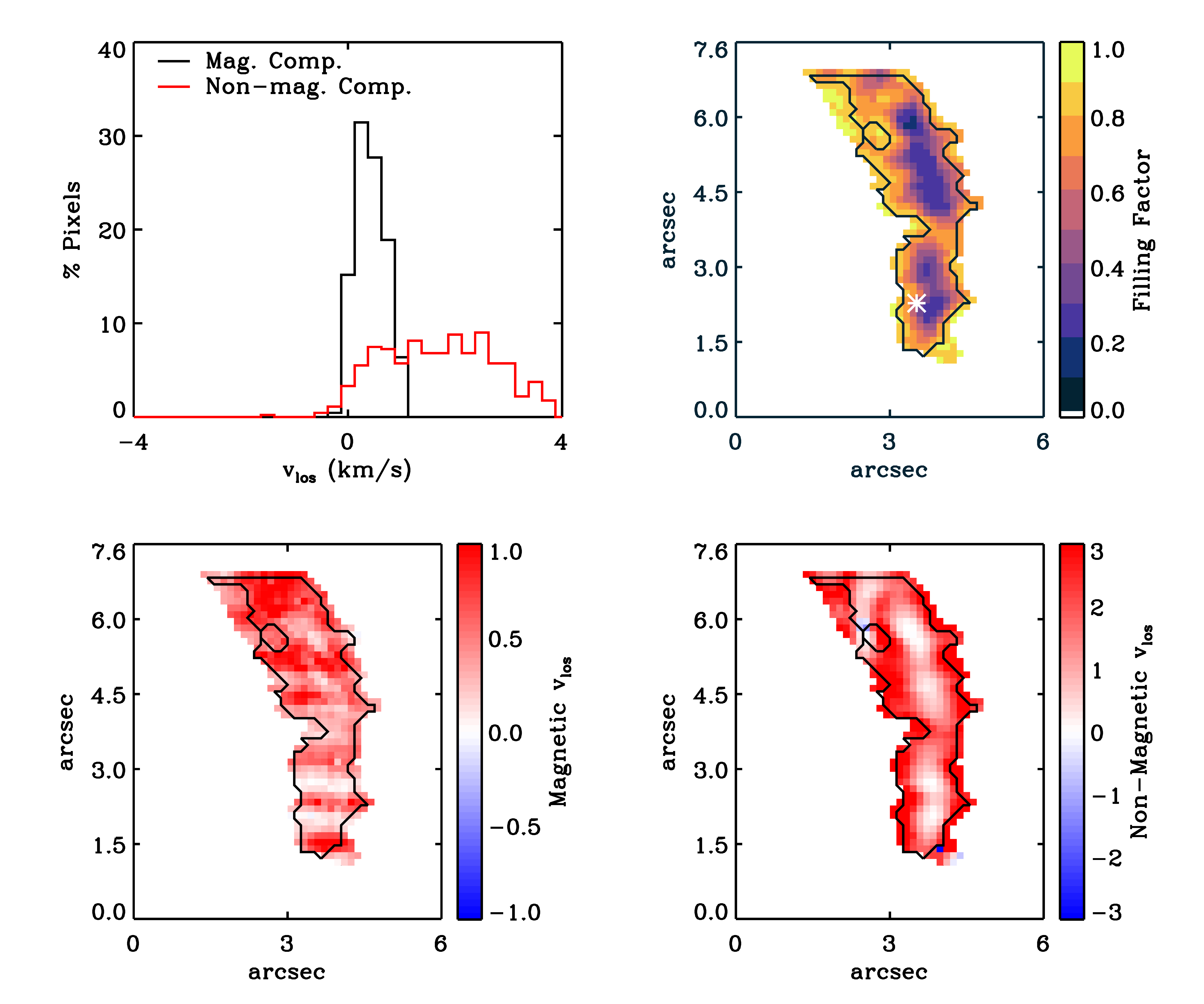}
    \caption{Same as Figure \ref{Fig:03may_012_comparacion_vv_lb2}.}
    \label{Fig:03may_012_comparacion_vv_lb2_appendix}
  \end{center}
\end{figure}

\begin{figure}[h] 
  \begin{center}
    \includegraphics[width=0.5\textwidth]{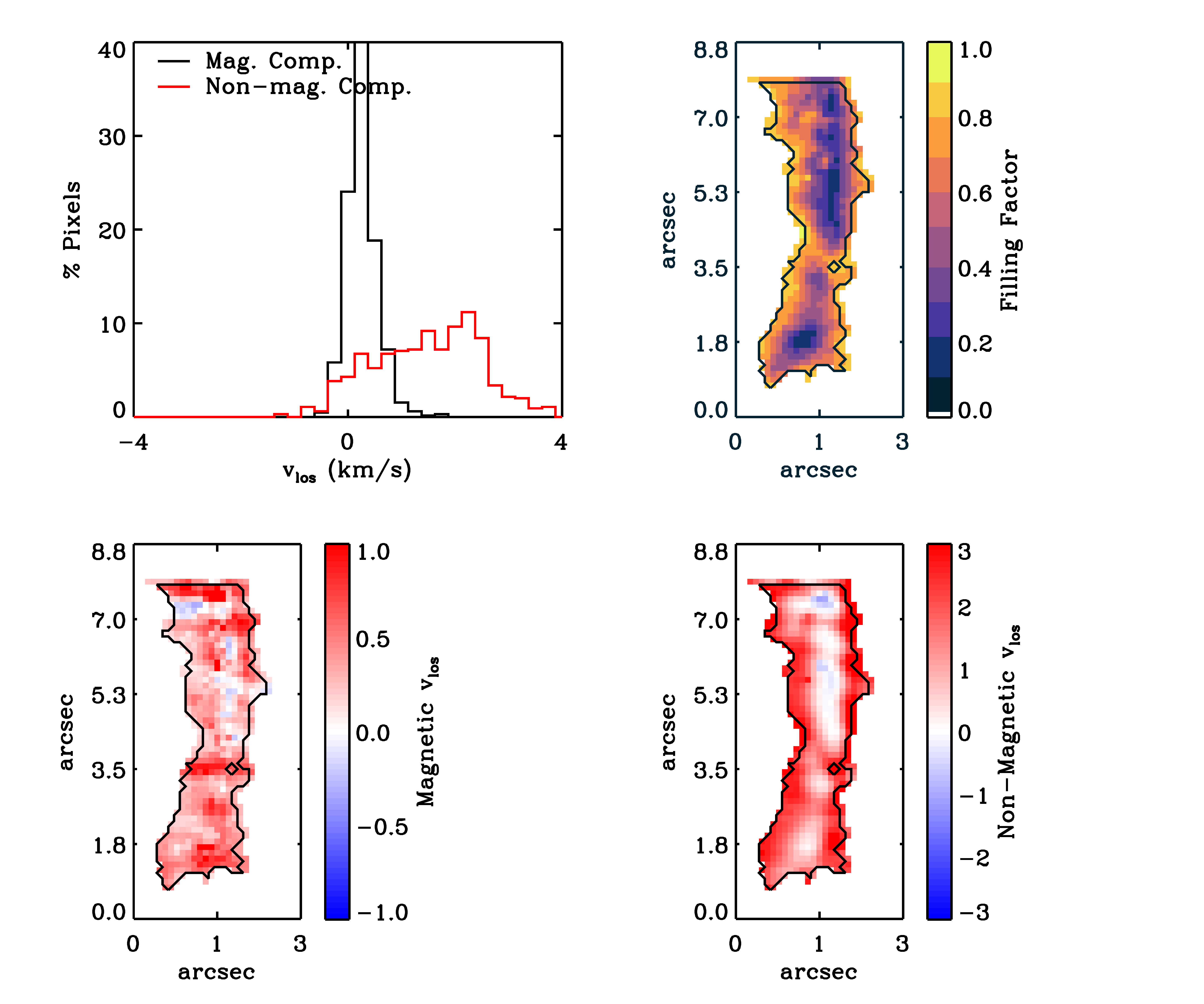}
    \caption{Same as Figure \ref{Fig:03may_012_comparacion_vv_lb2}, but for LB1 of Scan 3\_06.}
    \label{Fig:03may_027_comparacion_vv_lb2}
  \end{center}
\end{figure}

\begin{figure}[h] 
  \begin{center}
    \includegraphics[width=0.5\textwidth]{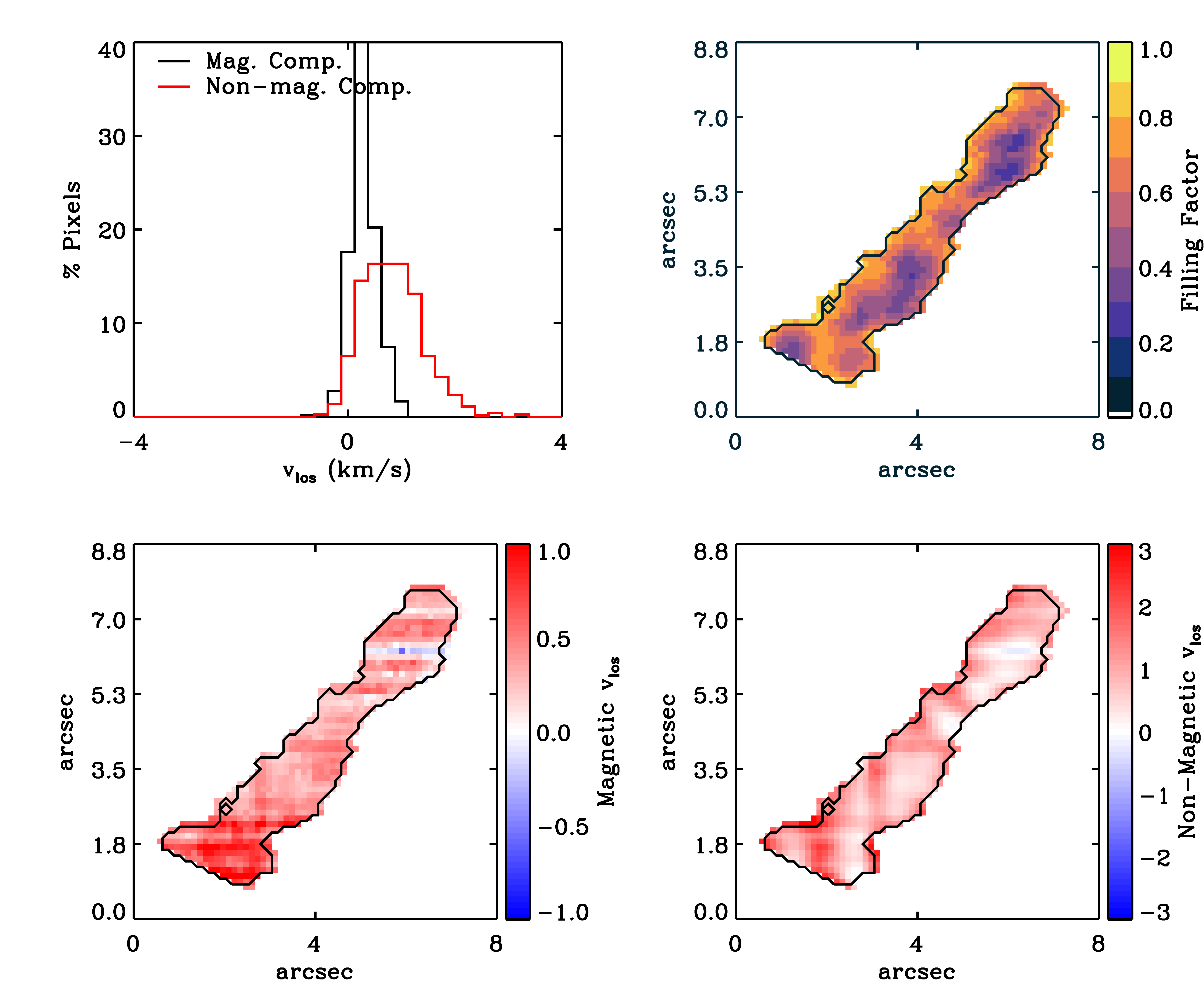}
    \caption{Same as Figure \ref{Fig:03may_012_comparacion_vv_lb2}, but for LB1 of Scan 5\_07.}
    \label{Fig:05may_002_comparacion_vv_lb2}
  \end{center}
\end{figure}

\clearpage

\subsection{Light Bridge 2}
\label{appA_lb2}

\begin{figure}[h] 
  \begin{center}
    \includegraphics[width=0.5\textwidth]{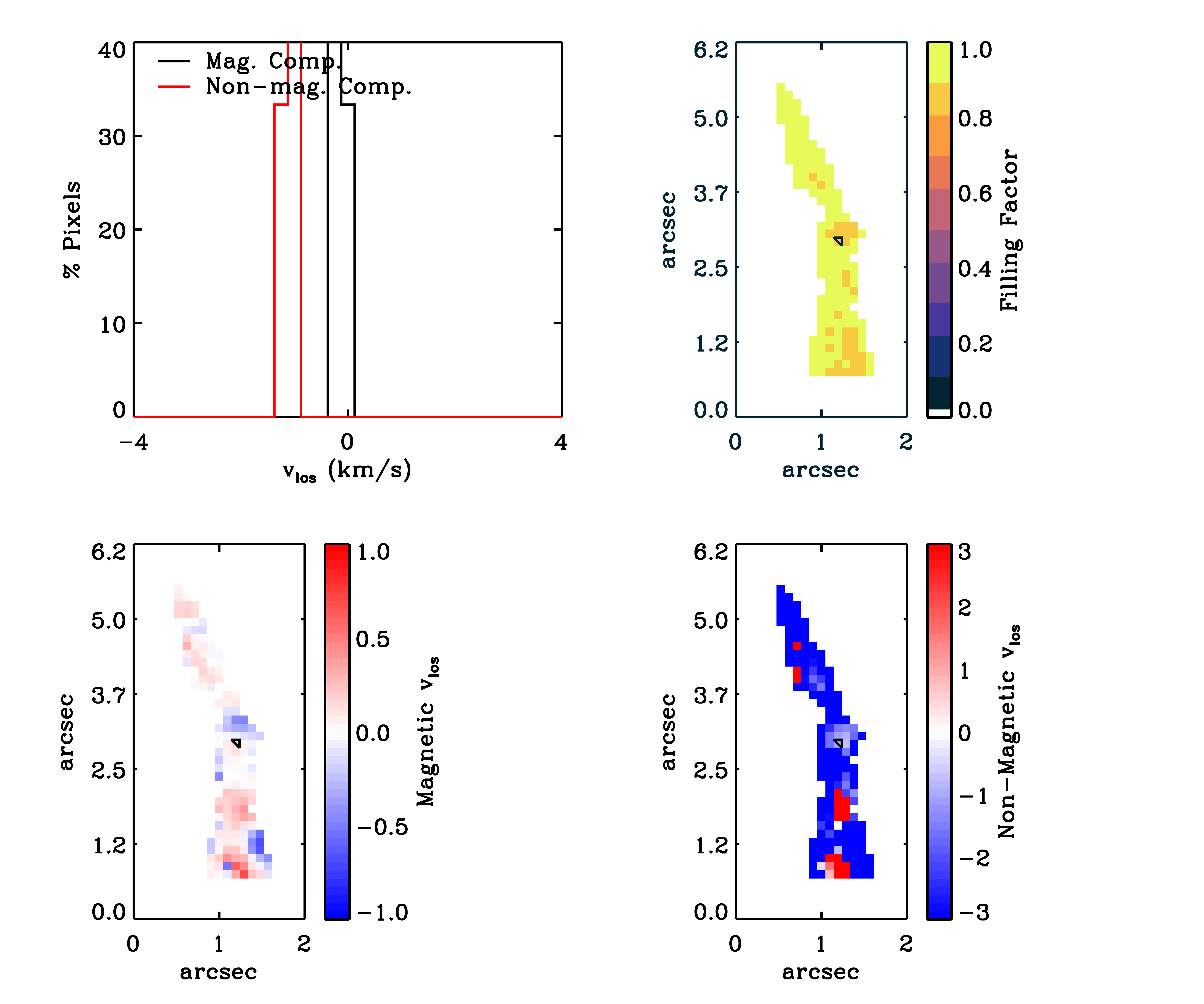}
    \caption{Same as Figure \ref{Fig:03may_012_comparacion_vv_lb2}, but for LB2 of Scan 2\_02.}
    \label{Fig:02may_017_comparacion_vv_lb3}
  \end{center}
\end{figure}

\begin{figure}[h] 
  \begin{center}
    \includegraphics[width=0.5\textwidth]{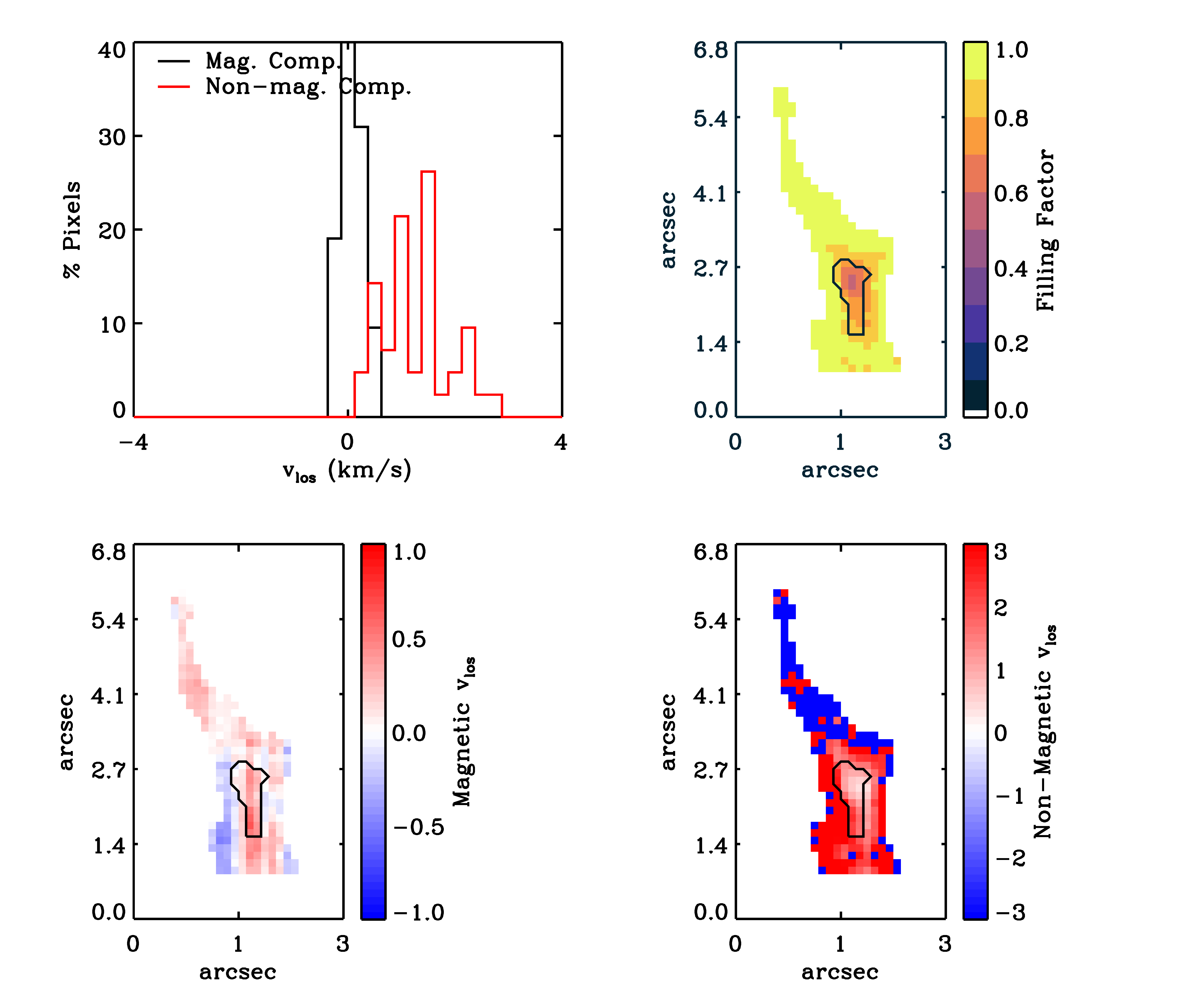}
    \caption{Same as Figure \ref{Fig:03may_012_comparacion_vv_lb2}, but for LB2 of Scan 3\_04.}
    \label{Fig:03may_007_comparacion_vv_lb3}
  \end{center}
\end{figure}

\begin{figure}[h] 
  \begin{center}
    \includegraphics[width=0.5\textwidth]{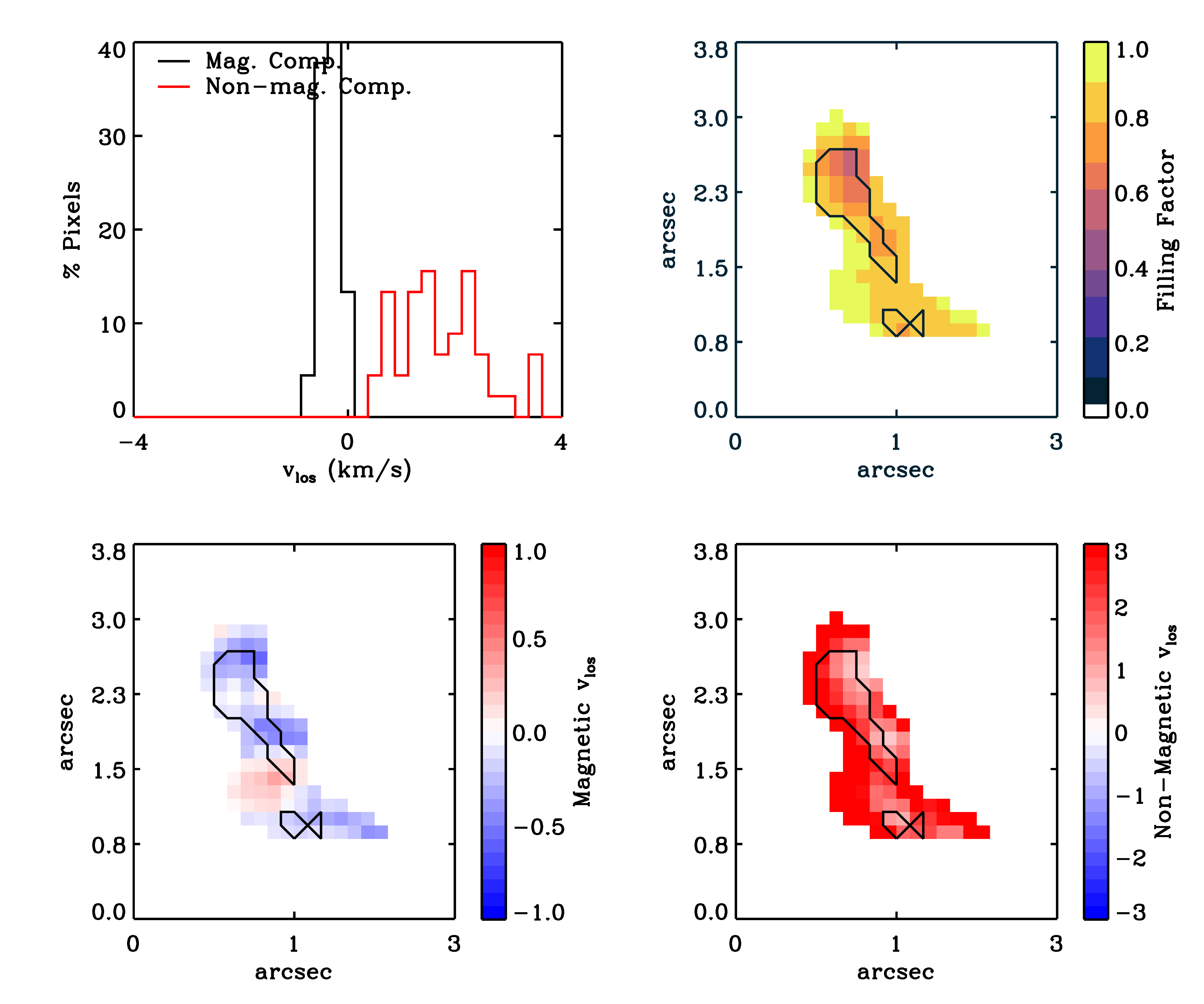}
    \caption{Same as Figure \ref{Fig:03may_012_comparacion_vv_lb2}, but for LB2 of Scan 3\_05.}
    \label{Fig:03may_012_comparacion_vv_lb3}
  \end{center}
\end{figure}

\begin{figure}[h] 
  \begin{center}
    \includegraphics[width=0.5\textwidth]{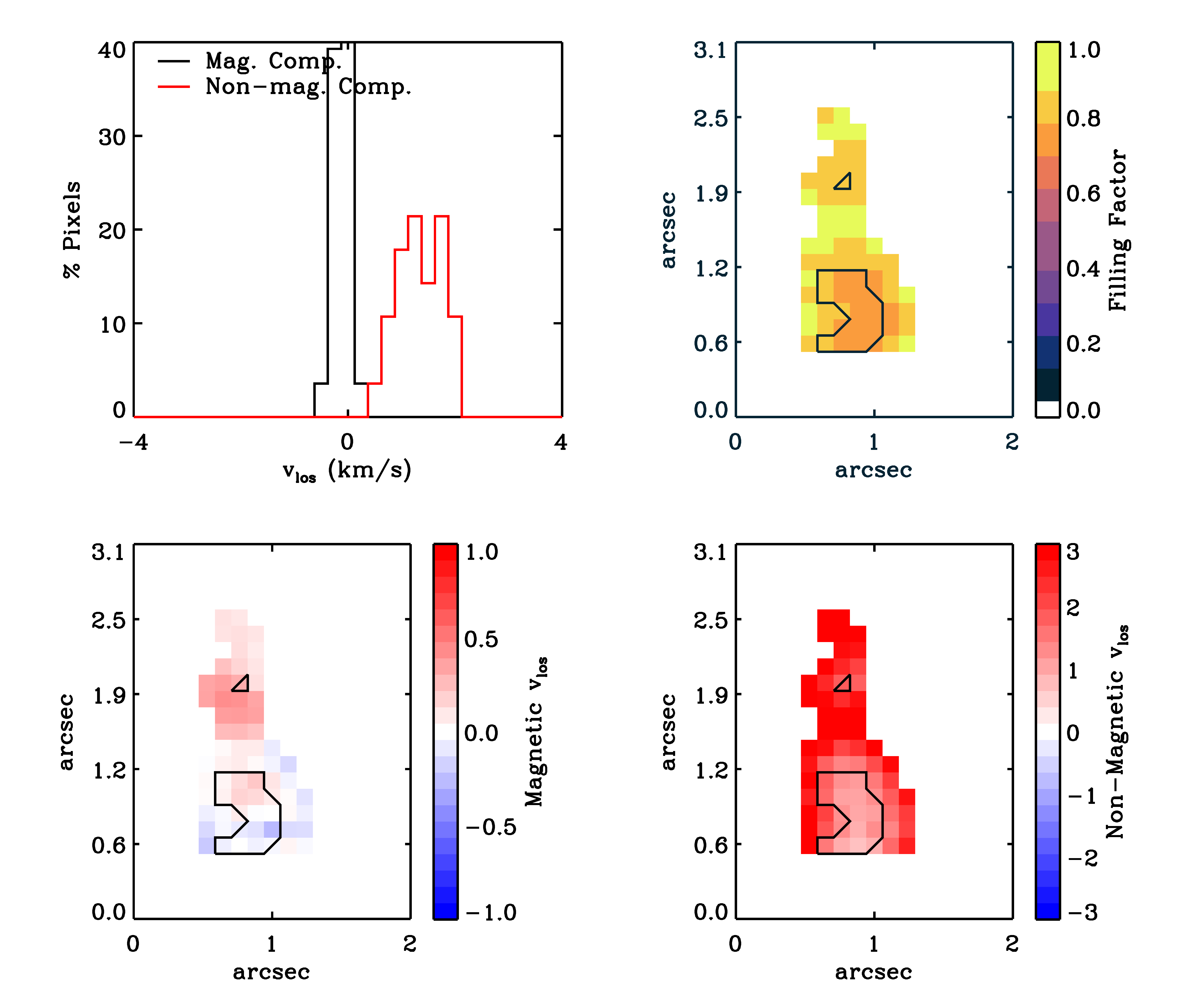}
    \caption{Same as Figure \ref{Fig:03may_012_comparacion_vv_lb2}, but for LB2 of Scan 3\_06.}
    \label{Fig:03may_027_comparacion_vv_lb3}
  \end{center}
\end{figure}

\clearpage

\subsection{Light Bridge 3}
\label{appA_lb3}

\begin{figure}[h] 
  \begin{center}
    \includegraphics[width=0.5\textwidth]{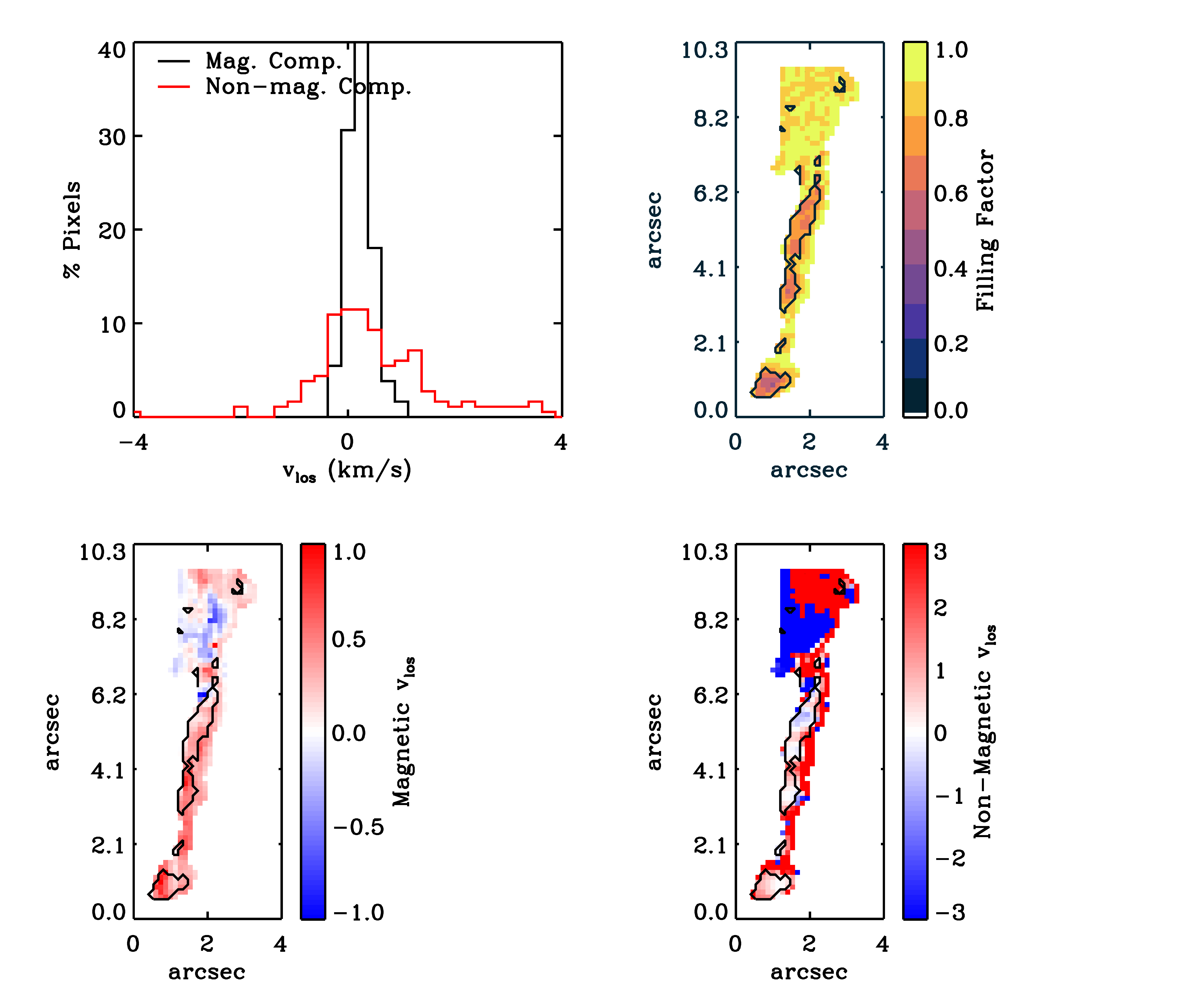}
    \caption{Same as Figure \ref{Fig:03may_012_comparacion_vv_lb2}, but for LB3 of Scan 2\_01.}
    \label{Fig:02may_001_comparacion_vv_lb1}
  \end{center}
\end{figure}

\begin{figure}[h]
  \begin{center}
    \includegraphics[width=0.5\textwidth]{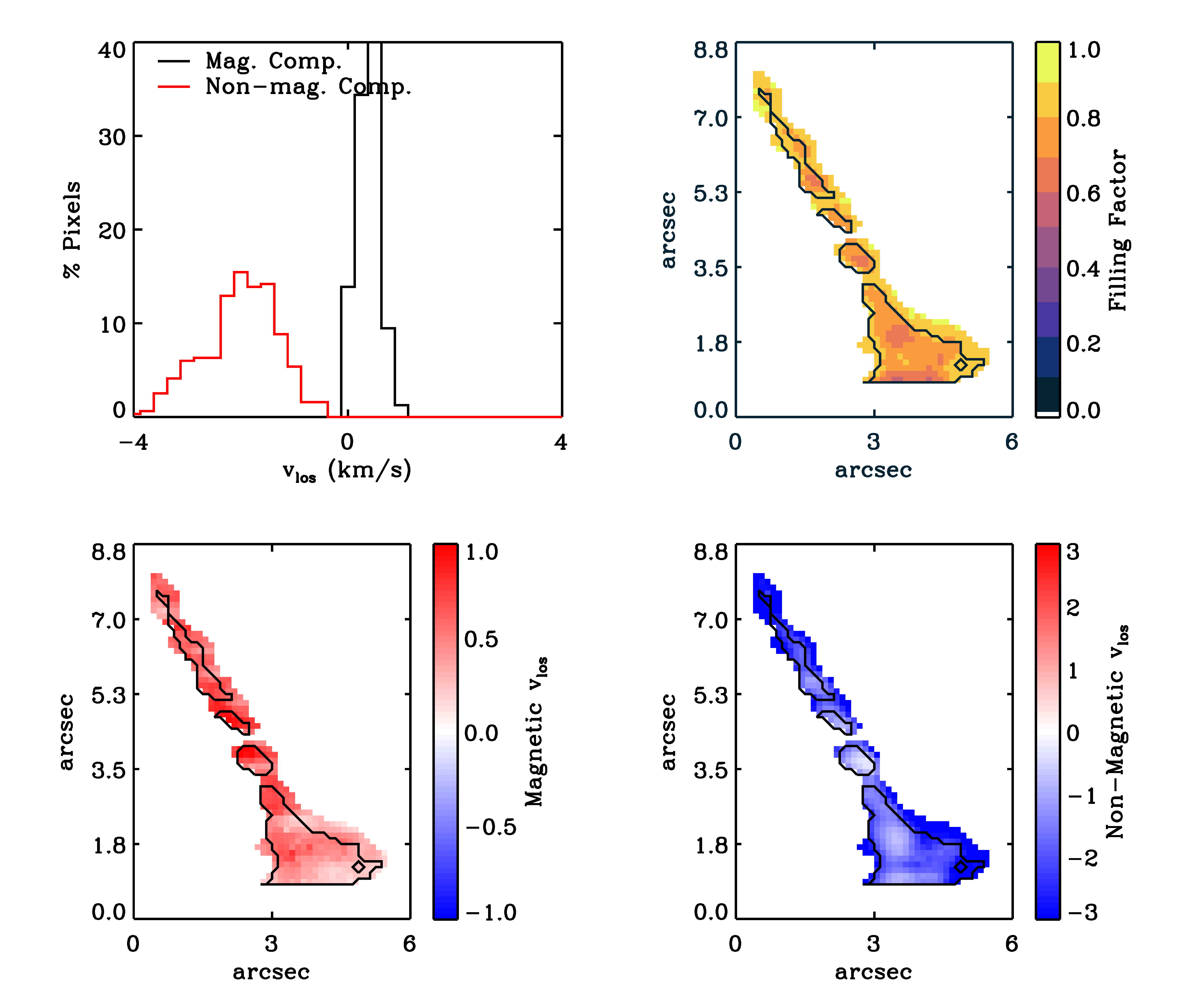}
    \caption{Same as Figure \ref{Fig:03may_012_comparacion_vv_lb2}, but for LB3 of Scan 2\_02.}
    \label{Fig:02may_017_comparacion_vv_lb1}
  \end{center}
\end{figure}

\begin{figure}[h] 
  \begin{center}
    \includegraphics[width=0.5\textwidth]{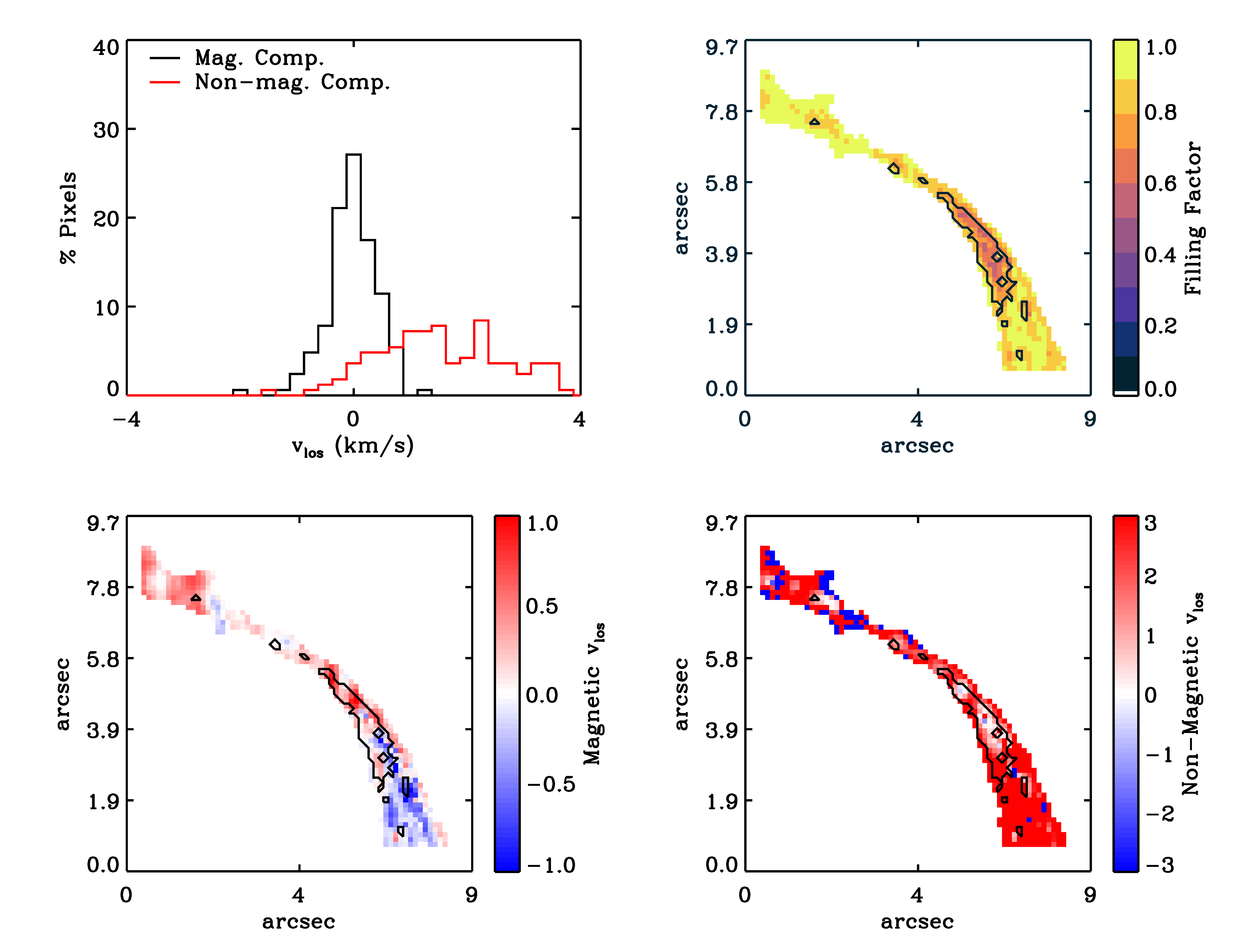}
    \caption{Same as Figure \ref{Fig:03may_012_comparacion_vv_lb2}, but for LB3 of Scan 3\_03.}
    \label{Fig:03may_004_comparacion_vv_lb1}
  \end{center}
\end{figure}

\begin{figure}[h]
  \begin{center}
    \includegraphics[width=0.5\textwidth]{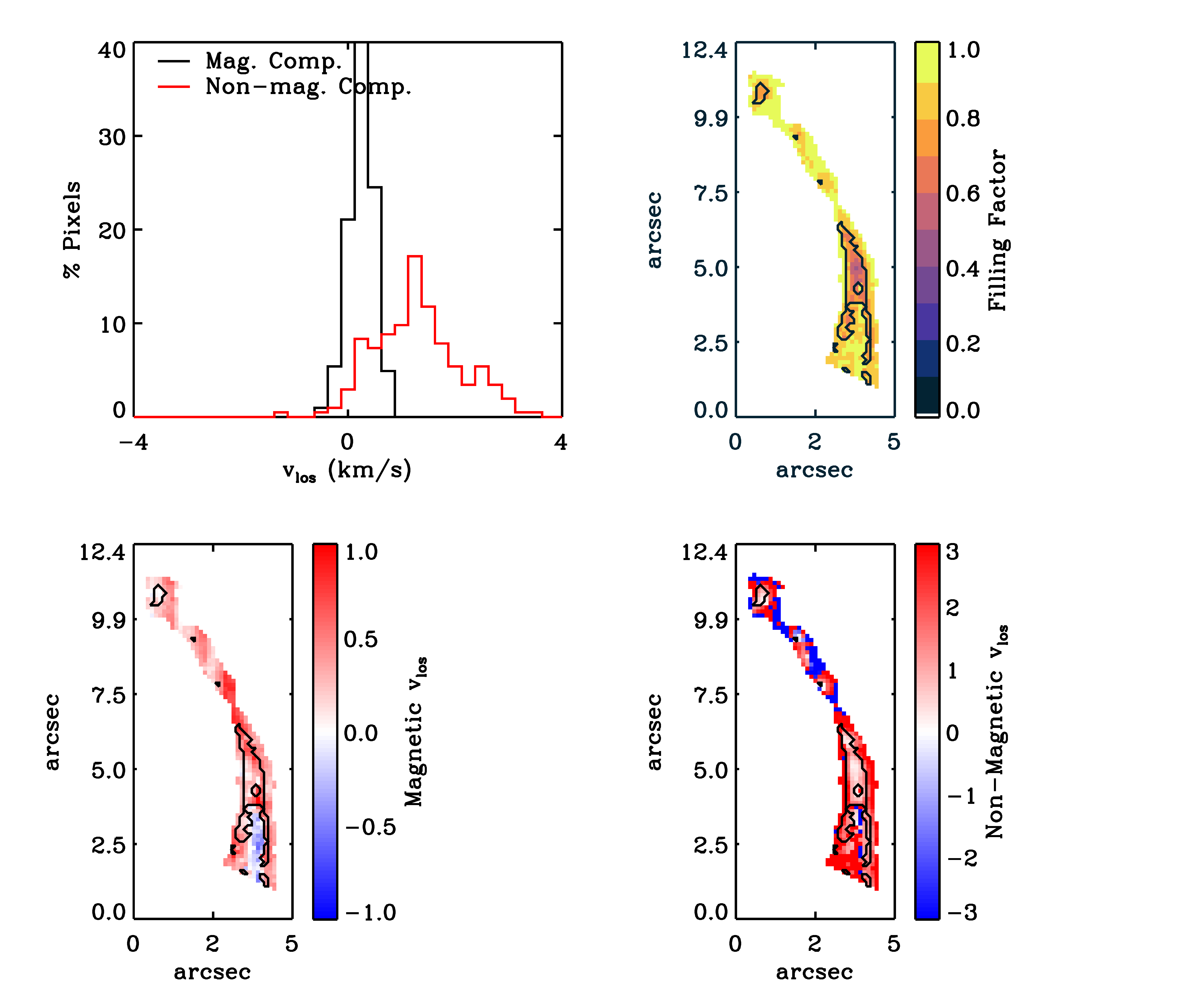}
    \caption{Same as Figure \ref{Fig:03may_012_comparacion_vv_lb2}, but for LB3 of Scan 3\_04.}
    \label{Fig:03may_007_comparacion_vv_lb1}
  \end{center}
\end{figure}

\begin{figure}[h] 
  \begin{center}
    \includegraphics[width=0.5\textwidth]{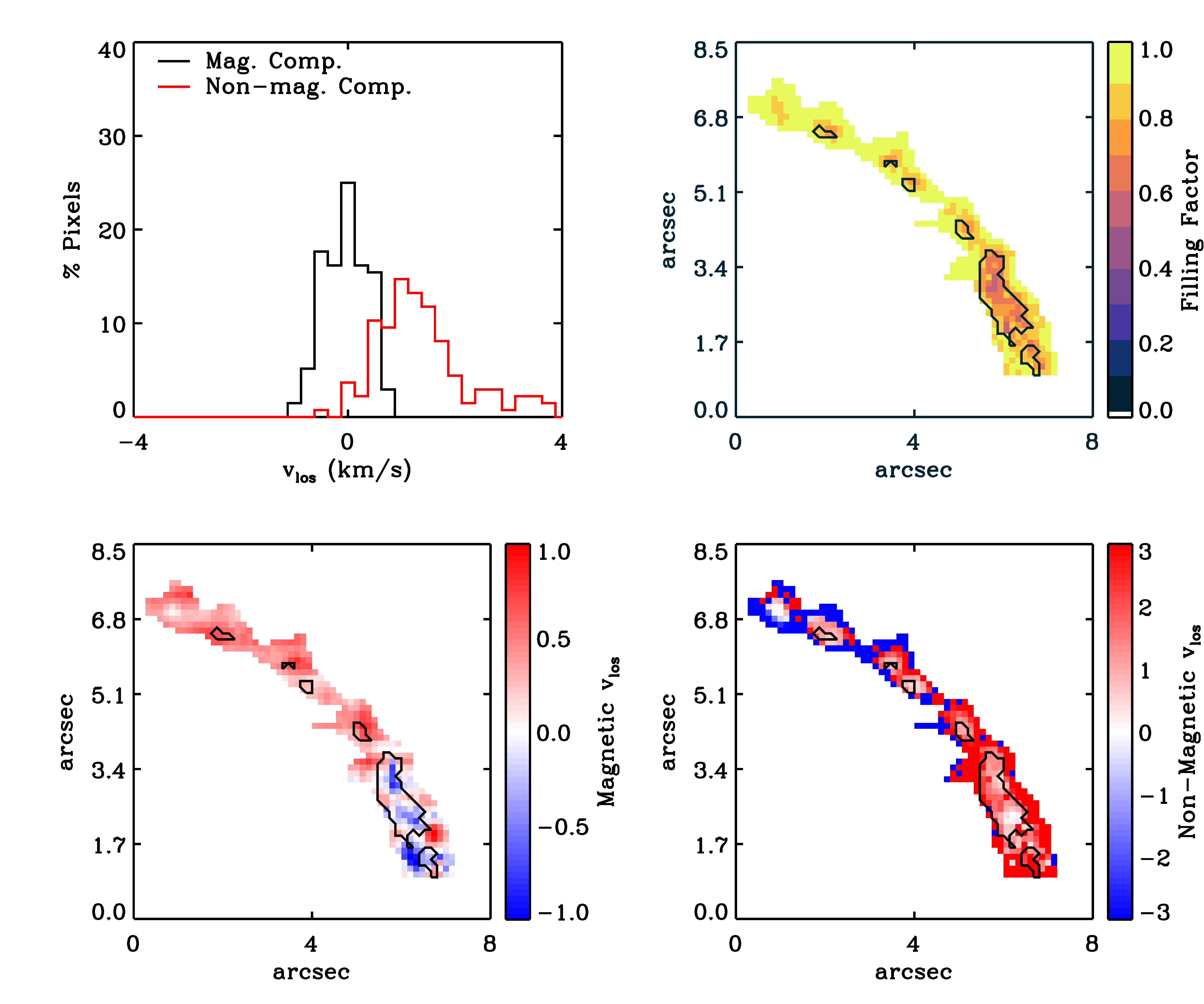}
    \caption{Same as Figure \ref{Fig:03may_012_comparacion_vv_lb2}, but for LB3 of Scan 3\_05.}
    \label{Fig:03may_012_comparacion_vv_lb1}
  \end{center}
\end{figure}

\begin{figure}[h] 
  \begin{center}
    \includegraphics[width=0.5\textwidth]{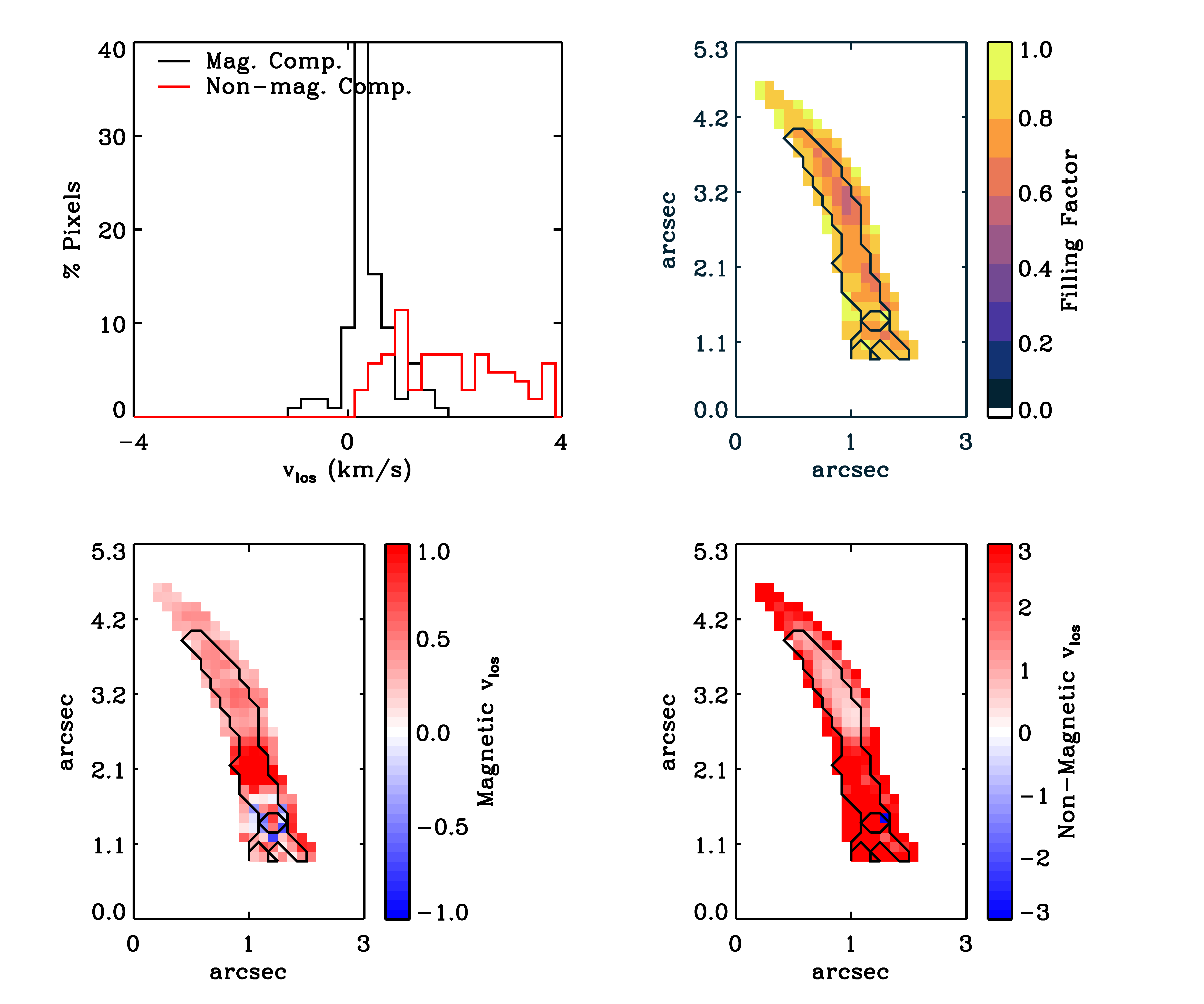}
    \caption{Same as Figure \ref{Fig:03may_012_comparacion_vv_lb2} but for LB3 of Scan 3\_06.}
    \label{Fig:03may_027_comparacion_vv_lb1}
  \end{center}
\end{figure}

\end{appendix}

\end{document}